\documentclass[]{interact}

\usepackage[caption=false]{subfig}

\usepackage{slashbox,multirow}

\usepackage[numbers,sort&compress]{natbib}
\bibpunct[, ]{[}{]}{,}{n}{,}{,}

\theoremstyle{plain}

\theoremstyle{definition}

\theoremstyle{remark}

\newcommand{\be}{\begin{equation}}
\newcommand{\ee}{\end{equation}}
\newcommand{\eq}[1]{Eq.~(\ref{#1})}
\newcommand{\fig}[1]{Fig.~\ref{#1}}
\def\bea{\begin{eqnarray}}
\def\eea{\end{eqnarray}}

\def\bra{\langle}
\def\ket{\rangle}
\def\vq{{\bf q}}

\def\vk{{\bf k}}
\def\vQ{{\bf Q}}
\def\vr{{\bf r}}
\def\qp{{\bf q}_{\parallel}}

\begin{document}

\articletype{REVIEW}

\title{Beyond on-site Hubbard interaction in charge dynamics of cuprate superconductors}

\author{
\name{Hiroyuki Yamase\textsuperscript{a}\thanks{CONTACT H. Yamase. Email: yamase.hiroyuki@nims.go.jp}}
\affil{\textsuperscript{a}Research Center of Materials Nanoarchitectonics (MANA),  
National Institute for Materials Science (NIMS), Tsukuba 305-0047, Japan}
}

\maketitle

\begin{abstract}
The study of high-temperature cuprate superconductors has long been dominated by models that primarily focus on short-range electron-electron interactions, such as the two-dimensional Hubbard and $t$-$J$ models. However, new insights into the charge dynamics reveal the indispensable and often overlooked role of the long-range Coulomb interaction, a factor critically important due to the layered structure of these materials.

In this review, we first present compelling evidence from resonant inelastic x-ray scattering data that highlights the significance of the long-range Coulomb interaction in cuprate charge dynamics, particularly around the in-plane momentum $\qp=(0,0)$. We show that these experimental observations are well-captured by the layered $t$-$J$-$V$ model, which extends the standard $t$-$J$ framework to include the long-range Coulomb interaction $V$ and the layered structure. 

This new perspective elucidates how charge dynamics renormalizes one-particle excitation properties, leading to several profound and often counterintuitive consequences. We demonstrate that the electron dispersion does not exhibit a sharp kink, and Landau quasiparticles persist in the low-energy limit despite a significant suppression of their spectral weight. We further show that while charge fluctuations alone cannot fully account for the pseudogap, they are a crucial component for understanding its formation. Additionally, we reveal that optical plasmon excitations generate fermionic quasiparticles, known as plasmarons, which give rise to a distinct, incoherent replica band.

We argue that accurately describing these plasmonic effects requires a three-dimensional theoretical approach. This perspective on plasmon excitations may offer a critically new clue to a long-standing puzzle: why multi-layer cuprate superconductors, containing more than two CuO$_{2}$ layers per unit cell,  consistently exhibit a higher critical temperature $T_{c}$ than their single-layer counterparts. Finally, we review the spin-fluctuation mechanism of superconductivity suffers from the {\it self-restraint effect} and show how important the screened Coulomb interaction is in the spin-fluctuation mechanism to realize high-$T_{c}$ superconductivity. 
\end{abstract}


\maketitle

\section{Introduction}
Since the discovery of high-temperature (high-$T_{c}$) cuprate superconductors \cite{bednorz86}, our understanding of these fascinating materials has been guided by models that emphasize short-range electron-electron interactions such as two-dimensional Hubbard and $t$-$J$ models \cite{anderson87}. These materials possess a layered structure, where ${\rm CuO_{2}}$ planes and the so-called block layers are stacked along the $c$ axis. The relevant electronic states are believed to be confined within the CuO$_{2}$ planes. At the heart of this physics are Cu$^{2+}$ ions, which have $(3d)^{9}$ electronic state with an unpaired electron with spin 1/2, while ${\rm O^{2-}}$ ions are fully occupied. The unpaired $3d$ electrons interact through the superexchange spin interaction $J$, which is mediated by the oxygen ions and leads to antiferromagnetic long-range order in the parent compound. Hence it is important to note that the functional form of $J$ is more complex \cite{matsukawa89} than the well-known form such as $J= 4 t^{2}/U$, where $U$ is the on-site Coulomb interactions  and $t$ is the hopping integral between nearest-neighbor sites. The parent compound is therefore  described as a charge-transfer-type antiferromagnetic Mott insulator  \cite{zaanen85}

Superconductivity emerges upon carrier doping, for example, by substituting atoms in the block layers. In the hole-doped case, the doped holes preferentially enter the oxygen sites and form a singlet state with the Cu spins, a composite particle known as the Zhang-Rice singlet \cite{fczhang88}. In this way, the oxygen degrees of freedom are integrated out, and the electronic state in the CuO$_{2}$ planes can be described by a single-band model such as the two-dimensional Hubbard and $t$-$J$ models on a square lattice. In the electron-doped case, doped electrons enter the Cu sites and the concept of the Zhang-Rice singlet is not necessary, yet the electronic state can still be described by the same fundamental models. 

\begin{figure}[ht]
\centering
\includegraphics[width=7cm]{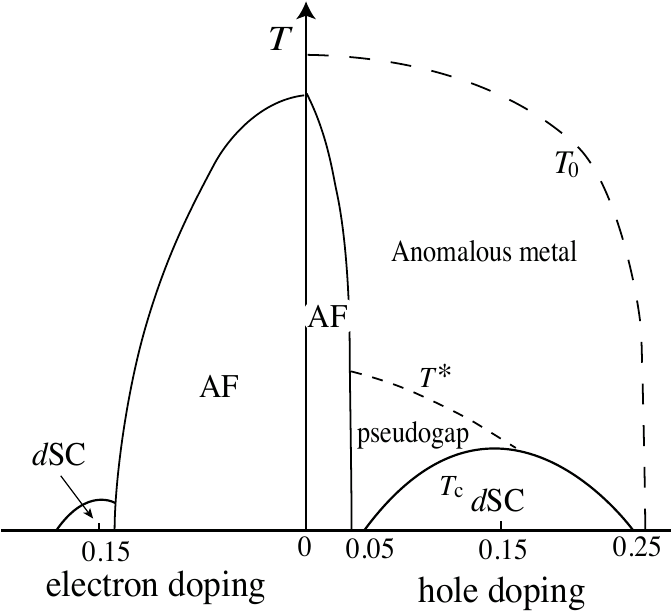}
\caption{Schematic phase diagram of the high-$T_{c}$ cuprate superconductors. Both hole and electron doping are possible. At half-filling, the system is an antiferromagnetic (AF) Mott insulator. The antiferromagnetism is suppressed by carrier doping, leading to a $d$-wave superconducting ($d$SC) state. On the hole-doped side, the pseudogap phase is realized below the temperature $T^{*}$, where a gaplike feature is observed even above the superconducting onset  temperature $T_{c}$. In a wide doping region, an anomalous metallic phase is realized, where the standard Fermi-liquid description does not hold. The actual phase diagram is more complicated as shown in Refs.~\cite{keimer15} and \cite{jjwen19}
}
\label{exp-phase}
\end{figure}

Figure~\ref{exp-phase} is a schematic phase diagram of high-$T_{c}$ cuprate superconductors. The diagram includes the antiferromagnetic Mott insulating state at zero doping, the $d$-wave superconducting phase that appears close to it, and the pseudogap phase, especially prominent on the hole-doped side. In the pseudogap phase, gap-like features are observed at temperatures far above the onset temperature ($T_{c}$) of superconductivity. Furthermore, a broad region of the phase diagram is occupied by an anomalous metallic phase, where  conventional metallic theories cannot be applied, with a prime example being the $T$-linear resistivity \cite{patel19,grissonnanche21,phillips22}. 

A significant body of work has been done for these single-band models to provide a conceptual understanding of this phase diagram. Early studies using slave-boson mean-field theory of the $t$-$J$ model successfully captured the qualitative feature of the phase diagram \cite{kotliar88a,fukuyama98a}. 
Fluctuations around the mean-field state were then investigated using random phase approximation (RPA), U(1), and SU(2) gauge theories \cite{lee06}. Together with various numerical techniques such as variational Monte Carlo \cite{yokoyama96,edegger07}, exact diagonalization (ED) \cite{dagotto94}, density-matrix renormalization group (DMRG) \cite{schollwoeck11}, and others \cite{foussats02}, we obtained a deep understanding of phenomena, including   $d$-wave superconductivity \cite{kotliar88a,suzumura88}, spin excitation spectra \cite{tanamoto94,brinckmann99,yamase01,brinckmann01,onufrieva02,li03,sega03,yuan05,yamase06,yamase07,yamase09,cpchen09,li16}, spin-charge stripe orders \cite{white98a,white99,white00}, orbital currents \cite{cappelluti99}, electronic nematic correlations \cite{yamase00a,yamase00b,miyanaga06,edegger06}, and particle-hole asymmetry of the phase diagram \cite{bejas12,bejas14}. The Hubbard model, which covers a higher energy region than the $t$-$J$ model, is believed to share essentially the same low-energy physics \cite{hybertsen90}.  Since superconductivity cannot be captured by a mean-field analysis in the Hubbard model, it was much later that the Hubbard model can also capture the above mentioned properties together with the superconductivity by developing dynamical mean-field theory (DMFT) \cite{georges96}, cellular dynamical mean-field theory (C-DMFT) \cite{maier05,tremblay06}, dynamical cluster approximation (DCA) \cite{maier05,tremblay06,leblanc15}, cluster perturbation theory \cite{maier05,tremblay06}, diagramatic extension of DMFT  \cite{rohringer18}, functional renormalization group theory \cite{metzner12,dupuis21},  and other sophisticated techniques \cite{leblanc15,schafer21,qin22}. The collective success of the $t$-$J$ and Hubbard models strongly implies that short-range electron-electron interactions such as on-site Coulomb interaction $U$ and the nearest-neighbor spin superexchange interaction $J$ as well as two dimensionality are  crucial for understanding high-$T_{c}$ superconductivity. 

Nevertheless, a complete consensus on the high-$T_{c}$ mechanism remains elusive. It is worth noting the recent achievement of superconductivity at 200--260 K in metal hydrides under high pressures 150--200 GPa \cite{drozdov15,drozdov19,somayazulu19}. The observation of an isotope effect in these materials \cite{drozdov15,drozdov19} suggests that they are driven by the conventional electron-phonon coupling mechanism, a distinct  category from cuprates \cite{ashcroft68,ashcroft04}. As of today, a scientific breakthrough for achieving room-temperature superconductors under ambient pressure, which would lead to technological revolution, has not yet been obtained. 

The large on-site Coulomb interaction $U$ and the resulting spin exchange interaction $J$ both originate from the long-range Coulomb interaction (LRC), which in continuum space behaves as $1/r$, where $r$ is the distance between  two electrons. Inside a material, however, the LRC is screened, typically decaying exponentially: $\frac{1}{r}{\rm e }^{-\kappa r}$, where $\kappa$ is known as Thomas-Fermi wavevector. This screening has often been cited to justify the use of effective  short-range interaction models, such as the Hubbard model, and may be a major reason why the effect of the LRC has not been studied extensively until recently. In addition, the strong singularity of the LRC in momentum space at $\vq={\bf 0}$, where it behave as $1/\vq^{2}$, is known to give rise to the optical plasmon, whose energy in cuprate is around $\sim 1$ eV \cite{uchida91}. This fact has also been used to justify to consider the LRC irrelevant to low-energy physics less than the superexchange interaction $J$, which is about 100-150 meV \cite{bourges97b}.

However, a key insight that has not been widely recognized in the high $T_{c}$  community is that for layered materials \cite{greco16}, plasmon excitations can form the so-called {\it plasmon band} as a function of the out-of-plane momentum $q_{z}$ around in-plane momentum $\qp=(0,0)$ (see Sec.~\ref{exp-LRC}). This band can even exhibit a gapless mode at  $\qp=(0,0)$ for a finite $q_{z}$ if the interlayer hoping integral is negligible. Hence, the plasmon from the LRC can be relevant to the low-energy physics. Furthermore, plasmon excitations renormalize the band structure and can generate new bands (see Sec.~\ref{section-self-energy}). 

Moreover, while the Coulomb interaction is the source of the spin exchange interaction between  nearest neighbor sites, it may also retain the sizable nearest-neighbor Coulomb interaction that can suppress superconductivity \cite{yamase23}.  These phenomena represent physics beyond the standard Hubbard and $t$-$J$ models and are characteristic features of the Coulomb interaction that have not been adequately explored in the history of high-$T_{c}$ cuprates.

In this review we highlight the importance of Coulomb interaction beyond the on-site Hubbard interaction for the understanding of the charge dynamics in high-$T_{c}$ cuprate superconductors. We first summarize in Sec.~2 the experimental evidence of the importance of LRC. This establishment was achieved through  quantitative comparisons with theory. Hence, in Sec.~2.1 we state what was a theoretical challenge to analyze the LRC. Sec.~2.2 explains the current situation about where experimental and theoretical studies are directed.  Having established that the charge dynamics of the $t$-$J$-$V$ model captures the resonant inelastic x-ray scattering (RIXS) data quantitatively, we next study in Sec.~3 how electrons gain the self-energy from charge fluctuations. To do so, we first review in Sec.~3.1 the leading-order large-$N$ formalism in the $t$-$J$-$V$ model in the case that the unit cell contains one CuO$_2$ plane. To compute the electron self-energy, we go into next-leading order, whose outcome is summarized in Sec.~3.2. The self-energy effects are presented in two parts: mainly high-energy part in Sec.~3.3 and low-energy part close to the Fermi energy in Sec.~3.4. In particular, the pseudogap issue is analyzed separately in Sec.~3.5. Finally, we comment a role of spin fluctuations in Sec.~3.6. In Sec.~4, we briefly mention that there is also bond-charge fluctuations from the nearest-neighbor {\it spin} exchange, leading to the dual structure of the charge excitation spectrum in momentum-energy space. As explained in Sec.~2.2, RIXS experiments are directed toward multilayer cuprates. However, possible application of the large-$N$ theory (in Sec.~3.1) to multilayer cuprates is not straightforward. We therefore formulate the bilayer lattice electron model in Sec.~5.1. A highly non-trivial aspect is the functional form of the LRC on the bilayer lattice, which is derived in Sec.~5.2. The analytical expression of the dynamical charge susceptibility is studied in Sec.~5.3, with emphasis not to mix up the concept of the even and odd modes in the bilayer electron systems. The charge dynamics is clarified in Sec.~5.4 through numerical calculations. Obtained results are compared with RIXS data for Y-based cuprates in Sec.~5.5. To highlight the specific feature of the LRC, the charge/spin susceptibility from nearest-neighbor interaction is derived in Sec.~5.6, where the concept of even and odd modes is valid. 

In cuprates spin fluctuations are the dominant source of superconductivity. However, we obtain the insight that there is the {\it self-restraint effect} of superconductivity from spin fluctuations in general and consequently the instantaneous spin interaction specific to cuprates does matter to superconductivity. Since the Coulomb interaction is also instantaneous, we recognize the important physics of the interplay of spin and nearest-neighbor Coulomb interaction. To explain this, we describe the model and formalism in Sec.~6.1. and present results in Sec.~6.2. 

In this review, the large-$N$ theory of the $t$-$J$ model and Eliashberg theory are employed. We apply those theories to cuprates, but emphasize that in Sec.~7 the theories are not limited to cuprates, but more general. In Sec.~8, given that the cuprate research is still going on, we present perspectives rather than conclusions.

\section{Evidence of the importance of LRC} \label{exp-LRC}
In continuum space, two particles at a distance $r$ interact with the LRC, which depends on $1/r$. The Fourier transform of this interaction to momentum space yields a strong singularity at $\vq={\bf 0}$, where its behavior is proportional to 
 $1/\vq^{2}$ in three dimensions. This singularity is the fundamental origin of collective charge excitations, known as optical plasmons. Early reports of optical plasmons in high-$T_{c}$ cuprate superconductors were made around 1990 using electron-energy loss spectroscopy (EELS) \cite{nuecker89,romberg90}. 

Around the same time, spin excitations were a major focus on research, with their spectral weight extensively mapped in momentum $\vq$ and energy $\omega$ space through neutron scattering experiments \cite{birgeneau06,fujita12}. The corresponding charge excitation spectrum was studied much later, following the  advent of resonant x-ray scattering (RXS) and RIXS \cite{ament11,degroot24}. An early map of charge excitations in $\vq$-$\omega$ space was reported by Ref.~\cite{ishii05}, which revealed a very broad feature developing roughy from $(\vq_{\parallel}, \omega)=({\bf 0},0)$ to high energy as the in-plane momentum transfer $\vq_{\parallel}$ was increased. 

\begin{figure}
\centering
\includegraphics[width=13cm]{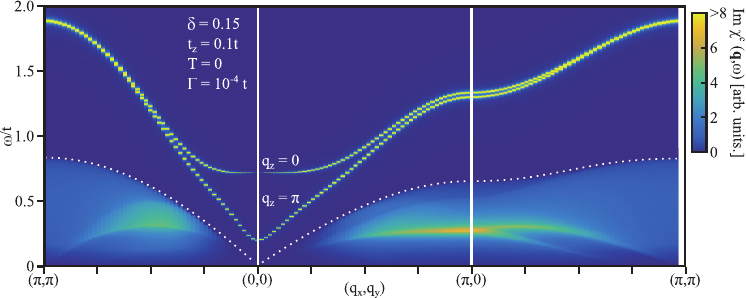}
\caption{Typical charge excitation spectrum along the symmetry axes for $q_z=0$ and $\pi$ computed in the large-$N$ theory of the layered $t$-$J$-$V$ model for electron-doping rate $\delta=0.15$ at zero temperature; the interlayer hopping integral is taken as $t_z=0.1t$. The dotted line denotes the upper boundary of a particle-hole continuum for $q_z=0$. Adapted from Ref.~\cite{greco16}, where the superexchang interaction $J/t=0.3$ and next nearest-neighbor hopping $t'/t=0.3$, and broadening parameter $\Gamma/t=10^{-4}$ were used (\copyright\, 2016 American Physical Society).
}
\label{plasmon1}
\end{figure}

\begin{figure}
\centering
\includegraphics[width=7cm]{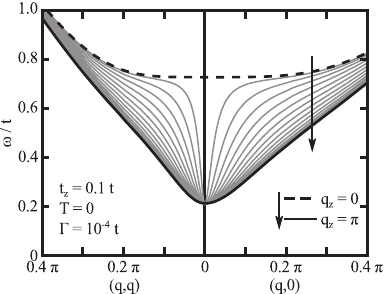}
\caption{Plasmon band characterized by a value of $q_{z}$ around $\qp=(0,0)$. Adapted from Ref.~\cite{greco16}, where the superexchang interaction $J/t=0.3$ and next nearest-neighbor hopping $t'/t=0.3$, and broadening parameter $\Gamma/t=10^{-4}$ were used (\copyright\, 2016 American Physical Society). 
}
\label{plasmon2}
\end{figure}

In 2014, a similar charge excitations spectrum was reported by different groups \cite{ishii14,wslee14}. In typical electron-doped cuprates like ${\rm Nd_{2-x}Ce_{x}CuO_{4}}$ (NCCO), it showed a distinct V-shaped dispersion with a gap at $\qp=(0,0)$. The origin of this signal became a subject of controversy. Refs.~\cite{ishii14,ishii17}, the same group that first reported the charge signals around $(\qp, \omega)=({\bf 0},0)$ \cite{ishii05}, confirmed their original interpretation that the signal originated from  incoherent particle-hole excitations and could not be classified as a collective mode. In contrast, Refs.~\cite{wslee14,dellea17} suggested it was a new collective mode stemming  from a hidden quantum critical point specific to electron-doped cuprates, because of their unsuccessful detection of a similar signal in hole-doped cuprates. This debate was resolved from a theoretical standpoint when Greco {\it et al.} included the LRC in the $t$-$J$ model---we refer to it as the layered $t$-$J$-$V$ model \cite{greco16}. By employing a large-$N$ technique, they demonstrated the existence of not only a well-known plasmon mode at $q_{z}=0$, but also acousticlike plasmon mode for $q_{z}\neq 0$ that collectively forms a V-shaped dispersion around $\qp=(0,0)$ as shown in \fig{plasmon1}. The former was linked to the plasmon mode observed by  EELS \cite{nuecker89,romberg90}, while the latter was identified with the charge mode under debate \cite{ishii14,wslee14,ishii17,dellea17}. The V-shaped dispersion exhibits a very strong $q_{z}$ dependence near $\qp=(0,0)$ as shown in \fig{plasmon2}, which is called as a plasmon band characterized by $q_{z}$ 

The V-shaped dispersion and plasmon bands were known long ago in the context of plasmons in layered electron gas model \cite{fetter74,grecu73,grecu75}. However, cuprates are strongly correlated electron system and moreover they are not modeled by an electron gas but an electron liquid. Hence, the further theoretical studies were evolved in terms of the $t$-$J$-$V$ model.

\begin{figure}[ht]
\centering
\includegraphics[width=13cm]{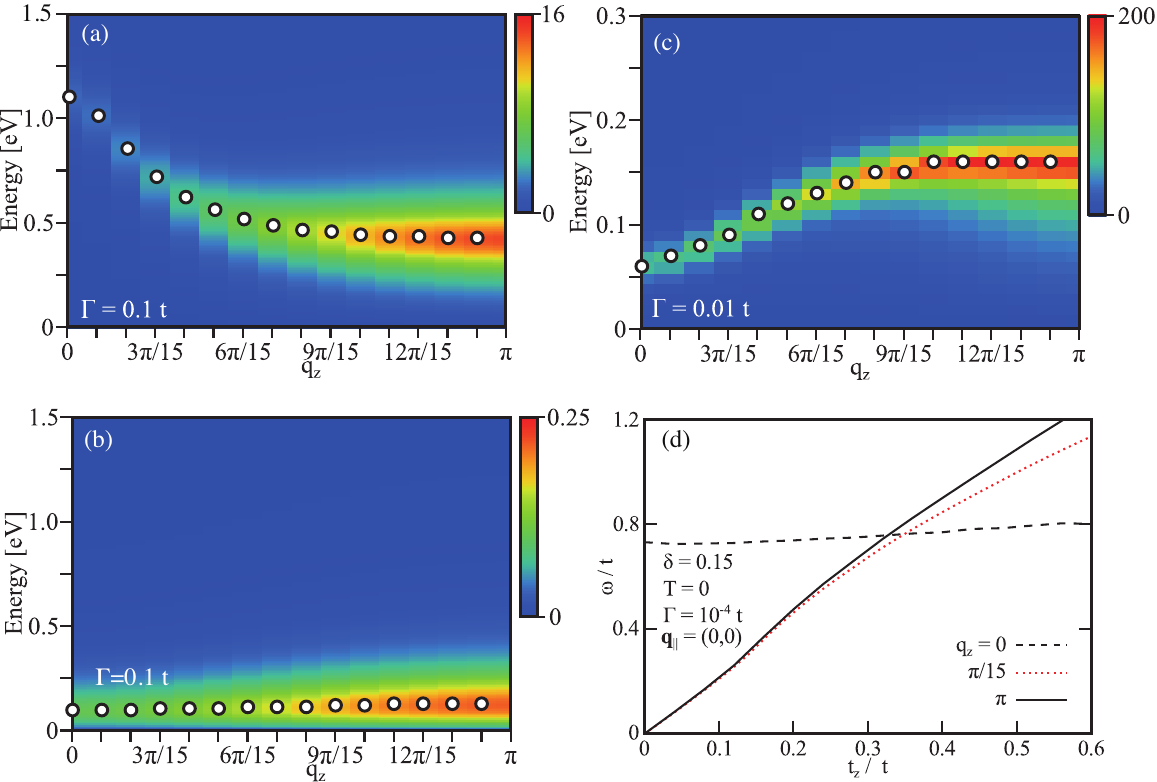}
\caption{(a) Plasmon scenario. Intensity map in the plane of $q_{z}$ and $\omega$. The open circles denotes the peak energy, which decreases with increasing $q_{z}$. (b) Incoherent particle-hole scenario. The $q_{z}$ dependence of the peak positions is negligible. (c) Short-range interaction scenario. At small $\qp=(0.02\pi, 0.02\pi)$, the peak  energy increases with $q_{z}$, the opposite dependence to the plasmon scenario (a).  (d) Plasmon energy as a function of $t_{z}$ for several choices of $q_{z}$. To define the peak position sharply, a broadening factor $\Gamma$ is taken to be smaller. The gap at $q_{z}=0$ corresponds to the optical plasmon energy. (a), (b), (c) are adapted from 
Ref.~\cite{greco19} [\copyright\, 2019 The Author(s)] 
and (d) from Ref.~\cite{greco16} (\copyright\, 2016 American Physical Society). 
}
\label{scenarios}
\end{figure}

The layered  $t$-$J$-$V$ model made a number of key predictions that were subsequently confirmed by experiments  \cite{greco19,greco20,nag20,yamase21c,hepting22,hepting23}. The four main predictions were (see \fig{scenarios}): i) The acousticlike plasmon energy should decrease with increasing $q_{z}$ at a small in-plane momentum $\qp$ [\fig{scenarios}(a)], contrary to the negligible $q_{z}$ dependence expected for incoherent particle-hole excitations argued in Refs.~\cite{ishii14,ishii17}  [\fig{scenarios}(b)]. ii) A short-range interaction model would yield a zero-sound mode whose energy increases with $q_{z}$ for sufficiently small $\qp$, showing the opposite dependence to the predicted plasmon behavior  [\fig{scenarios}(c)]. iii) The acousticlike plasmon energy at a finite $q_{z}$ should have a gap at $\qp=(0,0)$, which is proportional to the interlayer hopping integral $t_{z}$ [\fig{scenarios}(d)]. iv) These acousticlike plasmons should be present in both  electron-doped and hole-doped cuprates. 

\begin{figure}[ht]
\centering
\includegraphics[width=13cm]{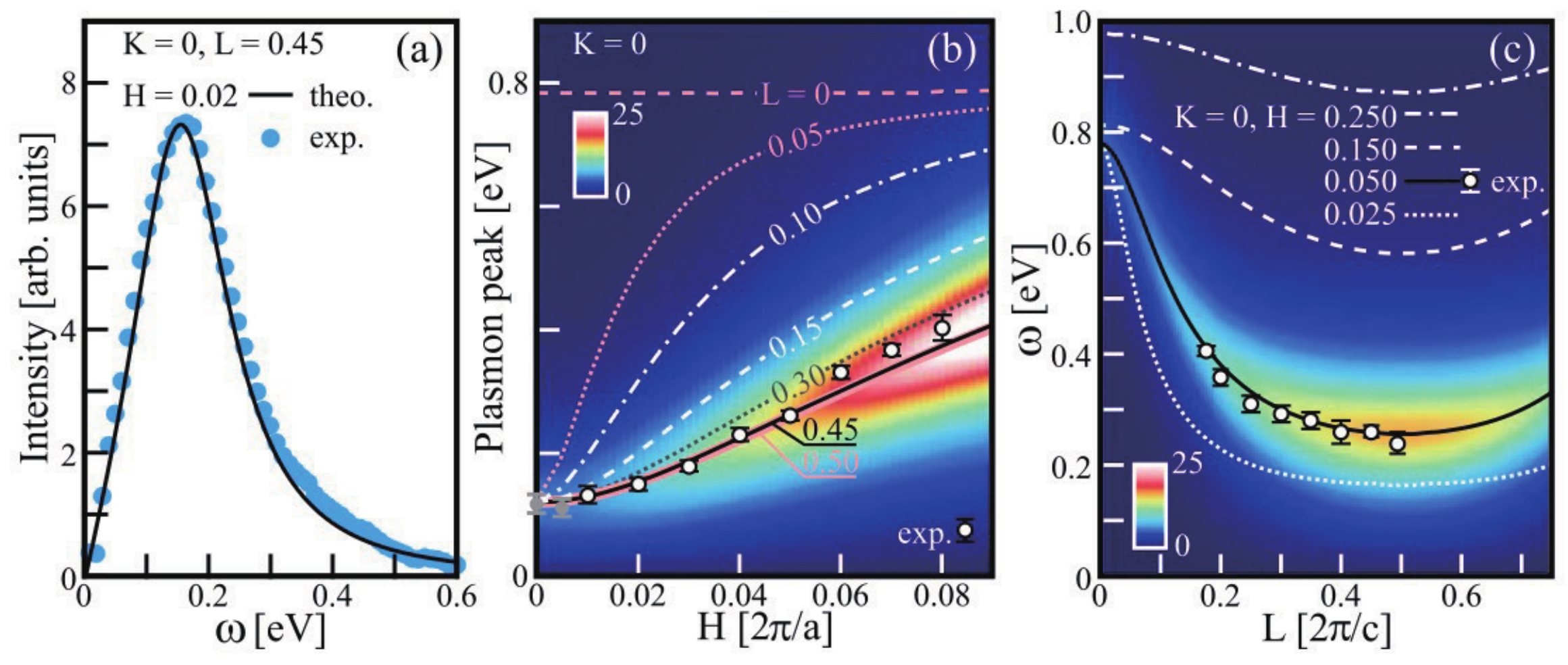}
\caption{(a) Imaginary part of the charge susceptibility Im$\chi_c({\bf q},\omega)$ for momentum $(0.02, 0, 0.45)$ (solid black line) computed in the layered $t$-$J$-$V$ model; here $H$, $K$, $L$ are units of $2 \pi /a$,  $2 \pi /a$,  $2 \pi /c$ with $a$ and $c$ being lattice constants along $x$ ($y$) and $z$ directions, respectively. Superimposed are experimental data (blue symbols), which correspond to the plasmon component in the RIXS raw data. The intensity of Im$\chi_c({\bf q},\omega)$ is scaled to fit the maximum of the RIXS data. (b) Computed intensity map of Im$\chi_c({\bf q},\omega)$ for momenta along the $(H,0,0.45)$ direction. The solid black line corresponds to the maxima of Im$\chi_c({\bf q},\omega)$. The other lines indicate the maxima of Im$\chi_c({\bf q},\omega)$ for different $L$. Experimental plasmon peak positions for momenta along the $(H,0,0.45)$ direction are superimposed as white and gray symbols. The former symbols correspond to peak positions used as an input for the fitting procedure for the $t$-$J$-$V$ model, while the latter were not included. (c) Computed intensity map and maxima for different $H$ along the $(0.05,0,L)$ direction. Experimental data are plotted by open circles. Adapted from Ref.~\cite{hepting22}  [\copyright\, 2022 The Author(s)]. 
}
\label{exp-RIXS}
\end{figure}

The predicted $q_{z}$ dependence of the plasmon energy was indeed observed in  the electron-doped cuprates \cite{hepting18} as shown in \fig{exp-RIXS}(c). Furthermore, a similar charge-excitation mode was also reported in hole-doped cuprates \cite{nag20,singh22a,hepting23,nag24,nakata25}. Given the small plasmon energy ($< 100$  meV) observed at $\qp=(0,0)$ in many cuprates NCCO \cite{hepting18}, ${\rm La_{2-x}Ce_{x}CuO_{4}}$ (LCCO) \cite{hepting18,lin20,hepting22,nag24}, ${\rm La_{2-x}Sr_{x}CuO_{4}}$ (LSCO) \cite{nag20,hepting22,singh22a,hepting23,nag24}, and ${\rm Bi_{2}Sr_{2}CuO_{6}}$ (Bi2201) \cite{nag20}, it had been refereed to as an ``acoustic plasmon''  \cite{hepting18,nag20,singh22a}. However, an electron-doped cuprates ${\rm Sr_{0.9}La_{0.1}CuO_{2}}$ (SLCO), with the $c$ axis lattice constant $c$ $(= 3.405$~\AA), which is smaller than the in-plane lattice constant $a=b$ $(= 3.960$~\AA) and is thus expected to yield a relatively large $t_{z}$, exhibits a  larger plasmon gap of $\sim 120$~meV at $\qp=(0,0)$. This demonstrates that these low-energy modes are not strictly acoustic \cite{hepting22}, but rather have a gap proportional to the interlayer hopping integral, which is a key prediction of the theory [see \fig{scenarios}(d)] \cite{greco16}. Now, many details observed in experiments \cite{hepting18,lin20,nag20,singh22a,hepting22,hepting23,nag24,nakata25} consistently support that the observed charge excitations around $\qp=(0,0)$ (\fig{exp-RIXS}) are plasmons with a finite $q_{z}$, a feature specific to the layered materials  \cite{greco16,greco19,greco20,yamase21c,yamase25,sellati25}. 

Despite this evidence, some \cite{mitrano18,husain19} have claimed that plasmons are heavily damped in the anomalous metallic region in cuprates because the metallic phase could be distinct so that a collective charge mode such as plasmons is not stable and decays immediately. However, a systematic study of plasmons in LSCO as a function of doping at $T=300$ K revealed a smooth evolution of plasmon signals as the system crosses the anomalous metallic phase \cite{hepting23}, which corroborates earlier experiments  \cite{uchida91}. Reference~\cite{hepting23} also studied the temperature dependence of plasmon signals across $T_{c}$ and found no substantial change. However, this observation does not necessarily imply that the plasmons are irrelevant to superconductivity; more measurements at in-plane momenta much closer to $(0,0)$ and with higher energy resolution, a similar scale as superconducting gap,  would be needed to conclude their relationship to superconducting instability. 

\subsection{Theoretical challenges}
Early theoretical studies predicted the presence of the acoustic plasmons in a layered electron gas model  (the interlayer hopping was neglected) \cite{grecu73,fetter74,grecu75}. Inspired by these low-energy plasmons, several authors proposed mechanisms for high-temperature superconductivity in a weak coupling schemes \cite{ruvalds87,cui91,ishii93,malozovsky93,varshney95,pashitskii08}, from plasmon-phonon mode coupling \cite{falter94,bauer09}, or as a constructive contribution in the presence of electron-phonon coupling \cite{kresin88,bill03}. In the context of cuprates, plasmons were also studied across the metal insulator transition in two dimension \cite{vanloon14} and it was argued theoretically that they can be detected by RIXS \cite{markiewicz08}.  

However, cuprates are not electron gas systems, but electron liquid ones.  In fact, 
cuprate superconductors are well-established doped Mott insulators \cite{anderson87}, where strong electron-electron correlation effects are of central importance. To accurately analyze the plasmons and systematically compare with experimental data, it is vital to go beyond the most widely studied $t$-$J$ and Hubbard models, which contain only short-range interactions. This presents a particularly difficult task. First,  various numerical techniques \cite{qin22} such as ED, DMRG, C-DMFT, DCA, and others employ a finite-size cluster where the LRC cannot be properly incorporated. Second, three-dimensional calculations are required to correctly capture the plasmon mode observed in experiments, but they come with significant numerical costs. 

To overcome these difficulties, the large-$N$ theory was proposed, incorporating the LRC into the $t$-$J$ model \cite{greco16}. This method does not invoke a cluster and thus there is no technical difficulty to incorporate the LRC. The strong correlation effect---non-double occupancy of electrons at any lattice site---is handled by invoking the Hubbard operators and a Lagrange multiplier \cite{foussats02}. By increasing the number of spin from 2 to $N$ the analysis can be performed systematically as $1/N$ expansion. It was the calculations using this layered  $t$-$J$-$V$ model that first showed the well-defined plasmon mode in doped Mott insulators \cite{greco16}. Subsequent work demonstrated that the $t$-$J$-$V$ model accurately reproduces experimental data by choosing appropriate parameters  for each material \cite{greco19,greco20,nag20,hepting22,hepting23,nag24}; see also \fig{exp-RIXS}. Later, it was shown that a variational-wave-function and large-$N$ approach can also capture the plasmon dispersion correctly \cite{fidrysiak21}. 

Some researchers believe that plasmon physics should also be captured by a weak coupling theory. While it is true that plasmons originate from the $1/\vq^{2}$ singularity of the LRC in the long-wavelength limit in both strong- and weak-coupling theories, the latter often struggles to reproduce systematically the observed doping, momentum, interlayer hopping dependences, and other features with a single set of parameters---moreover, the band width might need to be suppressed substantially by hand. In contrast, the $t$-$J$-$V$ model accurately reproduces experimental data up to the optical doping \cite{hepting23} once the parameter set is fixed for a given material. A detailed comparison between the $t$-$J$-$V$ model and weak-coupling RPA was made in light of  experimental data \cite{nag24}. 

The introduction of the LRC in the $t$-$J$ model, which is simpler than the three-band model \cite{emery87}, is already challenging. Nonetheless, a more ambition to invoke a three-band model is also possible theoretical approach.  As long as the concept of the Zhang-Rice singlet \cite{fczhang88} is valid in hole-doped cuprates, we would expect that essentially the same results are obtained as those in the $t$-$J$-$V$ model, where experimental data about plasmon excitations are well reproduced.

The charge degrees of freedom (holes) in the $t$-$J$ model are described by the Zhang-Rice singlets. Hence, the formalism is characterized in such a way that the LRC should vanish at half-filling. This aspect is not easy to incorporate in the $t$-$J$ model, but the large-$N$ theory can be formulated to capture this fundamental physics. 

There is another intricate issue about the Zhang-Rice singlets---they are composite objects extending the Cu and O sites. That is, the Zhang-Rice singlets cannot be regarded as a point charge in a strict sense. However, such composite objects can be regarded as a point particle in the long-distant scale, yielding the same LRC asymptotically as the present work.

\subsection {Plasmons in multi-layer cuprates}
The experiments described above were primarily performed on cuprates with single CuO$_{2}$ plane per unit cell. As mentioned previously, a distinguished feature of plasmons is the strong dependence on the out-of-plane momentum $q_{z}$,  which highlights their three-dimensional nature \cite{hepting18,greco19}. This feature provides a novel route to understanding the high-$T_{c}$ mechanism, because the critical temperature $T_{c}$ is known to increase up to 140 K, with the number of CuO$_{2}$ plane per unit cell (up to three layers) and then remain high around 110 K for more than four layers \cite{iyo07}. Thus, a deeper understanding of the charge dynamics along the $c$ axis could provide a valuable hint to the high-$T_{c}$ mechanism. In fact, it has been emphasized that analyzing plasmon excitations in cuprates provides valuable information on the interlayer hopping integral $t_{z}$, which is otherwise difficult to obtain in experiments \cite{hepting22}. 

RIXS studies of plasmon excitations are now moving to multilayer cuprates (more than two CuO$_{2}$ layers per unit cell). A pioneering work \cite{bejas24} was performed for typical  bilayer-cuprates (Y-based cuprates), where two plasmon modes, $\omega_{\pm}$, were predicted \cite{fetter74,griffin89}. The original work \cite{bejas24} suggested the observation of the $\omega_{+}$ mode (in-phase mode at $q_{z}=0$), but the subsequent studies  \cite{yamase25,sellati25} implied the $\omega_{-}$ mode (out-of-phase mode at $q_{z}=0$); see Sec.~\ref{RIXS-Y-base}. All these works were performed within a weak coupling scheme, and the available experimental data were very limited. Needless to say, a more careful analysis of strong correlation effects \cite{yamase26} and more detailed RIXS experiments are called for. 

Possible plasmons in trilayer cuprates ${\rm Bi_{2}Sr_{2}Ca_{2}Cu_{3}O_{10+\delta}}$ (Bi2223) were studied recently \cite{nakata25}. In this case, three distinct  modes are predicted at $q_{z}=0$---the $\omega_{\pm}$ and $\omega_{3}$ modes \cite{griffin89}. The $\omega_{+}$ mode corresponds to an in-phase charge oscillation, the $\omega_{-}$ mode is an out-of-phase charge oscillation between two outer planes with no oscillation in the inner plane, and the $\omega_{3}$ mode is an out-of-phase charge oscillation between the outer and inner planes, with the larger oscillation in the inner plane. RIXS data \cite{nakata25} showed primary  contributions come from the $\omega_{-}$ mode, but the contribution from the $\omega_{+}$ and $\omega_{3}$ modes also appeared to be non-negligible, though a clear distinction between them was beyond the experimental accuracy. 

Experimental data in multi-layer cuprates are limited, and strong-coupling theory has been just constructed \cite{yamase26}, which will enable a detailed and realistic  comparison with experimental data. What is clear now is that charge dynamics in multi-layer cuprates is fundamentally different from that in single-layer systems. Further studies of charge dynamics in multilayer cuprates are expected to provide a fresh perspective, different from the conventional wisdom based on the in-plane spin-fluctuation mechanism. Toward this goal, the LRC has been formulated for a bilayer lattice systems \cite{yamase25}; see Sec.~\ref{bilayer-section}. It is important to note that the functional form of the LRC is frequently invoked from electron-gas models, but the electron density in cuprates is far from the low-density limit, necessitating a lattice-based approach. More general formulae of the LRC applicable to multi-layer cuprates are currently under development.

\section{Electron self-energy from charge fluctuations} \label{section-self-energy}
In the preceding section,  we have established that a theoretical  framework combining the layered $t$-$J$ model with the LRC provides a robust description of charge dynamics in cuprates. We have demonstrated that this approach, validated by both the large-$N$  expansion \cite{greco16} and hybrid variational-large-$N$ techniques \cite{fidrysiak21}, accurately reproduces the observed plasmon dispersions. This is a significant achievement, as it confirms that the collective charge excitations in these materials are well-captured theoretically. 

With the dynamics of the bosonic charge fluctuations now understood, we next take a crucial step: investigating how these fluctuations influence electron properties of the system. This is accomplished by calculating the electron self-energy, which fundamentally describes how electron's behavior is modified by its interactions with the surrounding many-body environment. The self-energy is the key quantity that determines the renormalization of the electronic band structure and the lifetime of quasiparticles. Here, we will specifically explore how the charge fluctuations, which give rise to the plasmon modes, contribute to the electron self-energy.

For this analysis, we will proceed with the framework of the large-$N$ formalism \cite{greco16}. This choice is motivated by its systematic and controlled nature.  Given the extensive body of work already performed using this formalism, we can exploit its well-established machinery to gain a deep understanding of the feedback effect of charge dynamics on the single-particle properties of cuprates.

\subsection{Large-$N$ formalism of the $t$-$J$-$V$ model}
\begin{figure}[ht]
\centering
\includegraphics[width=8cm]{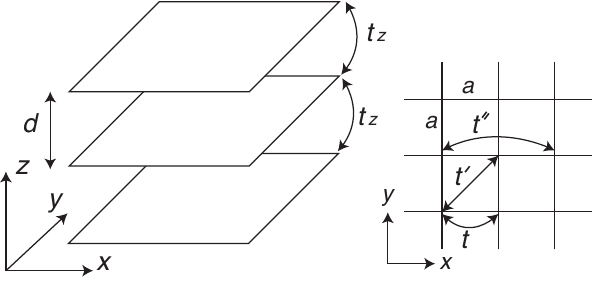}
\caption{Layered $t$-$J$-$V$ model. Each layer, separated by a distance $d$, forms a square lattice with lattice constant $a$, which is set to be 1 in the present review. The square lattice models the CuO$_{2}$ plane, where each site  corresponds to a Cu atom. The effects of the O atoms are implicitly included via the  Zhang-Rice singlet \cite{fczhang88}. The hopping integrals $t$, $t'$, $t''$, and $t_{z}$ are defined as shown in the figure.  
}
\label{singlelayer}
\end{figure}

To understand the feedback of plasmons on the electrons, we begin with a  microscopic model of single-layer cuprates, the layered $t$-$J$-$V$ model. This  model, which incorporates both interlayer hopping and the LRC (see \fig{singlelayer}), provides a comprehensive description of the underlying physics. The Hamiltonian is given by 
\begin{equation}
H = -\sum_{i, j,\sigma} t_{i j}\tilde{c}^\dag_{i\sigma}\tilde{c}_{j\sigma} + 
\sum_{\langle i,j \rangle} J_{ij} \left( \vec{S}_i \cdot \vec{S}_j - \frac{1}{4} n_i n_j \right)
+\frac{1}{2} \sum_{i \neq j} V_{ij} n_i n_j \,,
\label{tJV}  
\end{equation}
where $\tilde{c}^\dag_{i\sigma}$ ($\tilde{c}_{i\sigma}$) is 
the creation (annihilation) operator of an electron with spin $\sigma (=\uparrow, \downarrow)$  in the restricted Hilbert space that forbids double occupancy at any lattice site. $n_i=\sum_{\sigma} \tilde{c}^\dag_{i\sigma}\tilde{c}_{i\sigma}$ 
is the electron density operator, $\vec{S}_i$ is the spin operator, and 
the sites $i$ and $j$ run over a three-dimensional lattice. Figure~\ref{singlelayer} shows the single-layer lattice structure with various hopping parameters, including the crucial interlayer hopping $t_{z}$. 
$\langle i,j \rangle$ denotes the nearest-neighbor sites and the exchange interaction $J_{i j}=J$ is considered only inside the plane. We neglect the exchange term between the planes, which is much smaller than $J$ (Ref.~\cite{thio88}). $V_{ij}$ is the LRC on the three-dimensional lattice and its functional form is given in momentum space later [\eq{LRC}]. While cuprates are often considered two-dimensional, it is essential to use a layered, three-dimensional model to correctly capture the $q_{z}$ dependence of the LRC, which is responsible for the unique plasmon modes we discussed earlier

The primary theoretical challenge posed by the $t$-$J$-$V$ model (\ref{tJV}) is the treatment of the local constraint of no double occupancy at any lattice site. To overcome this, we employ a large-$N$ technique in a path integral representation using the Hubbard operators \cite{foussats02}.  This method  extends the number of spin components from 2 to an arbitrary integer $N$, allowing for a systematic expansion of physical quantities  in powers of $1/N$. This approach is particularly powerful because it treats all possible charge excitations on an equal footing \cite{bejas12,bejas14}, enabling an exclusive  analysis of charge dynamics at leading order. In fact, the present formalism yields results consistent with those obtained by dynamical DMRG method \cite{tohyama15,greco17} and ED \cite{merino03,bejas06}. However, the spin degrees of freedom, including superconductivity, appear at higher order.

We now present essential steps of the formalism, leaving a complete derivation to Ref.~\cite{yamase21a}.  At leading order, the electron dispersion $\varepsilon_{\vk}$ is given by a simple, renormalized form: 
\be
\varepsilon_{\vk} = \varepsilon_{\vk}^{\parallel}  + \varepsilon_{\vk}^{\perp} \,,
\label{xik}
\ee
where the in-plane and out-of-plane dispersions are given, respectively, by
\be
\varepsilon_{\vk}^{\parallel} = -2 \left( t \frac{\delta}{2}+\Delta \right) (\cos k_{x}+\cos k_{y})-
4t' \frac{\delta}{2} \cos k_{x} \cos k_{y} - 2t'' \frac{\delta}{2} (\cos 2 k_{x} +  \cos 2 k_{y})- \mu \,,\\
\label{Epara}
\ee
\be
\varepsilon_{\vk}^{\perp} = - 2 t_{z} \frac{\delta}{2} (\cos k_x-\cos k_y)^2 \cos k_{z}  \,. 
\label{Eperp}
\ee
Here we measure the in-plane momenta $k_x$ and $k_y$ and the out-of-plane momentum $k_z$ in units of $a^{-1}$ and $d^{-1}$, respectively; $a$ ($d$) is the  lattice constants in the plane (a distance between the planes); see \fig{singlelayer}.  While the dispersions are similar to non-interacting ones, the hopping integrals $t$, $t'$, $t''$, and $t_z$ are renormalized by a factor of $\delta/2$ where $\delta$ is the hole-doping rate. This renormalization is a direct consequence of the strong-electron correlations. The quantity $\Delta$ in \eq{Epara} is a mean-field value of the bond field and is given by 
\bea{\label {Delta}}
\Delta = \frac{J}{4N_s N_z} \sum_{\vk} (\cos k_x + \cos k_y) n_F(\varepsilon_\vk) \; , 
\eea
where $n_F(\varepsilon_\vk)$ is the Fermi distribution function, $N_s$ the total number of lattice sites on the square lattice, and $N_z$ the number of layers along the $z$ direction. For the given doping $\delta$, $\mu$ and $\Delta$ are determined self-consistently by solving \eq{Delta} and 
\be
(1-\delta)=\frac{2}{N_s N_z} \sum_{\vk} n_F(\varepsilon_\vk)\,.
\ee

Charge fluctuations are described by a $6 \times 6$ matrix of the bosonic propagators $D_{ab}$. Its inverse is given by the Dyson equation, 
\be
[D_{ab}(\vq,\mathrm{i}\nu_n)]^{-1} 
= [D^{(0)}_{ab}(\vq,\mathrm{i}\nu_n)]^{-1} - \Pi_{ab}(\vq,\mathrm{i}\nu_n)\,,
\label{dyson}
\ee
where $a$ and $b$ run from 1 to 6; $\vq$ is a three-dimensional wavevector and $\nu_n$ is a bosonic Matsubara frequency. 
$D^{(0)}_{ab}(\vq,\mathrm{i}\nu_n)$ is the bare propagator and is obtained as 
\begin{eqnarray}
[D^{(0)}_{ab}({\bf q},\mathrm{i}\nu_{n})]^{-1}= N \left(
 \begin{array}{cccccc}
\frac{\delta^2}{2} \left[ V(\vq)-J(\vq)\right]
& \frac{\delta}{2} & 0 & 0 & 0 & 0 \\
   \frac{\delta}{2}  & 0 & 0 & 0 & 0 & 0 \\
   0 & 0 & \frac{4}{J}\Delta^{2} & 0 & 0 & 0 \\
   0 & 0 & 0 & \frac{4}{J}\Delta^{2} & 0 & 0 \\
   0 & 0 & 0 & 0 & \frac{4}{J}\Delta^{2} & 0 \\
   0 & 0 & 0 & 0 & 0 & \frac{4}{J}\Delta^{2} \,.
 \end{array}
\right).
\label{D0}
\end{eqnarray}
Here $J(\vq) = \frac{J}{2} (\cos q_x +  \cos q_y)$; 
$V(\vq)$ is a key component, i.e., the LRC in momentum space for a layered system and is given by 
\be
V(\vq)=\frac{V_c}{A(q_x,q_y) - \cos q_z} \,,
\label{LRC}
\ee
where $V_c= e^2 d(2 \epsilon_{\perp} a^2)^{-1}$ and 
\be
A(q_x,q_y)=\alpha (2 - \cos q_x - \cos q_y)+1 \,.
\label{Aq}
\ee
The Coulomb interaction $V(\vq)$ is obtained by solving the Poisson's equation on the lattice \cite{becca96}; see Sec.~\ref{bilayer-section} for the bilayer lattice case. In \eq{Aq}, $\alpha=\frac{\tilde{\epsilon}}{(a/d)^2}$ and $\tilde{\epsilon}=\epsilon_\parallel/\epsilon_\perp$; $\epsilon_\parallel$ and $\epsilon_\perp$ are the dielectric constants parallel and perpendicular to the planes, respectively; $e$ is the electric charge of electron. Recalling that $a$ and $d$ are already known, we can extract $\epsilon_{\parallel}$ and $\epsilon_{\perp}$ by fitting $V_{c}$ and $\alpha$ 
through a comparison with RIXS data, which will be important when we discuss the Coulomb screening effect in the superconducting instability in Sec.~\ref{sc-section}. 

The $6 \times 6$ matrix is defined in a basis of a 6-component bosonic field,  
\be
\delta X^{a}_{i} = (\delta R_{i} \;,\; \delta \lambda_{i} ,\;  r_{i}^{{\eta}_{1}},\;r_{i}^{{\eta}_{2}},\; A_{i}^{{\eta}_{1}},\; A_{i}^{{\eta}_{2}})\,,
\label{deltaXa}
\ee
where the first component $\delta R_{i}$ is related to usual charge fluctuations; $\delta \lambda_{i}$ is the fluctuation of the Lagrange multiplier $\lambda_i$ associated with the local constraint that the double occupancy of electrons is forbidden at any lattice site; the remaining components describe fluctuations of the bond field, 
\be
\Delta_i^{\eta}=\Delta(1+r_i^\eta+iA_i^\eta)\,,
\label{staticDelta}
\ee
where $r_i^{\eta}$ and $A_i^{\eta}$ correspond to the real and imaginary parts of the bond-field fluctuations, respectively, and $\Delta$ is a static mean-field value. The index $\eta$ denotes the bond directions ${\eta}_{1}=(1,0)$ and ${\eta}_{2}=(0,1)$ on a square lattice.  As seen in Eqs. (3) and (4), there is a similarity to Gutzwiller mean-field approximation. However, we consider the fluctuations of the Lagrange multiplier $\delta \lambda_i$ in Eq. (11)---Our approximation is beyond the Gutzwiller mean-field approximation. 

The bosonic self-energy $\Pi_{ab}$ is computed as: 
\begin{eqnarray}
&& \Pi_{ab}(\vq,\mathrm{i}\nu_n)
            = -\frac{N}{N_s N_z}\sum_{\vk} h_a(\vk,\vq,\varepsilon_\vk-\varepsilon_{\vk-\vq}) 
            \frac{n_F(\varepsilon_{\vk-\vq})-n_F(\varepsilon_\vk)}
                                  {\mathrm{i}\nu_n-\varepsilon_\vk+\varepsilon_{\vk-\vq}} 
            h_b(\vk,\vq,\varepsilon_\vk-\varepsilon_{\vk-\vq}) \nonumber \\
&& \hspace{25mm} - \delta_{a\,1} \delta_{b\,1} \frac{N}{N_s N_z}
                                       \sum_\vk \frac{\varepsilon_\vk-\varepsilon_{\vk-\vq}}{2}n_F(\varepsilon_\vk) \; , 
\label{Pi}
\end{eqnarray}
where the $6$-component vertex $h_a$ is given by  
\bea
& h_a(\vk,\vq,\nu) = \left\{
                   \frac{2\varepsilon_{\vk-\vq}+\nu+2\mu}{2}+
                   2\Delta \left[ \cos\left(k_x-\frac{q_x}{2}\right)\cos\left(\frac{q_x}{2}\right) +
                                  \cos\left(k_y-\frac{q_y}{2}\right)\cos\left(\frac{q_y}{2}\right) \right];
                                                   \right. \nonumber \\
               & \hspace{5mm} \left. 1; -2\Delta \cos\left(k_x-\frac{q_x}{2}\right); -2\Delta \cos\left(k_y-\frac{q_y}{2}\right);
                         2\Delta \sin\left(k_x-\frac{q_x}{2}\right);  2\Delta \sin\left(k_y-\frac{q_y}{2}\right)
                 \right\} \, .
\label{vertex-h}
\eea

Charge fluctuation spectra are obtained by the analytical continuation in \eq{dyson}
\be
\mathrm{i}\nu_n \rightarrow \nu + \mathrm{i} \Gamma_{\rm ch}\,,
\label{gamma-ch}
\ee
where $\Gamma_{\rm ch} (>0)$ is infinitesimally small. By studying Im$D_{ab}(\vq, \nu)$ we can elucidate all possible charge dynamics in the layered $t$-$J$ model. In particular, the usual charge-charge correlation function is obtained from 
\be
{\rm Im}\chi_{c}(\vq, \nu) = N\left(\frac{\delta}{2}\right)^{2} {\rm Im} D_{11}(\vq, \nu)
\label{Imchic}
\ee
by taking $N=2$ in the end. 
The comparison of the experiments shown in \fig{exp-RIXS} is an outcome of the present large-$N$ theory 

The large-$N$ formalism reveals a {\it dual} aspect of the charge dynamics of the $t$-$J$-$V$ model (see Sec.~\ref{short-range-section}.1). On one hand, if we set $J=0$, we would obtain $\Delta=0$ and all fluctuations associated with the bond field vanish in \eq{D0}; note $\Delta \propto J$ in \eq{Delta}. The bosonic propagator $D_{ab}$ is reduced to a $2\times2$ matrix with $a,b=1,2$ and only usual charge fluctuations are active [see also \eq{deltaXa}]. In fact, the element $(1,1)$ of $D_{ab}$ is related to the usual charge-charge correlation function [see \eq{Imchic}] \cite{foussats02}.  $D_{22}$ and $D_{12}$ correspond to fluctuations associated with the non-double-occupancy condition and correlations between non-double-occupancy condition and charge density fluctuations, respectively. Since $V(\vq)$ is the LRC and usually $V(\vq)  \gg | J(\vq) |$ in interesting cases. Thus the charge dynamics is described by the $2 \times 2$ matrix  in \eq{D0}. 

When $J$ is finite,  bond-charge fluctuations become active and $a$ and $b$ take values from $1$ to $6$. Thus, both usual charge and bond-charge fluctuations are present for a realistic situation. Since the bond-charge excitations originate from the $J$-term, they are related to physics associated with the nearest-neighbor spin interaction. This feature shall be discussed briefly in Sec.~\ref{short-range-section}.1.

\subsection{Electron self-energy in the large-$N$ formalism of the $t$-$J$-$V$ model} 
As shown in Sec.~\ref{exp-LRC}, the $t$-$J$-$V$ model can quantitatively reproduce the charge excitations observed by RIXS. This success gives us confidence to take the next step: by using these charge fluctuations, we formulate the electron self-energy and understand how they renormalize electron properties. 

The self-energy is obtained by considering the next-to-leading order contribution in the $1/N$ expansion \cite{yamase21a}. At this order, the resulting imaginary part of the electron self-energy, denoted by Im$\Sigma_{\rm ch}$, arises solely from the coupling of electrons to the charge fluctuations we have just described. It is given in a compact form by the following expression \cite{bejas06}: 
\begin{equation}
{\mathrm{Im}}\Sigma_{\rm ch} ({\mathbf{k}},\omega) = \frac{-1}{N_{s} N_z}
\sum_{{\mathbf{q}}} \sum_{a,b} {\rm Im}D_{ab} (\vq,\nu) h_{a}(\vk,\vq,\nu) 
h_{b}(\vk,\vq,\nu) \left[
n_{F}( -\varepsilon_{\vk-\vq}) +n_{B}(\nu) 
\right]
\label{ImSig}
\end{equation}
where $\nu=\omega - \varepsilon_{\vk-\vq}$ and $n_B(\nu)$ is the Bose function. The equation directly links the quasiparticle lifetime to the spectrum of charge fluctuations described by the $6 \times 6$ matrix $D_{ab}$ and the vertex function $h_{a}$. The effects of spin fluctuations, which are integral part of the $t$-$J$ model, appear at higher order and are thus not present in this expression. This allows us to isolate and study the specific impact of charge fluctuations on electron properties. 
The form of Im$\Sigma_{\rm ch} (\vk,\omega)$ in \eq{ImSig} is structurally similar to a self-energy obtained from a Fock diagram in perturbation theory, but it also contains nontrivial Hartree contributions \cite{yamase21a}. 

The imaginary part of the self-energy Im$\Sigma_{\rm ch}$ provides a measure of the scattering rate of electrons by charge fluctuations. The real part of the electron self-energy $\mathrm{Re} \Sigma_{\rm ch} (\vk,\omega)$ is obtained from the imaginary part via the Kramers-Kronig relations and describes the renormalization of the electron band structure. These two components  fully define the electron self-energy $\Sigma_{\rm ch}({\bf k},\omega)$. With the full self-energy we can calculate the one-particle spectral function $A({\bf k},\omega)$, a quantity directly accessible in experiments such as angle-resolved photoemission spectroscopy (ARPES). The spectral function is derived from the electron Green's function, which is given by  
\be
A({\bf k},\omega)= -\frac{1}{\pi} \frac{{\rm Im}\Sigma_{\rm ch}({\bf k},\omega) - \Gamma_{\rm sf}}
{[\omega- \varepsilon_{\vk}-{\rm Re}\Sigma_{\rm ch} ({\bf k},\omega)]^2 
+ [-{\rm Im}\Sigma_{\rm ch} (\vk,\omega) +\Gamma_{\rm sf}]^2} \,,
\label{Akw}
\ee
where $\Gamma_{\rm sf}(>0)$ is a small, phenomenological broadening term that accounts for scattering processes not captured by charge fluctuations at this order \cite{prelovsek99}. This spectral function provides the complete picture about how the original dispersion is renormalized by Re$\Sigma_{\rm ch}$ and how much the original electron is damped by Im$\Sigma_{\rm ch}$.

This completes the theoretical framework for analyzing how charge fluctuations influence electron properties of the system. The next step is to use this framework to compute and analyze specific results, which will shed light on the pseudogap and other intriguing phenomena in cuprates. 

\subsection{Self-energy effect from charge fluctuations} 
Building on our formulation of the electron self-energy, we now apply this theoretical framework to investigate the renormalization effects mainly in the high-energy regime. Our primary goal is to compute the one-particle spectral function for two distinct physical scenarios: electron-doped and hole-doped cuprates. 

The numerical calculations for these two cases are performed with appropriate parameter sets chosen to best represent key features of each material. For the electron-doped cuprates, the calculations are carried out in the hole picture for computational convenience. This is a common practice in the $t$-$J$ model, because the model was developed for the hole-doped case. Subsequently, the  results are converted to the particle picture to allow for a direct and meaningful comparison with experimental data. In this way, our theoretical prediction can be tested against the results of spectroscopic technique such as ARPES. 

As inferred from the structure of the bare bosonic propagator $D_{ab}^{0}$ in \eq{D0}, the charge fluctuations are naturally separated into the distinct physical channels. The $2 \times 2$ sector, corresponding to indices $a,b=1,2$, captures the dynamics of the usual charge fluctuations. This is the channel that is most sensitive to the LRC and, as we have shown, is responsible for the plasmon modes. 
In this section, we focus on  $2 \times 2$ sector of the matrix ($a,b=1,2$) and clarify how one-particle property is renormalized by charge fluctuations. 
The remaining $a,b=3$--$6$ sector describes bond-charge fluctuations, which arise from nearest-neighbor spin interactions and dominate the low-energy physics. This physics shall be discussed briefly in the next section, Sec.~4.

\subsubsection{Renormalization of band structure in electron-doped case} 

\begin{figure}[t]
\centering
\includegraphics[width=9cm]{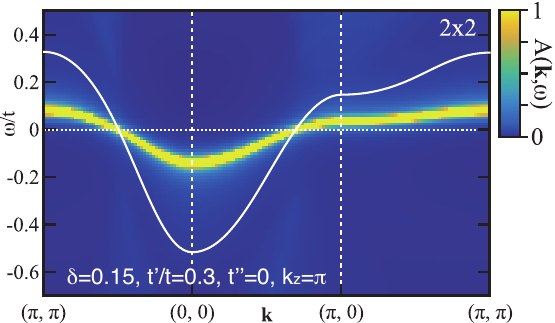}
\caption{Intensity map of $A(\vk,\omega)$ along the direction 
$(\pi,\pi)$--$(0,0)$--$(\pi,0)$--$(\pi,\pi)$ for $k_z=\pi$; $k_{z}$ dependence is very weak. The quasiparticle dispersion in the presence of the self-energy is in yellow. The white curve is the quasiparticle dispersion obtained in leading order theory without the self-energy [\eq{xik}]. See also \fig{Akw-ph}~(a) obtained after a particle-hole transformation. Adapted from Ref.~\cite{yamase21a}. (\copyright\, 2021 American Physical Society).
}
\label{QPdispersion}
\end{figure}
Having established that plasmon excitations are a key feature of the charge dynamics in electron-doped cuprates \cite{hepting18}, we now investigate their effect on the electronic band structure. For our calculations, we use a set of parameters optimized for this class of materials: $J/t=0.3$, $t'/t=0.3$, $t''/t=0$, $t_{z}/t=0.1$, $V_{c}/t=17$, $\alpha=4.5$, $N_{z}=10$,  $\Gamma_{\rm ch}/t=0.03$, $\Gamma_{\rm sf}/t=0.01$, and a doping rate $\delta=0.15$. 

Figure~\ref{QPdispersion} presents the quasiparticle dispersion in both the leading-order theory (white curve), as described by Eq. \eqref{xik}, and the fully renormalized dispersion (yellow curve) that includes the effects of the self-energy. The leading-order dispersion is already renormalized by mean-field factors, with a characteristic energy scale of $J$ $(=0.3t)$. The coupling to charge fluctuations, however, induces a further substantial renormalization. As shown by the yellow curve, which tracks the peak of the spectral function $A(\vk,\omega)$ [Eq.~\eqref{Akw}], the quasiparticle bandwidth is significantly suppressed to less than half of its original value. We interpret this dramatic reduction as a consequence of strong correlations---no double occupancy at any lattice site. Charges are mobile under this constraint, which may make the effective mass heavier. The resulting band width is severely suppressed. In addition, the electron dispersion does not exhibit ``kink'' behavior, in contrast to phonons \cite{lanzara01,zhou05} and magnetic fluctuations \cite{kaminski01,johnson01,gromko03,valla20}.

\begin{figure}[ht]
\centering
\includegraphics[width=12cm]{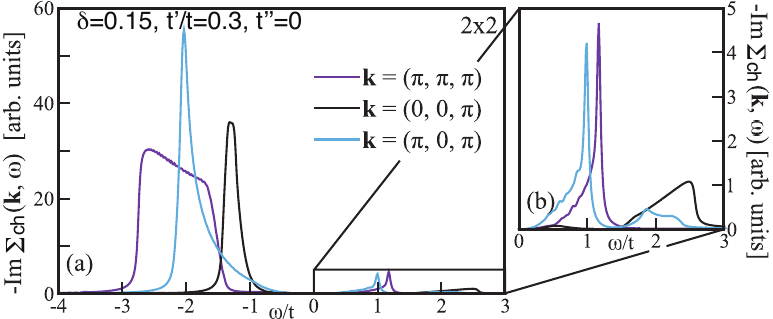}
\caption{Imaginary part of the electron self-energy, $-{\rm Im}\Sigma_{\rm ch} (\vk,\omega)$, as a function of $\omega$ for several choices of $\vk$. The positive energy region is magnified in the right panel. Results do not depend much on a value of $k_z$ and $k_z=\pi$  is taken as a representative one. Adapted from Ref.~\cite{yamase21a} (\copyright\, 2021 American Physical Society).  
}
\label{ImSig-asym}
\end{figure}

We next examine the imaginary part of the electron self-energy, Im$\Sigma_{\rm ch}$, which represents the scattering rate of the electrons. As shown in \fig{ImSig-asym}, Im$\Sigma_{\rm ch}$ exhibits a pronounced asymmetry with respect to $\omega=0$, which implies a substantial breaking of particle-hole symmetry. This asymmetry is not a trivial consequence of the underlying band structure (e.g., from the $t'$ hopping), but rather stems from the strong correlation effects captured by the local constraint of no double occupancy inherent in the $t$-$J$ model.

The mathematical reason for this particle-hole asymmetry can be traced back to the structure of the self-energy expression in \eq{ImSig}. Im$D_{ab}(\vq, \nu)$ is an odd function with respect to $\nu$. For $\nu > 0$, we have Im$D_{ab}(\vq, \nu) \geq 0$ for the diagonal component ($a=b$) and Im$D_{ab}(\vq, \nu) \leq 0$ for the off-diagonal components ($a \ne b$); see Figs.~1 and 8 in Ref.~\cite{bejas17}. The vertex function is given by $h_{1}(\vk, \vq, \nu) \propto \nu$ for a large $\nu$ and $h_{2}(\vk, \vq, \nu) =1$ [see \eq{vertex-h}]. The term $n_{F}( -\varepsilon_{\vk-\vq}) +n_{B}(\nu)$ acts as a selection rule as summarized in Table~\ref{selection-rule}. Considering these factors, we recognize that all components $a,b=1,2$ contribute additively to the summation for $\nu <0$ in \eq{ImSig} whereas the contribution from the diagonal parts is largely cancelled out by the off-diagonal components for $\nu > 0$. The off-diagonal components originate from the strong correlation effect related to the fluctuations of the Lagrange multiplier. Recalling that we have the relation $\nu = \omega - \varepsilon_{\vk-\vq}$ and the value of $ |\varepsilon_{\vk-\vq} |$ is less than $1t$ while the typical energy scale in \fig{ImSig-asym} is much larger than $1t$, the sign of $\nu$ becomes the same as that of $\omega$ independent of $\vk$ and $\vq$ in most cases. This explain the strong asymmetry of the Im$\Sigma(\vq, \omega)$ in \fig{ImSig-asym}. We do not have a physical explanation of this asymmetry because the negative off-diagonal component Im$D_{ab}(\vq, \nu)$ ($a \ne b$) does not have any physical meaning.  

Moreover, by associating the sign of $\omega$ to that of $\nu$, the selection rule in Table~\ref{selection-rule} serves to identify the origin of the structure of Im$\Sigma_{\rm ch}(\vq, \omega)$. For $\omega>0$, the peak around $\omega/t \approx 1$ for $\vk = (\pi, \pi, k_{z})$  comes from the coupling to plasmon excitations around $\vq \sim (0,0,q_{z})$. A long tail on a lower energy side of the peak originates from the acousticlike plasmons. The structure around $\omega/t \approx 2.5$ for $\vk=(0,0,k_{z})$ comes mainly from the coupling to charge excitations around $\vq \sim (\pi, \pi, q_{z})$. For an intermediate momentum such as $\vk =(\pi, 0, k_{z})$, Im$\Sigma_{\rm ch}(\vq, \omega)$ exhibits typically two structures around $\omega/t \approx 1$ and $2$. The former stems from plasmons and the latter from charge excitations around $\vq \sim (\pi, \pi, q_{z})$. For $\omega<0$, the major contribution becomes {\it vice versa}: plasmons around $\vq \sim (0,0, q_{z})$ and charge fluctuations around  $\vq \sim (\pi, \pi, q_{z})$ form the structure around $\omega/t=-1$ for $\vk=(0,0,k_{z})$ and $\omega/t = -2$ for $\vk = (\pi, \pi, k_{z})$, respectively. For an 
intermediate momentum $\vk=(\pi, 0, k_{z})$, essentially a single peak is realized around $\omega/t =-2$, with a sizable tail on the side of $\omega=0$. 

\begin{table}[tb]
\begin{center}
\begin{tabular}{c||c|c|c|c} 
\hline
\multirow{2}*{\backslashbox {$\vk$} {$\vq$}} 
 & \multicolumn{2}{c|} {$\sim (0,0,q_z)$} &  \multicolumn{2}{c} {$\sim (\pi,\pi,q_z)$}  \\
\cline{2-5}
 & {$\nu <0$\;} & {$\nu > 0$\; } & {$\nu<0$\;} & {$\nu>0$\;} \\
\hline \hline
$ \sim (0,0,k_z)$ & -1 &  0 & 0 & 1  \\ \hline
$ \sim (\pi,\pi,k_z)$ & 0 &  1 & -1 & 0 \\ \hline
$ \sim (\pi,0,k_z)$ & -1 or 0 & 0 or 1& -1 or 0&  0 or 1 \\ \hline
\end{tabular} 
\end{center}
\caption{Typical values of $n_{F}( -\varepsilon_{\vk-\vq}) +n_{B}(\nu)$ at $T=0$. 
They may depend sensitively on the precise values of $\vk$ and $\vq$ 
for $\vk \sim (\pi, 0, k_z)$,   
because the Fermi surface is located rather close to $\vk \approx (\pi,0,k_z)$. Adapted from Ref.~\cite{yamase21a} (\copyright\, 2021 American Physical Society). 
}
\label{selection-rule}
\end{table}

\begin{figure}
\centering
\includegraphics[width=8cm]{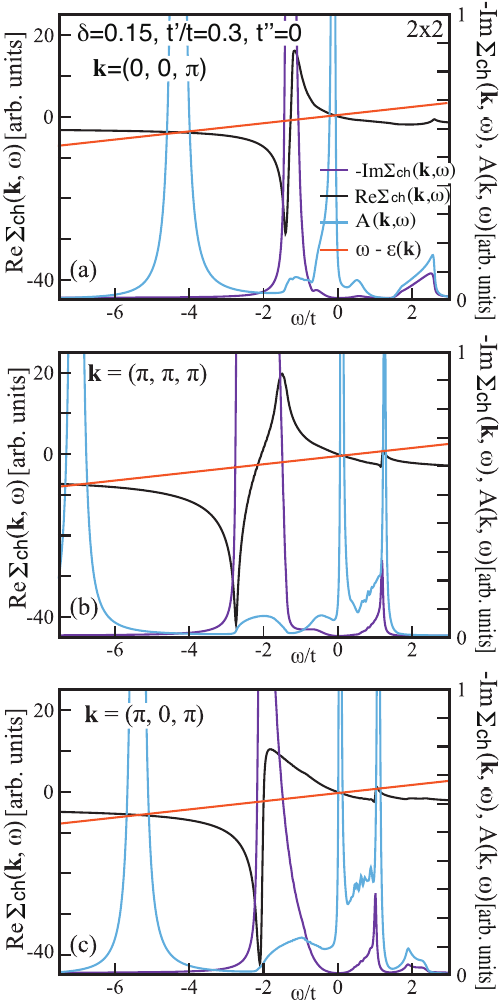}
\caption{Energy dependence of $- {\rm Im}\Sigma_{\rm ch} (\vk,\omega)$, ${\rm Re}\Sigma_{\rm ch} (\vk,\omega)$, and $A(\vk,\omega)$ as a function of $\omega$ at $\vk=(0,0,\pi)$ (a), $(\pi,\pi,\pi)$ (b), and  $(\pi, 0,\pi)$ (c). The line of $\omega - \varepsilon (\vk)$ is also drawn. The scales of $- {\rm Im}\Sigma_{\rm ch} (\vk,\omega)$ and $A(\vk, \omega)$ correspond to the right vertical axis and their units are taken differently to clarify their peak structure in the same panel. Adapted from Ref.~\cite{yamase21a} (\copyright\, 2021 American Physical Society). 
}
\label{ImReA}
\end{figure}

The combined effects of the real and imaginary parts of the self-energy are summarized in \fig{ImReA}, which plots -${\rm Im}\Sigma_{\rm ch} (\vk,\omega)$, ${\rm Re}\Sigma_{\rm ch} (\vk,\omega)$, and the spectral function $A(\vk,\omega)$ for representative momenta. A peak in $A(\vk,\omega)$ occurs when the quasiparticle condition,  
\be
\omega - \varepsilon_{\vk} - {\rm Re} \Sigma_{\rm ch} (\vk,\omega)=0
\label{QPcondition}
\ee
is satisfied. This condition is illustrated by the crossing point of the line $\omega - \varepsilon (\vk)$ with the curve Re$\Sigma_{\rm ch}(\vq, \omega)$. For most momenta, we find three solutions. The first, near $\omega=0$, corresponds to the renormalized quasiparticle band  shown in \fig{QPdispersion}. The second, located at negative energies around $\omega/t \approx -1$ to $-2$, exhibits a very broad structure due to the large magnitude of the imaginary part of the self-energy at these energies. The third  solution, found at deep negative energies ($\omega/t \approx -4$ to $-7$), yields a very sharp peak because the imaginary part of the self-energy approaches zero in this  region. This is a clear signature of a distinct, coherent band emerging from the coupling with charge fluctuations.

\begin{figure}[ht]
\centering
\includegraphics[width=9cm]{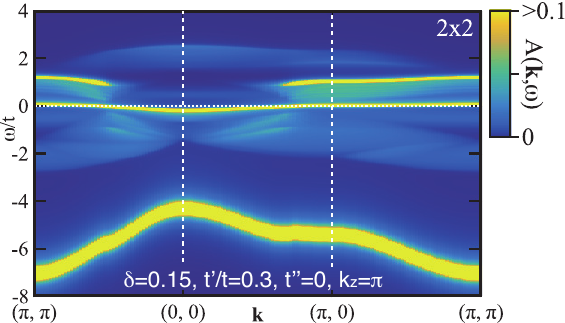}
\caption{Intensity map of $A(\vk,\omega)$ along the direction $(\pi,\pi)$--$(0,0)$--$(\pi,0)$--$(\pi,\pi)$; $k_z$ dependence is weak and $k_z=\pi$ is taken as a representative value. 
See also \fig{Akw-ph}~(b) obtained after a particle-hole transformation. Adapted from Ref.~\cite{yamase21a} (\copyright\, 2021 American Physical Society). 
}
\label{Akw-map}
\end{figure}

The comprehensive picture of the band renormalization is captured in the full spectral map of $A(\vk, \omega)$ shown in \fig{Akw-map}. While the main quasiparticle band appears as a relatively flat feature around $\omega=0$ in this  large energy window, it does disperse in the scale of $J$ $(=0.3t)$ as shown in \fig{QPdispersion}. The most striking features are the emergent secondary bands. A sharp dispersion is evident at deep negative energies, and another emergent band appears at $\omega/t \approx 1$. The former arises from the coupling to the usual charge fluctuations. The latter band is particularly pronounced for momenta around $\vk=(\pi,0,k_z)$ and $(\pi,\pi,k_z)$. The separation of this band from the main quasiparticle band is a direct result of the relatively high-energy scale of the plasmon modes that drive its formation. 

In addition to these prominent coherent features, the spectral function also reveals more subtle incoherent structures. For example, two weak band-like features emerge around $\omega/t \approx -1$ near $\vk = (0,0,k_z)$. These weak structures correspond to a peak in the imaginary part of the self-energy and are a consequence of the incoherent coupling to plasmons. Similarly, a weak, incoherent band is barely visible around $\omega/t \approx 2$ near $\vk \approx (0,0,k_z)$, originating from the coupling to charge excitations around $\vq \sim (\pi,\pi, q_z)$.

\subsubsection{Conversion to particle picture}
A crucial aspect of our theoretical approach for electron-doped cuprates is that our calculations are performed within the {\it hole} picture. This is a pragmatic choice, because the $t$-$J$ model is fundamentally defined in a restricted Hilbert space where double occupancy of electrons is forbidden. 

However, to enable a direct comparison with experimental data, which are almost universally presented and analyzed in the standard {\it particle} picture, it is essential to transform our theoretical results accordingly. This transformation allows us to present our finding in a physically transparent manner that can be immediately compared to  spectroscopic techniques such as ARPES. We perform this particle-hole transformation in momentum space. 

\begin{figure}[ht]
\centering
\includegraphics[width=8cm]{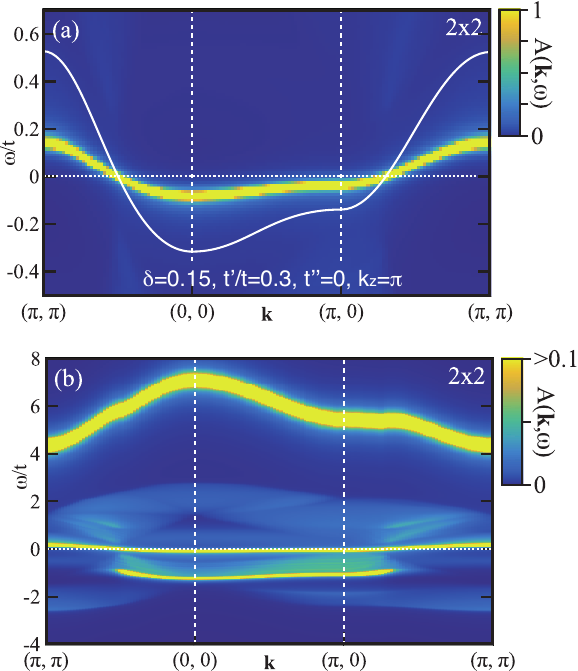}
\caption{Intensity maps of $A(\vk,\omega)$ along the direction $(\pi,\pi)$--$(0,0)$--$(\pi,0)$--$(\pi,\pi)$ after the particle-hole transformation (a) around the Fermi energy corresponding to \fig{QPdispersion}, and (b) in a larger energy region corresponding to \fig{Akw-map}, respectively.  Adapted from Ref.~\cite{yamase21a} (\copyright\, 2021 American Physical Society). 
}
\label{Akw-ph}
\end{figure}

The transformation is defined by the following changes:
\be
\tilde{c}_{\vk \sigma} \rightarrow \tilde{c}_{\vk+\vQ \sigma}^{\dagger} 
\hspace{5mm} {\rm and} \hspace{5mm} 
\tilde{c}_{\vk \sigma}^{\dagger} \rightarrow \tilde{c}_{\vk+\vQ \sigma} 
\label{ph-transform}
\ee
where $\vQ=(\pi,\pi,0)$ (Ref.~\cite{misc-ph}). This operation leads to a straightforward set of transformation for the bare energy, self-energy, and spectral function: 
\bea
&& \epsilon_{\vk} \rightarrow - \epsilon_{\vk+\vQ} \,,  \label{p-h1}\\
&&{\rm Re} \Sigma_{\rm ch}  (\vk, \omega) \rightarrow -{\rm Re} \Sigma_{\rm ch}  (\vk+\vQ, -\omega) \,,  \label{p-h2} \\
&&{\rm Im} \Sigma_{\rm ch}  (\vk, \omega) \rightarrow {\rm Im} \Sigma_{\rm ch}  (\vk+\vQ, -\omega) \,, \label{p-h3} \\
&&A(\vk, \omega)  \rightarrow A(\vk+\vQ, -\omega) \label{p-h4} \,.
\eea
Physically this transformation inverts both the energy axis ($\omega \rightarrow  -\omega$) and the momentum axes, shifting the origin of the Brillouin zone to $(\pi, \pi)$ and vice versa. It is important to note that the bosonic charge correlation function, which describes the dynamics of the charge fluctuations themselves, remains unchanged by this transformation. The transformed results for the quasiparticle dispersion (\fig{QPdispersion}) and the full spectral map (\fig{Akw-map}) are presented in Figs.~\ref{Akw-ph}(a) and (b), respectively.  

\subsubsection{Hole-doped case}
Having established the key features of the band renormalization for electron-doped cuprates, we now turn our attention to the hole-doped case. We use a set of parameters appropriate for this scenario, with second-nearest-neighbor hopping parameter $t'/t=-0.2$, while the other parameters remain identical to those used in the electron-doped case. This choice allows for a direct comparison, isolating the effects of the underlying band structure on electron properties.
 
\begin{figure}[th]
\centering
\includegraphics[width=8cm]{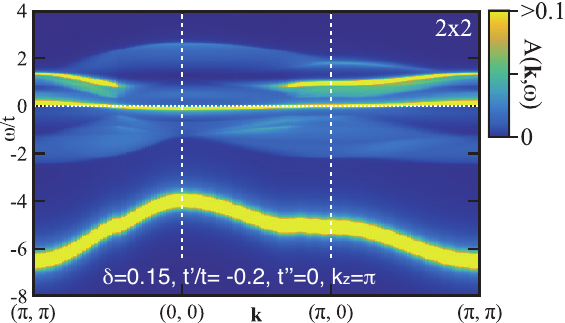}
\caption{Intensity map of $A(\vk,\omega)$ for parameters appropriate for hole-doped case (t'/t=-0.2) at a doping rate $\delta=0.15$. The map is plotted along the direction $(\pi,\pi)$--$(0,0)$--$(\pi,0)$--$(\pi,\pi)$. The $k_{z}$ dependence is weak, and $k_{z}=\pi$ is taken as a representative value. Adapted from Ref.~\cite{yamase21a} (\copyright\, 2021 American Physical Society). 
}
\label{h-cuprate}
\end{figure}

 Figure~\ref{h-cuprate} shows the intensity map of the spectral function $A(\vk,\omega)$ for the hole-doped case. A striking finding is that this result is qualitatively identical to the spectral map for the electron-doped case shown in \fig{Akw-map}. We find a main quasiparticle band near the Fermi energy, a sharp emergent band at deep negative energies, and an incoherent band at positive energies. This remarkable similarity holds even when the band parameters are varied further, indicating the robust features across different doping types.

This result is physically compelling. It demonstrates that while the microscopic details of the band structure may differ between electron- and hole-doped cuprates, the universal nature of charge fluctuations in the presence of the LRC,   which drives the plasmon modes, dictates a remarkably similar renormalization of the electronic band structure in the high-energy regime. This suggests that the renormalization effect by charge fluctuations is a fundamental and common feature in both classes of materials.

\subsubsection{Plasmarons: fermioninc quasiparticles}
The emergent incoherent band observed at high energies---$\omega \approx -1t$ in the electron-doped case [\fig{Akw-ph}(b)] and $\omega=1t$ in the hole-doped case (\fig{h-cuprate})---is a distinct signature of the plasmon-electron coupling. This band correspond to  fermionic quasiparticles known as plasmarons. A plasmaron is a composite excitation consisting of a bare electron (or hole) dressed by the optical plasmon mode. The remarkable feature of a plasmaron is its dispersive nature, which we now analyze in detail \cite{yamase23a}.   

\begin{figure}[tb]
\centering
\includegraphics[width=8cm]{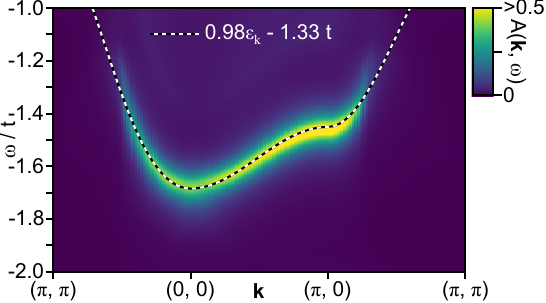}
\caption{Dispersion of plasmarons. It follows $0.98 \varepsilon_{\vk} - 1.33t$ (dashed curve) and in this sense, it is a replica band of $\varepsilon_{\vk}$. The parameters are chosen as $t'/t=0.3$, $t''=0$, $t_{z}/t=0.03$, $\alpha=2.9$, $V_{c}=18$,  $\delta=0.175$, $\Gamma_{\rm ch}/t=0.03$, $\Gamma_{\rm sf}/t=0.03$, and $t/2=0.5$ eV, for which the plasmons observed in electron-doped cuprate ${\rm La_{1.825}Ce_{0.175}CuO_{4}}$ are well captured in the present theory \cite{hepting22}. Adapted from Ref.~\cite{yamase23a} [\copyright\, 2023 The Author(s)]. 
}
\label{replica}
\end{figure}

The dispersion of plasmarons shown in \fig{replica} exhibits a clearly qualitative resemblance to the bare quasiparticle dispersion. A quantitative fit confirm this intuition: the plasmaron dispersion precisely follows the form $0.98 \varepsilon_{\vk} - 1.33t$. This behavior, where a low-energy band is mirrored at a higher energy, is a signature of a replica band, a concept well-established in the study of weakly correlated electron systems where electron-plasmon coupling is a dominant effect \cite{aryasetiawan96,kheifets03,tediosi07,markiewicz07a,polini08,hwang08,bostwick10,brar10,walter11,guzzo11,dial12,lischner13,caruso15,caruso15a,lischner15,jang17,zliu21}.  The factor of $0.98$ indicates a minor renormalization of the dispersion, while the energy offset of $-1.33t$ is related to, but not exactly equal to, the optical plasmon energy $\nu=1.15t$. This connection can be  understood intuitively from the imaginary part of the self-energy Im$\Sigma_{\rm ch} (\vk, \omega)$. The energy of the one-particle excitation $\omega$ is related to the charge fluctuation energy $\nu$ via $\omega=\nu+\varepsilon_{\vk-\vq}$ in \eq{ImSig}. Since the optical plasmon that gives rise to this band has its highest spectral weight at small momentum transfer $\vq$, the plasmaron dispersion essentially follows the bare band dispersion $\varepsilon_{\vk}$, shifting in energy by the plasmon frequency.

\begin{figure}[thb]
\centering
\includegraphics[width=8cm]{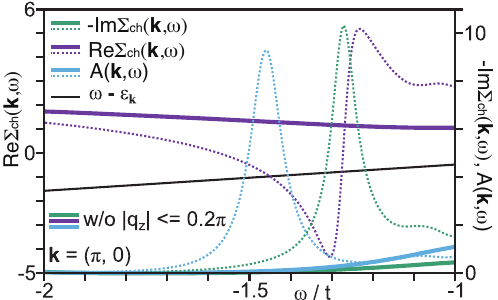}
\caption{Role of the optical plasmon for plasmarons. The self-energy and the spectral function are computed at $\vk=(\pi, 0)$ by removing a region $| q_{z}| \leq 2\pi/10$ in the $q_{z}$ summation in \eq{ImSig}, namely without contributions from the optical plasmon. They do not form any structure. Dotted curves are the corresponding results with the full $q_{z}$ summation. The line of $\omega-\varepsilon_{\vk}$ is also given. The contrast between the solid and dotted curves demonstrates the importance of the optical plasmon to forming the plasmarons. Adapted from Ref.~\cite{yamase23a} [\copyright\, 2023 The Author(s)]. 
}
\label{noqz0}
\end{figure}

The crucial role of the optical plasmon is numerically confirmed by removing its contribution from the calculation. As shown in \fig{noqz0}, we compute the self-energy by excluding the small $q_{z}$ region ($|q_{z}| \leq 2\pi/10$) from the summation in \eq{ImSig}. The result is striking: the sharp peak in Im$\Sigma_{\rm ch}$ completely disappears, and the resulting spectral function   $A(\vk, \omega)$ no longer forms any incoherent structure in that energy window. This contrast between the solid and dotted curves in \fig{noqz0} provides compelling evidence that the optical plasmon is the crucial ingredient for the formation of the plasmaron quasiparticles.

\begin{figure}[bt]
\centering
\includegraphics[width=8cm]{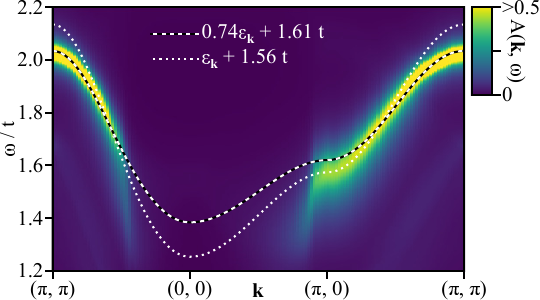}
\caption{Dispersion of plasmarons in hole-doped cuprates, especially ${\rm La_{2-x}Sr_{x}CuO_{4}}$. It follows $0.74 \varepsilon_{\vk} + 1.61t$ (dashed curve). The results may also be fitted by $\varepsilon_{\vk} + 1.56t$ (dotted curve). The parameters are chosen as $t'/t=-0.2$, $t''=0$, $t_{z}/t=0.01$, $\alpha=3.5$, $V_{c}=31$,  $\delta=0.16$, and $t/2=0.35$ eV, for which the plasmons observed in ${\rm La_{2-x}Sr_{x}CuO_{4}}$ are well captured in the present theory \cite{hepting22,uchida91}. Adapted from Ref.~\cite{yamase23a} [\copyright\, 2023 The Author(s)]. 
}
\label{hole-plasmaron}
\end{figure}

The concept of plasmarons provides a unifying framework for understanding the high-energy features in both electron- and hole-doped cuprates. As demonstrated in Sec. 3.3.2,  the electron- and hole-doped cases are connected via a particle-hole transformation shown in \eq{ph-transform}. Consequently, plasmarons appear at $\omega>0$ in the hole-doped case as seen in \fig{hole-plasmaron}, with a dispersion that follows $0.74 \varepsilon_{\vk} + 1.61t$. Despite the change in energy, the fundamental physical conclusion remains the same: this incoherent band represents   plasmaron excitations, originates from the coupling to the optical plasmon, and is essentially the same as the replica band observed in weakly correlated electron systems \cite{aryasetiawan96,kheifets03,tediosi07,markiewicz07a,polini08,hwang08,bostwick10,brar10,walter11,guzzo11,dial12,lischner13,caruso15,caruso15a,lischner15,jang17,zliu21}.

\subsubsection{Importance of strong correlations to plasmarons}
To precisely identify the physical origin of the plasmarons, we dissect the contributions of various components within the charge self-energy. As outlined by the sum of four components in \eq{ImSig}, the charge self-energy contains contributions from four sectors.  We can illuminate the specific contributions by introducing an auxiliary parameter $r (\geq 0)$ to control the weight of the terms associated with the local constraint. Our analysis focuses on a modified imaginary part of the self-energy:
\be 
{\rm Im}\Sigma_{\rm ch} (\vk, \omega; r) = {\rm Im}\Sigma_{11}+ r \times ({\rm Im}\Sigma_{22} + 2\, {\rm Im}\Sigma_{12}) \,,
\label{rSig}
\ee
where we have omitted the arguments for simplicity on the right-hand side of the equation and used the fact that Im$\Sigma_{12}$ is equal to Im$\Sigma_{21}$. The physical scenario of our full theory corresponds to the case where $r=1$. 

\begin{figure}[tb]
\centering
\includegraphics[width=14cm]{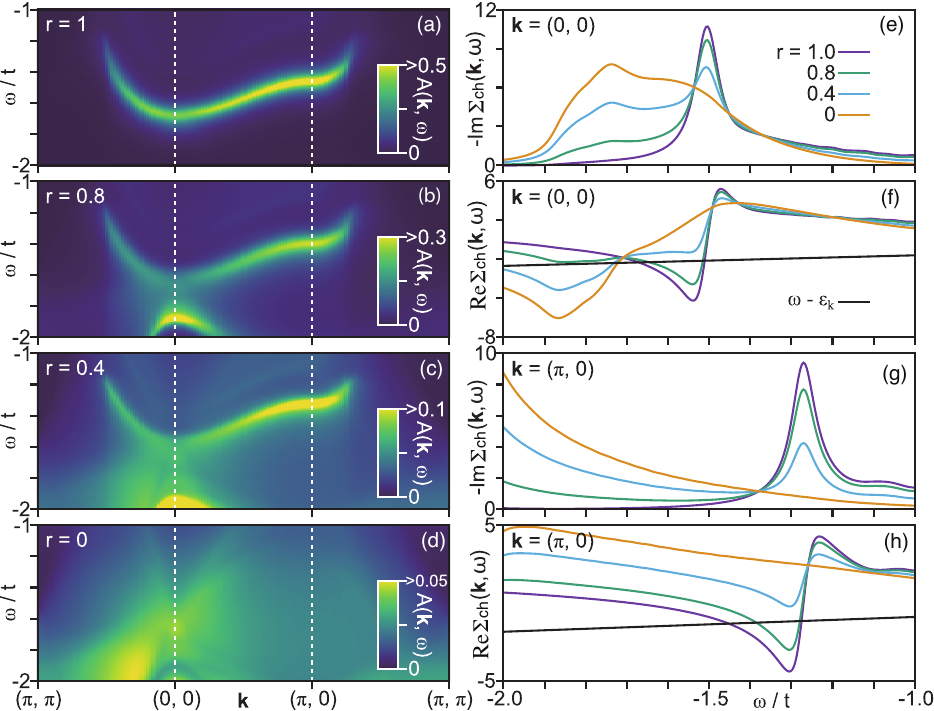}
\caption{Analysis of plasmarons in terms of \eq{rSig}. (a)--(d) Intensity map of the spectral function $A(\vk, \omega)$ computed with Im$\Sigma_{\rm ch} (\vk,\omega; r)$  [\eq{rSig}] in $-2 \leq \omega/t \leq -1$ for several choices of $r$  along the direction $(\pi, \pi)$--$(0,0)$--$(\pi,0)$--$(\pi,\pi)$ with $k_{z}=\pi$. The plasmaron band in (a) fades away upon decreasing $r$, indicating that fluctuations associated with the local constraint are crucially important to the plasmarons. Note a different color scale in each panel. (e)--(h)  Imaginary and real parts of $\Sigma_{\rm ch} (\vk, \omega)$ as a function of $\omega$ at $\vk=(0, 0)$ and $(\pi,0)$ for several choices of $r$. The peak of Im$\Sigma_{\rm ch} (\vk,\omega)$ in (e) and (g) is determined by the optical plasmon. The line of $\omega-\varepsilon_{\vk}$ is also shown in (f) and (h). The plasmaron energy is determined by its crossing point of Re$\Sigma_{\rm ch} (\vk, \omega)$ on the lower energy side when $r$ is close to 1. Adapted from Ref.~\cite{yamase23a} [\copyright\, 2023 The Author(s)]. 
}
\label{rAkw}
\end{figure}

We compute the spectral function $A(\vk, \omega)$ for various values of $r$, and the results are show in Figs.~\ref{rAkw}(a)--(d). A distinct color scale is used for each panel to highlight the subtle features that appear or disappear as $r$ is varied. As we progressively decrease $r$, the intensity of the incoherent band is substantially suppressed, fading away until it becomes nearly invisible for $r \lesssim 0.4$. This finding provides clear and compelling evidence that the formation of the plasmaron dispersion is critically driven by the self-energy components involving indices $a,b=2$. These components are a direct consequence of the strong correlation effects inherent in the restricted Hilbert  space of the $t$-$J$ model, namely the fluctuations associated with the local constraint.

The physics behind this behavior is further revealed by examining the self-energy itself. As shown in  Figs.~\ref{rAkw}(e)--(h), the imaginary part of the self-energy  ${\rm Im} \Sigma_{\rm ch} (\vk, \omega)$ exhibits a sharp peak at an energy responsible for a pronounced dip structure in the real part of the self-energy ${\rm Re}\Sigma_{\rm ch} (\vk, \omega)$ via the Kramers-Kronig relations. The plasmaron dispersion emerges as a response peak in the spectral function, which is formed when the real part of the denominator, $\omega - \varepsilon_{\vk}-{\rm Re}\Sigma_{\rm ch} (\vk, \omega)$, vanishes. This condition can be satisfied at two distinct energies: one close to the peak of Im$\Sigma_{\rm ch}$ and another at a lower energy corresponding to the tail of the dip in Re$\Sigma_{\rm ch}$. The plasmaron peak itself forms at this lower energy, where the damping, controlled by Im$\Sigma_{\rm ch}$, becomes small. This intricate interplay between the real and imaginary parts of the self-energy is the precise mechanism that gives rise to the incoherent, yet resonant, plasmaron dispersion. 

 A more detailed analysis reveals the unique roles played by the individual components of the self-energy. We have confirmed that the Im$\Sigma_{22}$  
term is the primary driving force behind the formation of the incoherent band, providing the dominant spectral weight for its existence. In contrast, the Im$\Sigma_{12}$ term, which accounts for the coupling between the electrons and the local constraint fluctuations, acts to sharpen the plasmaron dispersion by reducing the overall magnitude of the imaginary part of the self-energy. Together, these two components---both originating from strong correlation effects that double occupancy is forbidden at any lattice site---are essential for creating and defining the distinct characteristics of the plasmaron excitation.

\subsubsection{Resolution of puzzling behavior} 
Our understanding, however, appears to present an apparent paradox.  
In our theoretical framework, plasmons are defined by the poles of the charge-charge correlation function Im$D_{11}$, which correspond to the zeros of the determinant of the full correlation matrix. As a result, all components of the Im$D_{ab}$ matrix share these same poles and therefore describe the very same plasmon branches (see Figs.~1 and 8 in Ref.~\cite{bejas17}  for explicit calculations). Nevertheless, why did our earlier analysis (see \fig{rAkw}) demonstrate that the plasmaron is dominantly generated by the Im$\Sigma_{22}$ component, with other terms playing a much smaller role? The key to resolve this puzzle lies not in the poles of the correlation matrix, which are shared by all components, but in the specific form of their numerators. 

Let us examine the explicit from of the matrix components of $D_{ab}(\vq, \nu)$ in the limit of small momentum transfer $\vq \rightarrow {\bf 0}$, where the LRC $V(\vq)$ diverges as $\vq^{-2}$. 
\be
D_{ab}(\vq, \nu)=\frac{1}{\mathcal D} \left( 
\begin{array}{cc}
-\Pi_{22}(\vq, \nu) & \Pi_{12}(\vq, \nu)  - \frac{N\delta}{2} \\
 \Pi_{12}(\vq, \nu)  - \frac{N\delta}{2} \quad  &  -\Pi_{11}(\vq, \nu)  + \frac{N\delta^{2}}{2}\left( V(\vq) -J(\vq) \right) 
 \end{array}
\right) \,.
\ee
Here $\mathcal{D}$ is the determinant of the inverse matrix $[D_{ab}(\vq, \nu)]^{-1}$ and $\Pi_{ab}$ terms are bosonic self-energy with appropriate vertex functions $h_{a}(\vk, \vq, \nu)$; see \eq{Pi}. 
In the limit of $\vq \rightarrow {\bf 0}$, the dominant term in the matrix is the one that contains the divergent LRC. This leads to remarkable scaling behavior: 
\be
{\rm Im} D_{22}(\vq, \nu) \sim \frac{V(\vq)^{2}  {\rm Im}\Pi_{22}}{({\rm Re}\mathcal{D})^{2} + ({\rm Im}\mathcal{D})^{2}}\,. 
\label{ImDq0}
\ee
While all components of the correlation matrix diverge at the plasmon poles where $\mathcal{D} \rightarrow 0$, the numerator of Im$D_{22}$ diverges with as  $V(\vq)^{2}$. In contrast, the numerators of Im$D_{11}$ and Im$D_{12}$ behave with lower powers of $V(\vq)$, specifically as $V(\vq)^{0}$ and $V(\vq)^{1}$, respectively. Consequently, in the small-$\vq$ limit that governs the optical plasmon, Im$D_{22}$ becomes the overwhelmingly dominant component of the correlation matrix. This explains precisely why the Im$\Sigma_{22}$ component of the self-energy is the primary driver of the plasmarons. This result demonstrates how the long-range nature of the Coulomb interaction, though its $\vq^{-2}$ divergence, selects a specific physical channel---the fluctuations associated with the local constraint---to mediate the formation of the plasmarons.

\subsubsection{Plasmarons in weak-coupling theory} 
The mathematical structure of our theory has a formal correspondence with weak-coupling theories, which allows us to draw valuable comparisons. The form of our self-energy term Im$\Sigma_{22}$ in the small-momentum limit $\vq \rightarrow {\bf 0}$, bears a striking resemblance to the standard RPA expression for the self-energy in a weakly correlated electron gas: 
\be
{\rm Im} \Sigma^{\rm RPA} (\vk, \omega) =\frac{-1}{N_{z}N_{s}} \sum_{\vq} {\rm Im} D^{\rm RPA}(\vq, \nu) \left[
n_{\mathrm{F}}( -\varepsilon_{\vk-\vq}^{\rm RPA}) +n_{\mathrm{B}}(\nu) \right] \,.
\label{RPA}
\ee
Here ${\rm Im} D^{{\rm RPA}}(\vq, \nu) = V(\vq)^{2} {\rm Im} \Pi^{{\rm RPA}}(\vq, \nu)$ is the imaginary part of the screened Coulomb interaction and $\nu=\omega - \varepsilon_{\vk-\vq}^{\rm RPA}$. This mathematical parallel indicates that, in principle, plasmarons can also exist in weakly correlated electron systems \cite{aryasetiawan96,kheifets03,tediosi07,markiewicz07a,polini08,hwang08,bostwick10,brar10,walter11,guzzo11,dial12,lischner13,caruso15,caruso15a,lischner15,jang17,zliu21}.  However, in many such systems, plasmarons are often heavily overdamped, leading to very little spectral weight \cite{kheifets03,markiewicz07a,hwang08,lischner13,caruso15,caruso15a,lischner15}. The major challenge is that the imaginary part of the self-energy at the plasmon energy is often too large, which broadens the plasmaron peak and makes it difficult to observe.     

A key factor that can mitigate this issue is a relatively small electronic bandwidth, which effectively enhances correlation effects. This point was recently discussed in the context of plasmaron observation in the correlated semimetal SrIrO$_{3}$ \cite{zliu21}. In our present $t$-$J$-$V$ model, the bandwidth is intrinsically very small on the order of $t\delta/2$, with $\delta$  typically ranging from 0.1 to 0.2. This naturally puts the system in a regime where plasmarons can be more readily stabilized. 

However, our model presents a second crucial mechanism that goes beyond a simple bandwidth effect. The three components of the imaginary self-energy, ${\rm Im}\Sigma_{11}$, ${\rm Im}\Sigma_{12}$, ${\rm Im}\Sigma_{22}$, are all comparable in magnitude. Critically, the sign of ${\rm Im}\Sigma_{12}$ is opposite to that of the other two components in the energy region of interest ($\omega<0$); see the discussion in the context of \fig{ImSig-asym}. This results in a destructive interference, where the total imaginary part of the self-energy Im$\Sigma_{\rm ch}$ is substantially reduced due to the partial cancellation of these terms. This reduction in the imaginary part of the self-energy is precisely what is needed to form a sharp, distinct plasmaron peak in the spectral function, because it minimizes the damping of the quasiparticle. 

This synergy between a small bandwidth and the internal cancellation of the self-energy allows the system to satisfy the resonance condition, $\omega - \varepsilon_{\vk} -{\rm Re}\Sigma_{\rm ch}(\vq, \omega)$, even when the total 
self-energy is small. Consequently, these two features work constructively to host robust plasmaron excitations, making them a more stable and observable phenomenon in this strongly correlated electron system compared to their often-elusive counterparts in a simple weak-coupling picture.

\subsubsection{Plasmarons in cuprates}
Our theoretical predictions---plasmarons--- offer direct experimental verification, particularly through spectroscopic techniques. The most direct and robust test would be in electron-doped cuprates, for which our model is directly applicable. Given that the energy of the plasmarons is controlled by the optical plasmon, and that the optical plasmon energy is typically around 1 eV in cuprates \cite{uchida91}, ARPES is the ideal tool for this task. Specifically, we predict that ARPES should reveal an emergent incoherent band approximately 1 eV below the main quasiparticle dispersion especially along the direction $(0,0)$--$(\pi,0)$ [see \fig{Akw-ph}(b) and \fig{replica}]. This energy region has not been studied in detail \cite{armitage10}. Given that plasmarons have been detected even in weakly correlated systems such as graphene \cite{bostwick10,brar10,walter11}, two-dimensional electron systems \cite{dial12,jang17}, and SrIrO$_{3}$ films \cite{zliu21}, there is a high possibility of revealing them in cuprates, which have a naturally small electronic bandwidth.
 
The plasmaron energy is expected to be tunable with carrier doping, a feature that can be crucial to confirm its origin. In experiments \cite{uchida91}, the optical plasmon energy increases with carrier doping up to 20 \% doping. Our theory predicts that the energy of the plasmaron band should follow the same tendency. This doping dependence provides a unique fingerprint to distinguish the plasmaron band from other features. While ARPES is the most powerful method for visualizing the full dispersion \cite{yamaguchi25}, other techniques, such as x-ray photoemission spectroscopy \cite{guzzo11} and tunneling spectroscopy \cite{brar10}, can also be used to detect the plasmaron as a satellite peak in the spectral function.

It is critically important to distinguish the plasmon-generated plasmarons from the effects of other bosonic excitations, such as phonons and magnetic fluctuations. The coupling to phonons \cite{lanzara01,zhou05} or magnetic fluctuations \cite{kaminski01,johnson01,gromko03,valla20} is well-known to produce a "kink" in the electron dispersion. In sharp contrast, we predict that plasmon-electron coupling does not produce a kink (see \fig{QPdispersion}), but instead generates a distinct, incoherent plasmarons (\fig{replica}). 

A replica band is also discussed in the polar electron-phonon coupling mechanism in TiO$_{2}$ \cite{moser13,verdi17,caruso18} and the interplay between the electron-phonon and electron-plasmon couplings was studied \cite{jalabert89}. Our calculations have confirmed that the plasmons appear at a much higher energy scale ($\approx 1$ eV) than phonons (typically $< 100$ meV), thus making it straightforward to distinguish between these two effects.

We emphasize the crucial role of including the three-dimensional nature of the LRC in our model. A purely two-dimensional model is incapable of capturing the optical plasmon that we have shown to be essential for the plasmaron formation. This highlights that while the interlayer hopping integral is not directly relevant to the plasmarons, the layered geometry of the cuprates is a prerequisite for a proper theoretical description.

\subsection{Fermi liquid in the presence of charge fluctuations} 
To investigate the effects of charge fluctuations on electron properties near the Fermi energy, we choose parameters, which describe the plasmon dispersion observed in LSCO \cite{hepting22}. Specifically, we employ $t'/t=-0.20$, $J/t=0.3$, $t_{z}/t=0.01$, $V_c/t=31$, $\alpha=3.5$, and set the broadening parameters to $\Gamma_{\rm ch}=\Gamma_{\rm sf}=0.03t$. The number of layers is set to $N_{z}=10$, and we adopt $t=1$ as our energy unit. In LSCO, the plasmon mode with a finite out-of-plane momentum $q_z$ has low energy, below 55 meV, at the in-plane zone center \cite{hepting22}. This low-energy plasmon, along with gapless continuum spectrum, creates a favorable environment for plasmon excitations to effectively renormalize electron properties near the Fermi surface. For this analysis, we will focus on a small energy window around $\omega=0$. While the Fermi surface in our layered model depends on the out-of-plane momentum $k_{z}$, our conclusions are robust against variations in $k_{z}$, so we present results for the representative case of $k_{z}=0$.

\subsubsection{Fermi-liquid behavior} 
\begin{figure}[t]
\centering
\includegraphics[width=14cm]{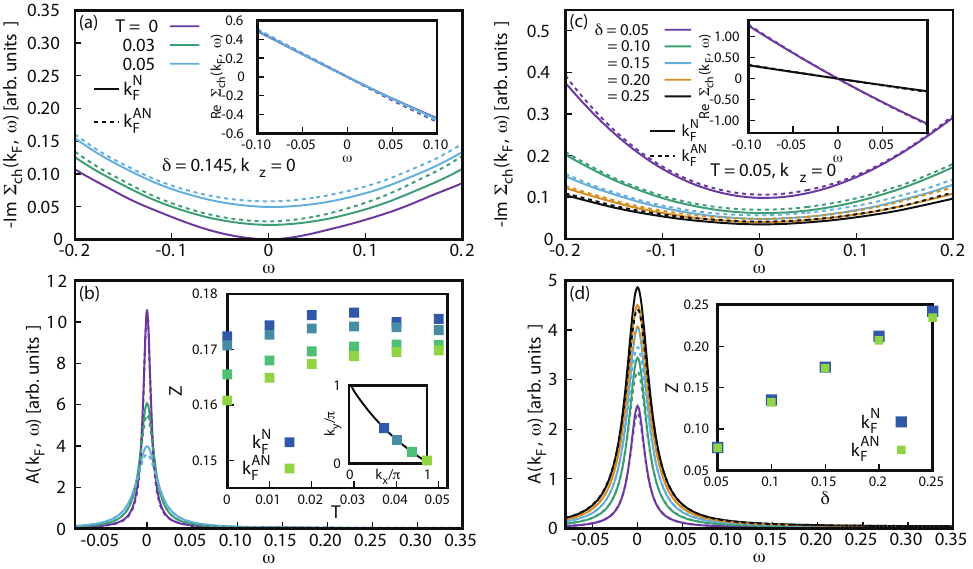}
\caption{Electron properties at the Fermi momenta. (a) Imaginary part of the electron self-energy at the nodal point $\vk_{F}^{N}$ and the antinodal point $\vk_{F}^{AN}$ on the Fermi surface for $T=0.03$ and $0.05$ at $\delta=0.145$; results at $T=0$ are shown only for $\vk_{F}^{N}$. The Fermi momenta are defined in panel (b). The inset is the corresponding real part of the self-energy at $T=0$ and $0.05$. (b) Corresponding spectral function for the temperatures in panel (a). The inset describes the quasiparticle weight $Z$ as a function of temperature at several choices of Fermi momenta. (c) Imaginary part of the self-energy at $\vk_{F}^{N}$ and $\vk_{F}^{AN}$ for $\delta=0.05, 0.10, 0.15, 0.20, 0.25$ at $T=0.05$. The inset shows the corresponding real part of the self-energy at $\delta=0.05$ and 0.25. (d) Spectral function for the doping rates in panel (c). The inset is the doping dependence of $Z$ at $\vk_{F}^{N}$ and $\vk_{F}^{AN}$. Adapted from Ref.~\cite{yamase24} (\copyright\, 2024 American Physical Society). 
}
\label{T-d-depend}
\end{figure}

Despite the presence of low-energy plasmon and gapless particle-hole excitations, our calculations show that the system retains a Fermi-liquid character. As shown in Fig.~\ref{T-d-depend}(a), the imaginary part of the electron self-energy from charge fluctuations,  Im$\Sigma_{\rm ch}(\vk_{F}, \omega)$, vanishes at energy $\omega=0$ and temperature $T=0$ with a characteristic quadratic dependence. Specifically, we find that Im$\Sigma_{\rm ch}(\vk_{F}, \omega) \sim \omega^{2}$ and Im$\Sigma_{\rm ch}(\vk_{F}, 0) \sim T^{2}$.  
The real part of the self-energy Re$\Sigma_{\rm ch}(\vk_{F}, \omega)$ shown in the inset of Fig.~\ref{T-d-depend}(a) exhibits a linear dependence with a negative slope at $\omega=0$, which is a typical feature of a Fermi liquid. The resulting spectral function $A(\vk_{F}, \omega)$ displays a single, sharp peak at $\omega=0$, as seen in Fig.~\ref{T-d-depend}(b). These results collectively demonstrate that charge fluctuations do not drive the system into a non-Fermi-liquid state.

While the system remains a Fermi liquid, charge fluctuations cause a substantial renormalization of the quasiparticles. The quasiparticle weight, defined as $Z=(1-\frac{\partial {\rm Re}\Sigma_{\rm ch}(\vk_{F}, \omega)}{\partial \omega} |_{\omega=0})^{-1}$, is significantly reduced to be around $Z \approx 0.17$ and shows only a weak dependence on temperature (see the inset of Fig.~\ref{T-d-depend}(b)). This implies that charge fluctuations are highly effective at suppressing the coherent quasiparticle weight, leaving only a tiny fraction to carry the low-energy excitations. Interestingly, as shown in Fig.~\ref{T-d-depend}(b), the spectral peak becomes sharper at lower temperatures despite the decrease in $Z$. 

Furthermore, the effect of charge fluctuations is remarkably isotropic. As shown in Figs.~\ref{T-d-depend}(a) and (b), the self-energy shows only a minimal difference between the nodal ($\vk_{F}^{N}$) and antinodal ($\vk_{F}^{AN}$)  
directions on the Fermi surface. This is a crucial finding, as it means charge fluctuations renormalize the quasiparticle properties in an essentially $s$-wave-like manner.

The right panels of Fig.~\ref{T-d-depend} illustrate the doping dependence of the self-energy at a fixed temperature of $T=0.05t$. In \fig{T-d-depend}(c), we find that the Fermi-liquid behavior persists across the entire doping range from $\delta=0.05$ to $0.25$, with Im$\Sigma_{\rm ch}(\vk_{F}, \omega)$ retaining its $\sim \omega^{2}$ dependence around $\omega=0$ for all doping rates. As  doping decreases, the magnitude of Im$\Sigma_{\rm ch}(\vk_{F}, 0)$ increases, and the slope of Re$\Sigma_{\rm ch}(\vk_{F}, \omega)$ at $\omega=0$ becomes steeper. These changes lead to a dramatic reduction in the quasiparticle weight $Z$ as the system approaches half-filling. The value of $Z$  drops from $0.24$ at $\delta=0.25$ to just $0.08$ at $\delta=0.05$ [see the inset of Fig.~\ref{T-d-depend}(d)]. Consequently, while the spectral function maintains a single peak, its total weight is strongly suppressed at lower doping. This counterintuitive trend—charge fluctuations have a stronger effect closer to half-filling where the charge degrees of freedom tend to be quenched—highlights the profound impact of electron correlations on the low-energy physics.

\subsubsection{Implication for overdoped cuprates}
Our theoretical findings on the electron self-energy from charge fluctuations provide a consistent framework for interpreting several key experimental observations in the overdoped regime of cuprates. In this region, the system is believed to behave as a Fermi liquid, where spin fluctuations become very weak \cite{thurston89,wakimoto07} but charge fluctuations persist. Our calculations show that the imaginary part of the self-energy from charge fluctuations,  Im$\Sigma_{\rm ch}$, is essentially isotropic along the Fermi surface and exhibits a characteristic $\sim T^{2}$ dependence as temperature approaches zero. These features are in remarkable agreement with experimental transport data, which reveal an isotropic scattering rate \cite{jawad06,jawad07,french09}  and a dominant $T^{2}$ dependence of the resistivity \cite{nakamae03,cooper09,harada22,jawad06,jawad07,french09}. Furthermore, our finding that Im$\Sigma_{\rm ch}$ decreases with increasing doping [Fig.~\ref{T-d-depend}(c)] directly mirrors the observed trend of decreasing resistivity with higher doping levels \cite{takagi92,timusk99}.
  
Beyond transport properties, our work provides a quantitative explanation for the observed mass enhancement. Our calculated quasiparticle weight $Z$, which approaches a value of approximately 0.25 in the overdoped regime, implies a mass enhancement factor ($m^{*}/m=1/Z$) of roughly 4. This result shows excellent consistency with values of 3.1 to 5.1 reported from quantum oscillation measurements in overdoped ${\rm Tl_2Ba_2CuO_{6+\delta}}$ \cite{vignolle08}.  The discrepancy with ARPES measurements for overdoped ${\rm Bi_2Sr_2CaCu_2O_{8+\delta}}$ \cite{johnson01}, which report a smaller value of $Z \approx 1.5$, may be attributed to the different energy scales probed by the two techniques. While quantum oscillations are sensitive to the low-energy properties precisely on the Fermi surface, ARPES integrates over a broader energy range, which may average out the strong self-energy effects localized near the Fermi level.
 
The combined consistency between our calculations and multiple independent experimental probes—including transport measurements and quantum oscillations—strongly suggests that charge fluctuations are a crucial scattering mechanism governing the low-energy physics of overdoped cuprates.

Our theory implies that the charge fluctuations are the major source to the mass enhancement in the overdoped regime. The coupling to spin-fluctuations is also additive to the mass enhancement, but we cannot estimate it in the present theory. Nonetheless, the present theory does not exclude other possibilities, for example, the possible mass enhancement factor by spin fluctuations.  

\subsection{Pseudogap property in charge fluctuations: underdoped cuprates} 
Our analysis demonstrated that realistic charge fluctuations do not destroy quasiparticles but rather induce a substantial suppression of the quasiparticle weight, with $Z$ ranging from 0.08 to 0.24, where $Z$ increases with doping. This Fermi-liquid behavior is a consistent outcome of our theory across various temperatures and Fermi momenta. Given that our model accurately captures the observed plasmon excitation spectra  \cite{greco19,greco20,nag20,hepting22,hepting23,nag24}, including others (see Sec.~4.1)  \cite{yamase15b,bejas17,yamase19}, we can be confident in the reliability of our calculated electron self-energy. However, this raises a fundamental question: if our model predicts Fermi-liquid behavior, what role do charge fluctuations play in the pseudogap state observed in underdoped cuprates, where the quasiparticle picture is notably absent?

To address this, we formally decompose the experimentally observed electron self-energy $\Sigma_{\rm ex}$ as a sum of contributions:
\be
\Sigma_{\rm ex} = \Sigma_{\rm ch} + \Sigma_{\rm pg} + \Sigma_{\rm others} \,. 
\label{selfex}
\ee
Here, $\Sigma_{\rm ch}$ is the contribution from the charge fluctuations we have calculated, and $\Sigma_{\rm pg}$ is the component responsible for the pseudogap feature in the spectral function. The term $\Sigma_{\rm others}$ encapsulates all other scattering contributions, such as those related to the strange metal state \cite{mitrano18,husain19,seibold21,caprara22}, marginal Fermi liquid behavior \cite{varma89}, and conventional Fermi-liquid corrections from other bosonic fluctuations \cite{carbotte11}. The idea behind this decomposition is the principle of superposition of the quantum mechanics. But possible interference among the charge fluctuations, pseudogap fluctuations, and other fluctuations are neglected in the present phenomenological analysis. For the sake of a clear and focused analysis, we will neglect the $\Sigma_{\rm others}$. 

\subsubsection{A minimal model of the pseudogap self-energy} 
By using realistic experimental data for $\Sigma_{\rm ex}$, we can estimate the contribution from the pseudogap $\Sigma_{\rm pg}$ by modeling it with a simplified form that has been shown to be consistent with various microscopic pseudogap theories \cite{norman07} (see Appendix~A for details):
\be
\Sigma_{\rm pg}(\vk, \omega) = \frac{c_{\vk}^{2}}{\omega + i \Gamma_{\vk}} \,. 
\label{selfPG}
\ee
This expression captures the essential features of the pseudogap, where $c_{\vk}$ represents a momentum-dependent gap-like parameter and $\Gamma_{\vk}$ is a broadening term. We shall focus on the antinodal Fermi momentum, simplifying the notations to $c_{\vk_{F}^{AN}}=c$ and $\Gamma_{\vk_{F}^{AN}}=\Gamma$. An important, yet often overlooked, aspect of this model is that the interplay between $c$ and $\Gamma$ is crucial for generating a true pseudogap.

\begin{figure}[tb]
\centering
\includegraphics[width=8cm]{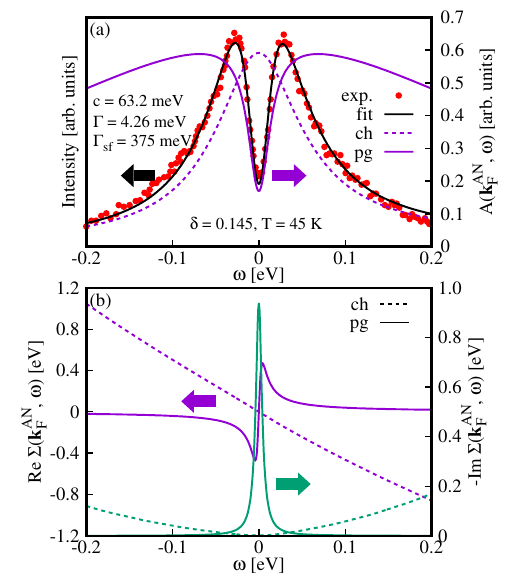}
\caption{Self-energy consistent with experimental data extracted from Ref.~\cite{kuspert22}. (a) Typical spectral function observed in underdoped cuprates at the antinodal region on the Fermi surface, showing the pseudogap, namely the suppression of the spectral weight at $\omega=0$. The solid black curve is a fitting in terms of Eqs.~(\ref{selfex}) and (\ref{selfPG}); we use $t/2=0.35$ eV \cite{hepting22,misc-factor2a}. 
The spectral functions in two different conditions, with only $\Sigma_{\rm ch}$ and with only $\Sigma_{\rm pg}$, are also plotted. 
(b) $\Sigma_{\rm ch}$ and $\Sigma_{\rm pg}$ used in the fitting to the experimental data in (a). Adapted from Ref.~\cite{yamase24} (\copyright\, 2024 American Physical Society).  
}
\label{PG-fit}
\end{figure}

Figure~\ref{PG-fit}(a) shows a recent experimental spectral function for LSCO \cite{kuspert22}, to which we fit our model by tuning the parameters $c$ and $\Gamma$ as well as a broadening of the spectral function $\Gamma_{\rm sf}$ [see \eq{Akw}] to reproduce the experimental data [\fig{PG-fit}(a)]. For a doping rate of $\delta=0.145$ and a temperature of $T=45$ K  (which corresponds to $T=0.0055t$ using $t/2=0.35$ eV  \cite{hepting22,misc-factor2a}), we find that our combined self-energy, $\Sigma_{\rm ex} \approx \Sigma_{\rm ch}$ + $\Sigma_{\rm pg}$ successfully reproduces the experimental data. The corresponding self-energy contributions are shown in Fig.~\ref{PG-fit}(b). We find that Im$\Sigma_{\rm pg}$ has a sharp peak at $\omega=0$ to generate the pseudogap, with its corresponding real part Re$\Sigma_{\rm pg}$ 
exhibiting a steep positive slope that effectively counteracts the negative slope of Re$\Sigma_{\rm ch}$. While we have used $\Gamma_{\rm sf}=0.03$ in Fig.~\ref{T-d-depend}, we obtain $\Gamma_{\rm sf}=0.536$ to get a better fit especially to the tails away from the Fermi energy in \fig{PG-fit}(a). This large $\Gamma_{\rm sf}$ may also reflect  broadening due to the other contributions $\Sigma_{\rm others}$. A surprising and significant finding is that despite the already small quasiparticle weight from charge fluctuations ($Z \approx 0.17$ at this doping), a very pronounced peak in Im$\Sigma_{\rm pg}$ is still necessary to reproduce the experimental pseudogap.

\subsubsection{Condition of pseudogap formation} 
\begin{figure}[tb]
\centering
\includegraphics[width=8cm]{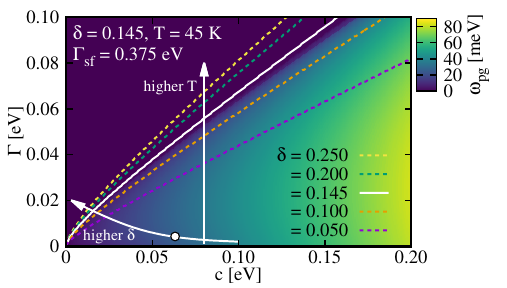}
\caption{Condition to realize the pseudogap in the presence of charge fluctuations in the plane of $c$ and $\Gamma$ in \eq{PG-eq}; we use $t/2=0.35$ eV \cite{hepting22,misc-factor2a}. The pseudogap ($\omega_{\rm pg} \ne 0$) is realized below the white curve.  Similar curves are also superimposed for other doping rates. To be consistent with the pseudogap observed experimentally, $c$ and $\Gamma$ may depend on doping and temperature as shown by arrows schematically. The solid circle corresponds to the values of $c$ and $\Gamma$ used in \fig{PG-fit}. Adapted from Ref.~\cite{yamase24} (\copyright\, 2024 American Physical Society). 
}
\label{PG-cond}
\end{figure}

Our analysis yields a crucial insight: the presence of charge fluctuations makes it more {\it difficult}  to form a pseudogap. We can quantify this by mapping the condition for a pseudogap to appear in the  $c$-$\Gamma$ plane, as shown in \fig{PG-cond}, where $\omega_{\rm pg}$ is a half distance of double peaks of $A(\vk, \omega)$. The pseudogap ($\omega_{\rm pg} \ne 0$) is realized only below a certain boundary, which can be approximated by the condition (see Ref.~\cite{yamase24} for an analytical understanding):  
\be
\Gamma^{2} <  2 Z_{\rm FL} c^{2} \,.
\label{PG-eq} 
\ee
Here $Z_{\rm FL}$ is the Fermi-liquid quasiparticle weight in the {\it absence} of $\Sigma_{\rm pg}$ (in the present case $Z_{\rm FL}=0.17$). This condition reveals a key result: the smaller the Fermi-liquid quasiparticle weight is, the more stringent the condition on $c$ and $\Gamma$ becomes for a pseudogap to form. In other words, a smaller $Z_{\rm FL}$ requires a more intense pseudogap contribution to destroy the remaining quasiparticles. This is a counterintuitive finding that highlights the complex interplay between different scattering mechanisms.

\begin{figure}[t]
\centering
\includegraphics[width=10cm]{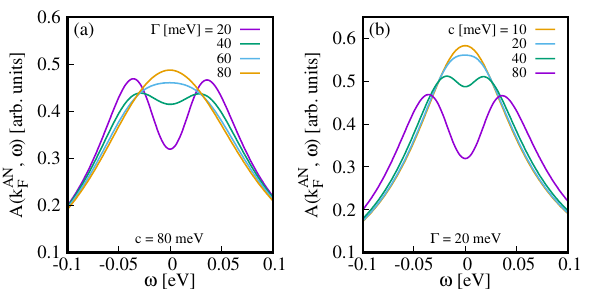}
\caption{Evolution of the spectral function at the antinodal Fermi momentum in the phase diagram shown in \fig{PG-cond}. (a) Several choices of $\Gamma$ at a fixed $c$ and (b) several choices of $c$ at a fixed $\Gamma$. Adapted from Ref.~\cite{yamase24} (\copyright\, 2024 American Physical Society). 
}
\label{Akw-Gc}
\end{figure}

Furthermore,  to be consistent with experiments, $c$ and $\Gamma$ should exhibit a special doping and temperature dependence as sketched with arrows in \fig{PG-cond}: the gap tends to be closed with decreasing $c$ and to be filled with increasing $\Gamma$---the former feature like a {\it gap-closing} may be caused mainly by increasing doping [see \fig{Akw-Gc}(a)] \cite{damascelli03} and the latter one like a {\it gap-filling} by increasing temperature [see \fig{Akw-Gc}(b)] \cite{norman98a,kanigel07,damascelli03}. These considerations lead us to sketch the arrows in \fig{PG-cond}. The microscopic origin of $c$ and $\Gamma$ is a challenge for understanding the pseudogap in cuprates. 

\subsubsection{Coherent and incoherent peaks}  

\begin{figure}[tb]
\centering
\includegraphics[width=8cm]{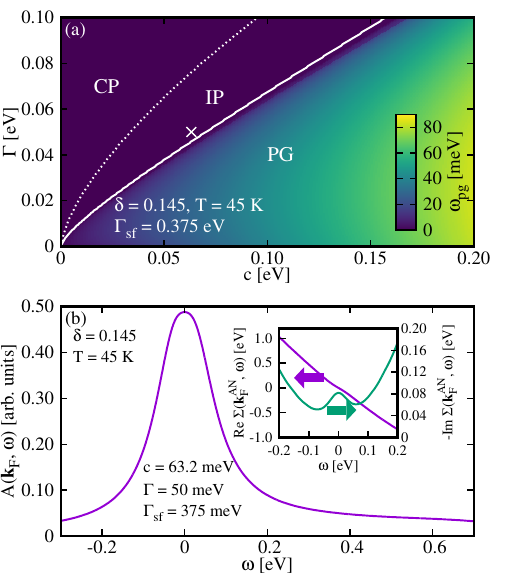}
\caption{Coherent and incoherent single peaks. (a) Reproduction of \fig{PG-cond}, but focusing on the results for $\delta=0.145$. The non-pseudogap region above the white curve is divided into two regions, where a coherent or an incoherent single peak is realized. (b) Spectral function at the antinodal Fermi momentum in the IP region marked by the cross in (a). The inset shows that the real part of the self-energy has a negative slope but the imaginary part has a peak at $\omega = 0$. Adapted from Ref.~\cite{yamase24} (\copyright\, 2024 American Physical Society).
} 
\label{incoherent}
\end{figure}

The condition for the pseudogap also defines a transition to states with a single peak at $\omega=0$. As shown in Fig.~\ref{incoherent}(a), the non-pseudogap region can be further divided into two distinct regimes: a coherent peak (CP) and an incoherent peak (IP) state. The CP is a Fermi-liquid feature as we have already discussed in Fig.~\ref{T-d-depend}, while the IP state shown in \fig{incoherent}(b) is a unique crossover where the imaginary part of the self-energy has a peak at $\omega=0$, but the real part retains a negative slope, as shown in the inset \cite{misc-IP}. This IP state, which intervenes between the pseudogap and the coherent Fermi-liquid states \cite{keimer15}, is particularly intriguing and may be related to the strange metal phase observed in cuprates.

\subsubsection{Disentangling self-energy contributions} 
Recently charge fluctuations were proposed to be responsible for a strange metal and the marginal Fermi-liquid phenomenology \cite{mitrano18,husain19,seibold21,caprara22}. This perspective is not supported by the present theory. 
Our results demonstrate that charge fluctuations play a crucial, but often overlooked, role in the low-energy physics of cuprates. While they are a key source of scattering, they do not, by themselves, drive the system into a non-Fermi liquid or pseudogap state. Instead, they act in conjunction with the pseudogap-generating mechanism to produce the experimentally observed spectrum. A major challenge remains in identifying the microscopic origin of the parameters $c$ and $\Gamma$, which must exhibit a specific doping and temperature dependence to be consistent with experiments. Our findings underscore the importance of disentangling the various self-energy contributions from experimental data. A possible approach is to assume that the pseudogap vanishes at the nodal Fermi momentum, allowing one to extract the Fermi-liquid contributions from the total self-energy. Subtracting these from the total self-energy at the antinodal momenta would then provide an estimate for the pseudogap self-energy, a procedure that could be applied to both numerical \cite{gunnarsson15,dong19} and experimental data.

\subsection{A role of spin fluctuations}
In the present large-$N$ theory of the $t$-$J$ model, the charge fluctuations appear already at leading order whereas spin fluctuations do at next-leading order. In this sense, in the present theory, the spin fluctuations are not relevant. Typical energy scale of spin fluctuation is less than 100 meV. Hence, we would not expect that the renormalization in high-energy region (Sec.~3.3) is affected by spin fluctuations. But they may be relevant to the low-energy physics especially in Sec.~3.5. The effect of spin fluctuations enters in $\Sigma_{\rm pg}$ or/and $\Sigma_{\rm others}$. We cannot specify a role of spin fluctuations, but the interplay with charge fluctuations yields intriguing results as shown in Figs.~\ref{PG-fit}--\ref{incoherent}.

\section{Charge fluctuations from a nearest-neighbor spin interaction} \label{short-range-section}
\begin{figure}[th]
\centering
\includegraphics[width=12cm]{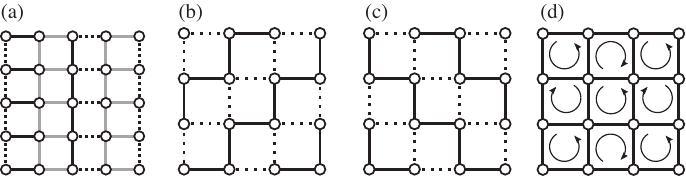}
\caption{Sketch of various bond-charge orders in real space. The black and dotted lines denote stronger and weaker bonds, respectively; the gray lines correspond to bonds with intermediate strengths. 
(a) and (b) $d$-bond states with $\vq=(0.5\pi,0)$ and $(\pi,\pi)$, respectively.  (c) $s$-bond states with $\vq=(\pi,\pi)$. (d) $d$CDW with $\vq=(\pi,\pi)$, where staggered circulating currents flow in each plaquette. This state is also refereed to as flux phase. Adapted from Refs.~\cite{bejas14} [\copyright\, 2014 The Author(s)] and \cite{yamase15b} (\copyright\, 2015 EPLA) for (b)--(d) and (a), respectively.  
}
\label{bond-order}
\end{figure}

In this section, we consider charge fluctuations especially from bond-charge fluctuations, which are known to arise from a nearest-neighbor spin-spin interaction \cite{affleck88a, marston89,sachdev91,vojta99,yamase00b,vojta02,yamase21c}. As such, they are a distinct form of charge fluctuations that go beyond the on-site Hubbard interaction. We consider three specific types of bond-charge fluctuations:  the $d$-wave bond-charge order ($d$-bond), the $s$-wave bond-charge order ($s$-bond), and the $d$-wave charge-density-wave ($d$CDW). Their real-space ordering patterns are depicted in \fig{bond-order}. The $d$-bond state with $\vq=0$ is particularly significant, as it corresponds to the electronic nematic order \cite{yamase00a,yamase00b,metzner00}.  We put $t=1$ as the energy unit.

\subsection{Dual structure of the charge-excitation spectrum} 
There are two different charge excitations: one is plasmons from the LRC and the other is bond-charge fluctuations from the nearest-neighbor spin exchange interaction. Hence the full charge excitation spectrum becomes a superposition of those excitations.

\begin{figure}
\centering
\includegraphics[width=8cm]{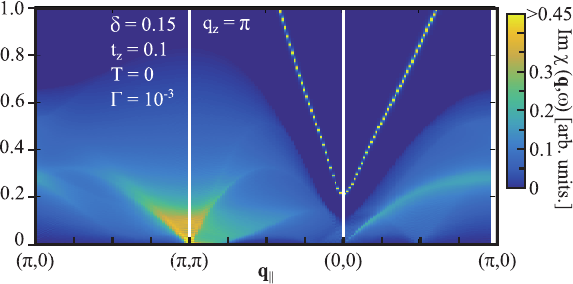}
\caption{$\qp$-$\omega$ map of the superposition of excitation spectra of 
$\chi_c$, $\chi_{d{\rm bond}}$, $\chi_{s{\rm bond}}$, and  $\chi_{d{\rm CDW}}$ 
along the symmetry axes $(\pi,0)$--$(\pi,\pi)$--$(0,0)$--$(\pi,0)$ for $q_z=\pi$ at three different doping rates: 
(a) $\delta=0.15$, (b) $\delta=0.20$, and (c) $\delta=0.25$. Adapted from Ref.~\cite{bejas17}  (\copyright\, 2017 American Physical Society).  
}
\label{qw-map-sum-pi}
\end{figure}

Figure~\ref{qw-map-sum-pi} reveals a striking dual structure---the usual charge fluctuations $\chi_c$ and three types of bond-charge fluctuations: $d$-bond ($\chi_{d{\rm bond}}$), $s$-bond ($\chi_{s{\rm bond}}$), and  $d$CDW ($\chi_{d{\rm CDW}}$). This structure is characterized by two distinct features: a gapped V-shaped dispersion that extends to high energies around $\qp =(0,0)$, and a continuum spectrum confined to the low-energy region, typically on the scale of the superexchange interaction $J$ $(=0.3)$. 

The origin of these two features is completely different. As we demonstrated in Sec.~\ref{exp-LRC}, the high-energy V-shaped dispersion arises exclusively from the plasmon mode with a finite $q_z$. This mode is governed by the $2 \times 2$ sector of \eq{dyson} and receives no contribution from the bond-charge excitations. On the other hand, the low-energy spectrum is a rich superposition of the bond-charge fluctuations, which have much larger spectral weight, along with the particle-hole continuum from $\chi_{c}$. Specifically, we observe a large spectral weight in the very low-energy region around $\qp=(\pi,\pi)$ [see also Figs.~\ref{bond-order}(b)--(d)]. Unfortunately, this momentum region is often inaccessible to RIXS. 

A peak along the direction $(0,0)$--$(\pi,0)$ originates from individual fluctuations associated with $d$CDW. The comprehensive study of these bond-charge excitations, including the effect of superconducting gap, can be found in Ref.~\cite{zafur24}. 

There is also a cusplike feature of the continuum at $\qp=(0.5\pi,0)$ and $\omega=0$, which becomes barely discernible in \fig{qw-map-sum-pi}. This is associated with a subleading $d$-wave bond-charge ordering tendency [see \fig{bond-order}(a)], which was analyzed in Refs.~\cite{yamase15b,bejas17,li17,yamase19,zafur24} in detail as an explanation of charge-order tendency reported in NCCO \cite{da-silva-neto15,da-silva-neto16,da-silva-neto18}. 

\section{Bilayer charge excitations}\label{bilayer-section}
Up to Sec.~4, we have studied a model, which contains one CuO$_{2}$ plane in the unit cell. As discussed in Sec.~2.2, it is known that the superconducting onset temperature is enhanced with increasing the number of CuO$_{2}$ planes and $T_c$ becomes higher than typical single-layer cuprates. Toward bridging the superconducting mechanism of cuprates, it is important to analyze the three-dimensional charge dynamics in the bilayer system and clarify the difference to the single-layer case. However, the extension to bilayer systems is not straightforward. A critical first step is to derive the functional form of the LRC for a bilayer lattice systems, as this was not known. To address this challenge, we step away from the specific $t$-$J$-$V$ model used so far and review the derivation of the LRC on a generic bilayer lattice, and then perform calculations within the RPA to highlight the key difference from the single-layer case \cite{yamase25}. 

\subsection{Hamiltonian}
\begin{figure}[th]
\centering
\includegraphics[width=8cm]{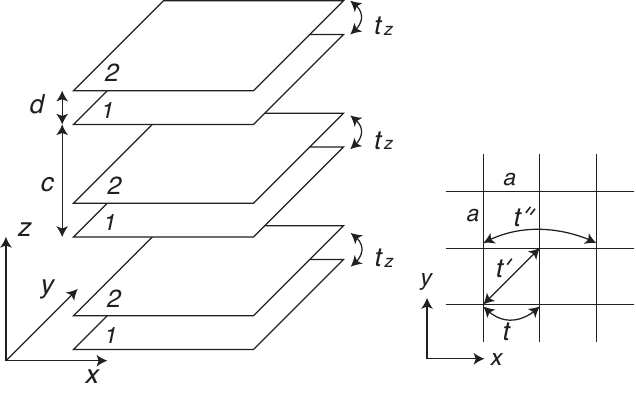}
\caption{Bilayer model. Each layer forms a square lattice and the hopping integrals are considered up to third nearest-neighbor sites, $t$, $t'$, and $t''$ (right figure). The unit cell contains two layers, 1 and 2. The intrabilayer hopping is given by $t_{z}$. While the interbilayer hopping is neglected, the LRC is three dimensional and present between different layers. The lattice constants are given by $a$, $a$ and $c$ along the $x$, $y$, and $z$ directions, respectively, and the intrabilayer distance is $d (\leq c/2)$. 
}
\label{bilayer}
\end{figure}

We consider an electron system interacting with the LRC on a bilayer square lattice stacked along the $z$ axis with the intrabilayer distance $d$ and interbilayer distance $c$ as shown in \fig{bilayer}; we can assume $0< d \leq c/2$ without losing generality. 
The total Hamiltonian consists of two terms: a kinetic term and an interaction term, 
\be
\mathcal{H}=\mathcal{H}_{0}+\mathcal{H}_{I} \,.
\ee
The kinetic term $\mathcal{H}_{0}$ describes the electron dispersion on the bilayer lattice and is given by 
\be
\mathcal{H}_{0}= \sum_{\vk, \sigma}  
\begin{pmatrix} 
c_{1 \vk \sigma}^{\dagger}   c_{2 \vk \sigma}^{\dagger} 
\end{pmatrix}
\begin{pmatrix}
 \xi_{\vk}  &  \varepsilon_{\vk} {\rm e}^{i k_{z} d} \\
\varepsilon_{\vk}^{*} {\rm e}^{-i k_{z} d} & \xi_{\vk} 
\end{pmatrix}
\begin{pmatrix}
c_{1 \vk \sigma} \\
c_{2 \vk \sigma} 
\end{pmatrix} \,,
\label{H0}
\ee
where $c_{1 \vk \sigma}^{\dagger}$ ($c_{2 \vk \sigma}^{\dagger}$) and $c_{1 \vk \sigma}$ ($c_{2 \vk \sigma}$) are creation and annihilation operators for  electron with momentum $\vk$ and spin $\sigma$ on layer 1 (2). The in-plane and out-of-plane dispersion, $\xi_{\vk}$ and $\varepsilon_{\vk}$, respectively, are defined as 
\be
\xi_{\vk} = -2 t (\cos k_{x}a + \cos k_{y}a ) -4 t' \cos k_{x}a \cos k_{y}a  
-2t'' (\cos 2k_{x}a + \cos 2k_{y}a ) - \mu \, , 
\label{xiplane} 
\ee
\be
 \varepsilon_{\vk} = - t_{z} (\cos k_{x}a - \cos k_{y}a)^{2}  \, .
\label{xiperp}
\ee
The form factor $\cos k_{x}a - \cos k_{y}a$ for the intrabilayer hopping integral $t_{z}$ is consistent with local density approximation (LDA) band calculations \cite{andersen95}. We have neglected any hopping integral between adjacent bilayers, as its magnitude is expected to be much smaller than $t_{z}$. A role of the interbilayer hopping will be discussed in Sec.~5.5. 

The interaction term $\mathcal{H}_{I}$ describes the density-density interaction between electrons on the lattice. The lattice sites in the layer 1 in \fig{bilayer} are specified by $\vr_{i}= n_{x} a \hat{{\bf x}} + n_{y} a \hat{{\bf y}}  +  n_{z} c \hat{{\bf z}}$ with  the unit vectors $\hat{{\bf x}}$, $\hat{{\bf y}}$, and $\hat{{\bf z}}$ along the $x$, $y$, and $z$ directions, respectively, and with integers $n_{x}$, $n_{y}$, and $n_{z}$. The lattice sites on the layer 2 are then described by $\vr_{i}+{\bf d}$ with ${\bf d}=(0,0,d)$. Hence in real space, this interaction is given by 
\bea
&&H_{I}=\frac{1}{2} \sum_{i j} \left[ n_{1}(\vr_{i}) V( \vr_{j} - \vr_{i}) n_{1}(\vr_{j})  \right. \nonumber \\ 
&&\hspace{20mm} + \left. n_{2}(\vr_{i}+{\bf d}) V( \vr_{j} - \vr_{i}) n_{2}(\vr_{j} +{\bf d})  \right.  \nonumber \\
&& \hspace{20mm} + \left. n_{1}(\vr_{i}) V( \vr_{j} + {\bf d} - \vr_{i}) n_{2}(\vr_{j} +{\bf d}) \right.   \nonumber \\
&& \hspace{20mm}  + \left. n_{2}(\vr_{i}+{\bf d}) V( \vr_{j} - \vr_{i} - {\bf d}) n_{1}(\vr_{j}) \right] \,,  
\eea
where $n_{1}$ and $n_{2}$ are electron density operators for layers 1 and 2, respectively. Upon performing a Fourier transform, 
\bea
&& n_{1}(\vr_{i}) =  \frac{1}{N_{0}} \sum_{\vq} n_{1}(\vq) {\rm e}^{i \vq \cdot \vr_{i}} \,, \\
&& n_{2}(\vr_{i}+{\bf d}) = \frac{1}{N_{0}} \sum_{\vq} n_{2}(\vq) {\rm e}^{i \vq \cdot (\vr_{i} +{\bf d}) } \,, 
\eea
we obtain the interaction Hamiltonian in momentum space
\be
\mathcal{H}_{I} = \frac{1}{2N_{0}} \sum_{\vq} 
\begin{pmatrix}
n_{1}(\vq) \; n_{2}(\vq) 
\end{pmatrix}
\begin{pmatrix}
V(\vq) \; V^{'}(\vq) \\
V^{'}(-\vq)  \; V(\vq) 
\end{pmatrix}
\begin{pmatrix}
n_{1}(-\vq)  \\
n_{2}(-\vq) 
\end{pmatrix} \,,
\label{HI}
\ee
where $N_{0}$ is the total number of lattice sites; $n_{1}(\vq)= \sum_{\vk, \sigma} c_{1 \vk  \sigma}^{\dagger}  c_{1 \vk + \vq \sigma}$ and $n_{2}(\vq)= \sum_{\vk, \sigma} c_{2 \vk \sigma}^{\dagger}  c_{2 \vk+\vq \sigma}$ are the electron density operators of the momentum $\vq$ in each layer. The Fourier transform of the interaction part is defined by 
\bea
 &&V({\bf l}) = \frac{1}{N_{0}} \sum_{\vq} V(\vq) {\rm e}^{i \vq \cdot {\bf l}} 
 \label{vq} \,,\\
 &&V({\bf l}+ {\bf d}) = \frac{1}{N_{0}} \sum_{\vq} V^{'}(\vq) {\rm e}^{i \vq \cdot ({\bf l} + {\bf d})} 
 \label{v'q} \, ,
 \eea
with ${\bf l} = \vr_{j} -\vr_{i}$.

\subsection{Lattice LRC on a bilayer system}
It is well-established that the LRC in a continuum space is given by $V({\bf l}) \propto  \frac{e^{2}}{| {\bf l} |}$, which in momentum space takes the form 
$V(\vq) \propto \frac{e^{2}}{q^{2}}$ in three dimensions; $e$ is the electric charge. While the procedure for calculating the LRC on a Bravais lattice is known \cite{becca96},  a comprehensive expression for $V(\vq)$ and $V'(\vq)$ in a bilayer system, or more generally multilayer system, is not readily available in the literature beyond the pioneering work by Fetter \cite{fetter74,griffin89}. 

To fill this gap, we begin with the Poisson equation, generalized to account for spatial anisotropy: 
\be
\epsilon_{\parallel} \left(  
\frac{\partial^{2}}{\partial x^{2}} +  \frac{\partial^{2}}{\partial y^{2}} \right) G(\vr) +
\epsilon_{\perp} \frac{\partial^{2}}{\partial z^{2}}G(\vr)  = -\delta(\vr) \,,
\label{Poisson}
\ee
where $\epsilon_{\parallel} (\epsilon_{\perp})$ is the dielectric constant parallel (perpendicular) to the layer. On the lattice, we replace the second partial derivative with a finite difference approximation: 
\bea
&&\frac{\epsilon_{\parallel}}{a^{2}} \left\{ \left[ 
G(\vr_{i} + a \hat{\bf x}) - 2 G(\vr_{i}) +  G(\vr_{i} - a \hat{\bf x})  \right] \right. \nonumber \\
&&\hspace{5mm} +\left. \left[ G(\vr_{i} + a \hat{\bf y}) - 2 G(\vr_{i}) +  G(\vr_{i} - a \hat{\bf y}) 
 \right] \right\}  \nonumber \\
&&\hspace{0mm} +  \epsilon_{\perp} \left[ h_{1} G(\vr_{i} + c \hat{\bf z}) + h_{2} G(\vr_{i} + d \hat{\bf z}) +
 h_{3} G(\vr_{i}) +  h_{4} G(\vr_{i} - (c-d) \hat{\bf z} ) \right] \nonumber \\
&& \hspace{0mm}  =- \frac{1}{N_{0}}\sum_{\vq}{\rm e}^{i \vq \cdot \vr_{i}} \,.
\eea
To ensure the equation is valid even for $d=0$, we evaluate the second derivative in the $z$ direction using four points. The coefficients, $h_{1}$, $h_{2}$, $h_{3}$, and $h_{4}$ are determined uniquely by a Taylor expansion, yielding 
\bea
&&h_{1}= \frac{2 (c-2d)}{c(2c-d) (c-d)} \,,  \label{h1}\\
&&h_{2}= \frac{2}{c(c-d)} \,, \\ 
&&h_{3}= -\frac{4}{c(c-d)} \,, \\ 
&&h_{4}= \frac{2 (c+d)}{c(2c-d) (c-d)}  \label{h4} \,.
\eea
Taking the the Fourier transform of this equation gives: 
\bea
&& \left[ \frac{2 \epsilon_{\parallel}} {a^{2}} (2-\cos q_{x}a - \cos q_{y}a) - \epsilon_{\perp} (h_{1} {\rm e}^{i q_{z} c} + h_{3}) \right] G(\vq) \nonumber \\
&& - \epsilon_{\perp} {\rm e}^{i q_{z}d} ( h_{2} + h_{4} {\rm e}^{-i q_{z}c} ) G^{'}(\vq) = 1 
\label{Poisson1} \,.
\eea
The presence of $G^{'}(\vq)$ suggests a need for second independent equation. We obtain this by considering the Poisson equation at a site on the second layer, namely at ${\bf r}_{i}+ {\bf d}$, 
\bea
&&\frac{\epsilon_{\parallel}}{a^{2}} \left\{ \left[ 
G(\vr_{i} + {\bf d} + a \hat{\bf x}) - 2 G(\vr_{i} + {\bf d}) +  G(\vr_{i} + {\bf d} - a \hat{\bf x})
 \right] \right. \nonumber \\
 &&\hspace{5mm} + \left.\left[ 
 G(\vr_{i} + {\bf d} + a \hat{\bf y}) - 2 G(\vr_{i} + {\bf d}) +  G(\vr_{i} + {\bf d} - a \hat{\bf y})
 \right] \right\} \nonumber \\
&&\hspace{0mm} + \epsilon_{\perp} \left[ h_{5} G(\vr_{i} + c \hat{\bf z}) + h_{6} G(\vr_{i} + d \hat{\bf z}) +
 h_{7} G(\vr_{i}) +  h_{8} G(\vr_{i} - (c-d) \hat{\bf z} ) \right]  \nonumber \\
 && =0 \,.
 \label{Possion2}
\eea
The right hand side becomes zero because $\delta(\vr_{i} + {\bf d})=0$ in the condition $0 < d \leq c/2$. The Taylor expansion shows that $h_{5}=h_{4}$, $h_{6}=h_{3}$, $h_{7}=h_{2}$, and $h_{8}=h_{1}$. Fourier transforming this second equation gives: 
\bea
&& \left[ \frac{2 \epsilon_{\parallel}} {a^{2}} (2-\cos q_{x}a - \cos q_{y}a) 
- \epsilon_{\perp} (h_{1} {\rm e}^{-i q_{z} c} + h_{3}) \right] G^{'}(\vq) \nonumber \\
&& - \epsilon_{\perp} {\rm e}^{-i q_{z}d} ( h_{2} + h_{4} {\rm e}^{i q_{z}c} ) G(\vq) = 0 
\label{Poisson2} \,.
\eea
Combining Eqs.~(\ref{Poisson1}) and (\ref{Poisson2}), we arrive at a $2 \times 2$ matrix equation
\be
\begin{pmatrix}
A_{11} \; A_{12} \\
A_{12}^{*}  \;  A_{11}^{*} 
\end{pmatrix}
\begin{pmatrix}
G(\vq) \\
G^{'}(\vq) 
\end{pmatrix} 
= 
\begin{pmatrix}
1  \\
0
\end{pmatrix} \,,
\ee
where 
\bea
&&A_{11}= \frac{2 \epsilon_{\parallel}}{a^{2}} (2 - \cos q_{x}a - \cos q_{y}a ) -   
\epsilon_{\perp}(h_{1} {\rm e}^{iq_{z} c} + h_{3}) \,, 
\label{A11} \\
&& A_{12}= -\epsilon_{\perp} {\rm e}^{iq_{z} d} ( h_{2} + h_{4} {\rm e}^{-iq_{z} c} ) \,. 
\label{A12}
\eea
Solving this yields the solutions for the bilayer system in momentum space: 
\bea
&&G(\vq)= \frac{A_{11}^{*}}{ |A_{11}|^2 -  |A_{12}|^2} \,, \\ 
&&G^{'}(\vq)= \frac{-A_{12}^{*}}{ |A_{11}|^2 -  |A_{12}|^2} \,.
\eea

These solutions are directly related to the diagonal and off-diagonal components of the lattice  LRC: 
\be
V(\vq)=\frac{e^{2}}{a^{2}c} {\rm Re} G(\vq)\,, \quad V^{'}(\vq)=\frac{e^{2}}{a^{2}c} G^{'}(\vq)  \,,
\label{Vq-define}
\ee
where the factor $1/(a^{2}c)$ accounts for the volume of the unit cell. The real part of $G(\vq)$ is taken for $V(\vq)$ because only the even component with respect to $\vq$ contributes to the diagonal part of the interaction in \eq{HI}. 

For convenience, we present the explicit expressions for $V(\vq)$ and $V^{'}(\vq)$ as computed from the above results: 
\bea
&& \hspace{-8mm} V(\vq)=\frac{V_{c}}{{\rm det}\tilde{V}}\left[
\alpha(2-\cos q_{x}a - \cos q_{y}a) - \frac{1}{2} \tilde{h}_{3} - \frac{1}{2} \tilde{h}_{1} \cos q_{z}c \right]
\label{Vq} \,, \\
&&  \hspace{-8mm} V^{'}(\vq)=\frac{1}{2} \frac{V_{c}}{{\rm det}\tilde{V}}\left[ 
\tilde{h}_{2} \cos q_{z}d + \tilde{h}_{4} \cos q_{z}(c-d)  -i \tilde{h}_{2}\sin q_{z}d + i   \tilde{h}_{4}\sin q_{z}(c-d)  \right] 
\label{Vqp}  \,, \\
&& \hspace{-8mm} {\rm det}\tilde{V} = \left[ \alpha ( 2 - \cos q_{x}a - \cos q_{y}a) \right]^{2} 
- \alpha ( 2 - \cos q_{x}a - \cos q_{y}a) (\tilde{h}_{1} \cos q_{z}c + \tilde{h}_{3})  \nonumber \\
&& \hspace{12mm} + \frac{6 c^{2}}{(c-d)(2c-d)} (1 - \cos q_{z}c) \,.
\eea
Here $\tilde{h}_{i} = c^{2} h_{i}$ with $i=1,2,3,4$ are the dimensionless coefficients given in Eqs.~(\ref{h1})--(\ref{h4}). We have introduced two key parameters: $V_{c}= \frac{e^{2} c}{2 a^{2} \epsilon_{\perp}}$, which has the dimension of energy, and the anisotropy parameter $\alpha= \frac{c^{2} \epsilon_{\parallel}}{a^{2}\epsilon_{\perp}}$, which is the ratio of the dielectric constants scaled by the lattice geometry. These notations are consistent with those used in the single-layer case [see Eqs.~(\ref{LRC}) and (\ref{Aq})]  \cite{greco19,nag20,greco20,hepting22,zinni23,hepting23,nag24}.

\subsection{Dynamical charge susceptibility on a bilayer lattice} 
With the bilayer Hamiltonian now established, we can proceed to study the collective charge dynamics of the system. This requires a detailed examination of the dynamical charge susceptibility, $\kappa_{ij}(\vq, \omega)$, which describes the response of the electron system to an external charge perturbation. Given the two-layer structure, the susceptibility is naturally described by a $2 \times 2$ matrix, and we will compute it within RPA. 

The RPA charge susceptibility is given by 
\be
\kappa_{i j}(\vq, \omega) =  \kappa_{i j}^{0}(\vq, \omega)+  \sum_{l_{1}, l_{2}}\kappa_{i l_{1}}^{0}(\vq, \omega)  V_{l_{1} l_{2}}(\vq) \kappa_{l_{2} j} (\vq, \omega) \, , 
\label{biRPA}
\ee
where $i, j, l_{1}, l_{2}$ run over the two layers 1 and 2. The non-interacting susceptibility matrix $\kappa^{0}_{i j}(\vq, \omega)$ represents the simple bubble diagram and is expressed as 
\bea
&&\kappa_{11}^{0}(\vq, \omega) = \frac{1}{2N_{0}} \sum_{\vk} (g_{++} + g_{+-} + g_{-+} + g_{--}) \,, \\
&&\kappa_{12}^{0}(\vq, \omega) = \frac{1}{2N_{0}} \sum_{\vk} \frac{\varepsilon_{\vk} \varepsilon_{\vk+\vq}^{*} {\rm e}^{-i q_{z}d}}{ |\varepsilon_{\vk} | |\varepsilon_{\vk + \vq} |} (g_{++} - g_{+-} - g_{-+} + g_{--}) \,, 
\label{ko12}\\
&&\kappa_{21}^{0}(\vq, \omega) =  \frac{1}{2N_{0}} \sum_{\vk} \frac{\varepsilon_{\vk}^{*} \varepsilon_{\vk+\vq} {\rm e}^{i q_{z}d}}{ |\varepsilon_{\vk} | |\varepsilon_{\vk + \vq} |} (g_{++} - g_{+-} - g_{-+} + g_{--})  \,, 
\label{ko21} \\
&&\kappa_{22}^{0}(\vq, \omega) =\kappa_{11}^{0}(\vq, \omega) \,, 
\eea
where the sum is over the first Brillouin zone. The function $g_{\mu \nu} (\vk, \vq, \omega)$ are given by  
\be
g_{\mu \nu} (\vk, \vq, \omega)= \frac{f(\lambda_{\mu}(\vk)) - f(\lambda_{\nu}(\vk + \vq))} {\lambda_{\mu} (\vk) +\omega + i \Gamma - \lambda_{\nu}(\vk + \vq)} \, ,
\label{gmunu}
\ee
with $f(x)$ being the Fermi distribution function. The eigenenergies $\lambda_{\pm} (\vk) = \xi_{\vk} \pm | \varepsilon_{\vk} |$ correspond to the antibonding and bonding bands, respectively [see Eqs.~(\ref{xiplane}) and (\ref{xiperp})]. $\Gamma (>0)$ is a small broadening parameter used for  numerical convenience. The interaction matrix $V_{l_{1} l_{2}}$ is given by the components of the matrix in \eq{HI}. The total susceptibility of the system is then given by the sum of all components: 
\be
\kappa(\vq, \omega)= \frac{1}{2} \sum_{i j} \kappa_{i j}(\vq, \omega) \,.
\label{kqw}
\ee
While this is a compact representation, it is more illuminating to examine the explicit forms of the susceptibility, beginning with the non-interacting case. 

\subsubsection{Non-interacting case}
The total non-interacting charge susceptibility \eq{kqw} can be expressed in terms of the even and odd modes, which are decoupled from each other: 
\bea
&&\kappa^{0}(\vq, \omega) =\frac{1}{2} \left[
 \kappa^{0}_{11}(\vq,  \omega) + \kappa^{0}_{12}(\vq,  \omega)+ \kappa^{0}_{21}(\vq,  \omega)+\kappa^{0}_{22}(\vq,  \omega) 
 \right] \\
&&
\hspace{15mm} =\cos^{2} \frac{q_{z}d}{2}  \kappa^{0}_{\rm even}(\vq, \omega)+  \sin^{2} \frac{q_{z}d}{2} \kappa^{0}_{\rm odd}(\vq, \omega) \,,
\label{kqw0}
\eea
where
\bea
&& \kappa^{0}_{\rm even}(\vq, \omega)  = \frac{1}{N_{0}} \sum_{\vk} (g_{++} + g_{--}) \,,\\
&& \kappa^{0}_{\rm odd}(\vq, \omega)  = \frac{1}{N_{0}} \sum_{\vk} (g_{+-} + g_{-+} ) \, .
\eea
The functions $\kappa^{0}_{\rm even}(\vq, \omega)$ and $\kappa^{0}_{\rm odd}(\vq, \omega)$ represent the susceptibility from the intraband and interband  scattering processes, respectively. They are often referred to as the even and odd modes, which are selected by choosing $q_{z}d=0$ and $\pi$, respectively. 

\subsubsection{Inclusion of lattice LRC} \label{bilayerLRC}
\begin{figure}[t]
\centering
\includegraphics[width=14cm]{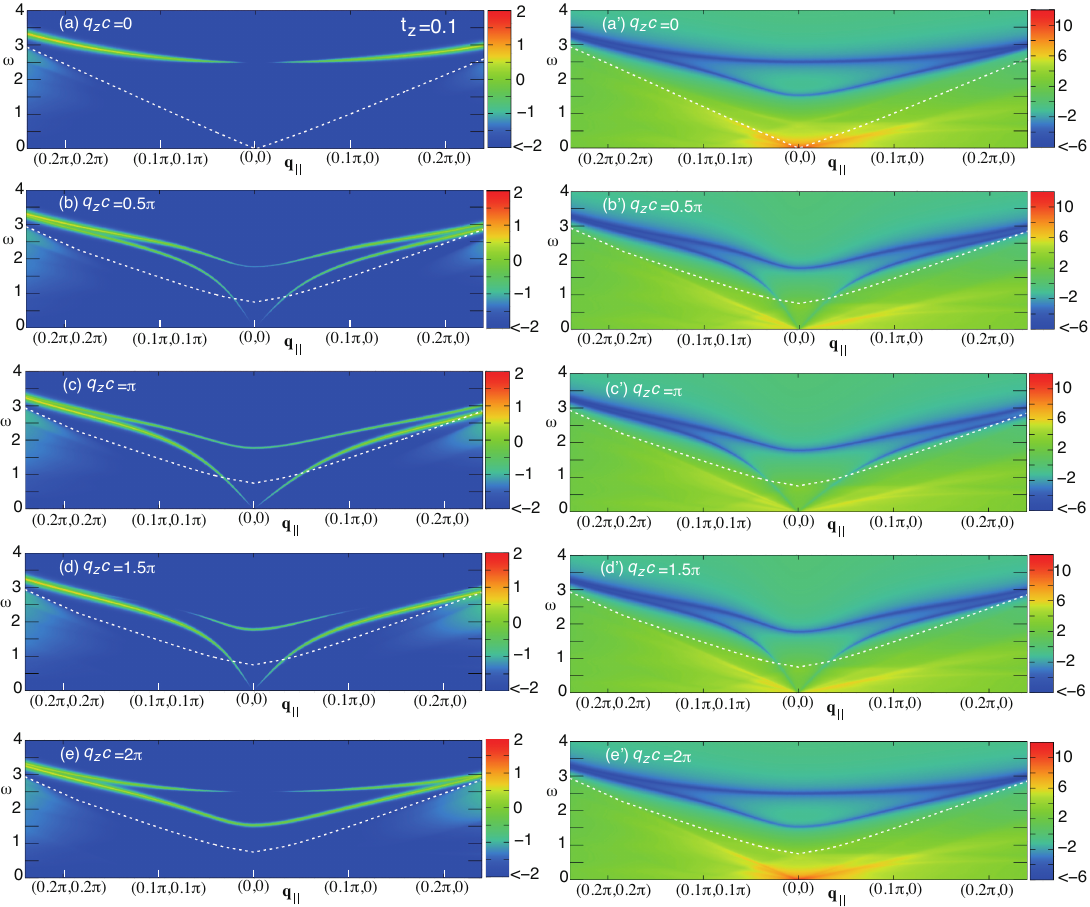}
\caption{Intensity maps of $\log_{10} | {\rm Im}\kappa(\vq, \omega)|$ (left panels) and $\log_{10}|\mathfrak{det}|^{2}$ given in \eq{kqw1} (right panels) for a sequence of $q_{z}c$ around a region of $\vq_{\parallel}=(0,0)$: (a) and (a') $q_{z}c=0$, (b) and (b') $q_{z}c=0.5\pi$, (c) and (c') $q_{z}c=\pi$, (d) and (d') $q_{z}c=1.5\pi$, and (e) and (e') $q_{z}c=2\pi$. The white dotted curve denotes the upper boundary of the particle-hole continuum. It goes to zero at $\vq_{\parallel} =(0,0)$ and $q_{z}c=0$ even for a finite $t_{z}=0.1$. To keep an appropriate contrast, the same color is used below -2 and -6 in the left panels and the right panels, respectively. Adapted from Ref.~\cite{yamase25} (\copyright\, 2025 American Physical Society). 
}
\label{kqw-00}
\end{figure}

The introduction of the lattice LRC, which includes both intra- and interbilayer interactions, significantly modifies the picture. Unlike the non-interacting case, the even and odd modes are now coupled, and the total susceptibility is no longer a simple sum. After solving the RPA equation, we obtain the full susceptibility 
\bea
&&\kappa(\vq, \omega) = \frac{1}{\mathfrak{det}} \left[ \cos^{2} \frac{q_{z}d}{2} \kappa^{0}_{\rm even}(\vq, \omega) + \sin^{2} \frac{q_{z}d}{2} \kappa^{0}_{\rm odd}(\vq, \omega)  \right. \nonumber \\
&& \left. \hspace{18mm} - \kappa^{0}_{\rm even}(\vq, \omega)\kappa^{0}_{\rm odd}(\vq, \omega) 
V^{''}(\vq) \right] \,,
\label{kqw1}
\eea
where 
\bea
&&\mathfrak{det} = \left[ 1  - \left( V(\vq) + V_{+}(\vq) \right) \kappa^{0}_{\rm even}(\vq,  \omega) \right] \nonumber  \\
&& \hspace{10mm} \times \left[ 1-  \left(V(\vq) - V_{+}(\vq) \right) \kappa^{0}_{\rm odd}(\vq,  \omega)
\right] \nonumber  \\
&&\hspace{12mm} +\kappa^{0}_{\rm even}(\vq, \omega) \kappa^{0}_{\rm odd}(\vq,  \omega) V_{-}(\vq)^{2} \,, 
\label{kqw1det} \\
&& V^{''}(\vq) = V(\vq) - \frac{V^{'}(\vq) + V^{'}(-\vq)}{2} \,, 
\label{Vpm} \\
&& V_{\pm}(\vq)= \frac{V^{'}(\vq) {\rm e}^{i q_{z}d} \pm V^{'}(-\vq) {\rm e}^{-i q_{z}d}}{2} \,.
\label{Vpp}
\eea
This result reveals a crucial physical insight: the even and odd modes are no longer fully decoupled, as seen in the presence of the coupling terms within the numerator and the denominator. This coupling arises directly from the LRC. 
However, there are two exceptional cases. i) The even mode is well defined at $q_{z}c=0$ [see also Eqs.~(\ref{kqw1qzc0-1}) and (\ref{kqw1qzc0-2})] because the $\omega_{-}$ mode vanishes at any $\vq_{\parallel}$. ii) The odd mode is well defined at specific points $\vq_{\parallel}=(0,0)$ and $q_{z}c=2n\pi$ with $n\ne 0$, where the $\omega_{+}$ mode has zero intensity [see Fig.~\ref{kqw-00}(e)].

The origin of $\mathfrak{det}$ in \eq{kqw1det} is readily understood. The RPA susceptibility \eq{biRPA} is written in a $2 \times 2$ matrix form 
\be
\hat{\kappa} = \hat{\kappa}^{0} + \hat{\kappa}^{0} \hat{V} \hat{\kappa}  \,,
\ee
where $\hat{V}$ corresponds to the interaction matrix given in \eq{HI}. We then obtain 
\be
\hat{\kappa} = (1-\hat{\kappa}^{0} \hat{V})^{-1} \hat{\kappa}^{0}  \,. 
\label{kappamatrix}
\ee
$\mathfrak{det}$ in \eq{kqw1det} is given by 
\be
\mathfrak{det} = {\rm det} (1- \hat{\kappa}^{0} \hat{V}) \,.
\label{detk}
\ee

\subsection{Charge dynamics on a bilayer lattice}
The charge excitation spectrum is governed by the imaginary part of the susceptibility, Im$\kappa(\vq, \omega)$ in \eq{kqw1}. As expected, plasmons are manifested as sharp peaks in Im$\kappa$ where the determinant vanishes, i.e., $\mathfrak{det}=0$. In practical numerical calculations, we can identify plasmon modes by searching for the minimum value of $| \mathfrak{det} |$. 

Figures~\ref{kqw-00}(a)--(e) show $\vq$-$\omega$ maps of charge excitation spectrum, focusing on plasmon modes around $\vq_{\parallel}=(0,0)$ for a sequence of $q_{z}c$ values. The square of the modulus of the denominator of \eq{kqw1}, namely \eq{detk} is also shown on the right hand side of the figure to highlight where plasmon resonances can occur. 

At $q_{z}c=0$ [\fig{kqw-00}(a)], only the higher-energy optical plasmon mode, denoted as $\omega_{+}$ mode, is realized, while the lower-energy $\omega_{-}$ mode is absent. This behavior is a direct consequence of the interaction matrix components given by Eqs.~(\ref{Vpm}) and (\ref{Vpp}) at $q_{z}c=0$:  
\bea
&& V^{''}(\vq) = V(\vq) - V^{'}(\vq) \,, \\
&& V_{+}(\vq) = V^{'}(\vq)\,, \\
&& V_{-}(\vq)=0 \,.
\eea
Substituting these into \eq{kqw1} yields a simplified total susceptibility:  
\bea
&& \kappa(\vq, \omega) = \frac{\kappa^{0}_{\rm even}(\vq, \omega) [1- \kappa^{0}_{\rm odd} (\vq, \omega) V^{''}(\vq) ]} {\left[1-\kappa^{0}_{\rm even}(\vq, \omega) (V(\vq)+ V^{'}(\vq)) \right]
 \left[ 1-\kappa^{0}_{\rm odd}(\vq, \omega) (V(\vq) - V^{'}(\vq)) \right] } \,, 
 \label{kqw1qzc0-1}\\
 &&\hspace{10mm} = \frac{\kappa^{0}_{{\rm even}}(\vq, \omega)} 
 {1-\kappa^{0}_{\rm even}(\vq, \omega) (V(\vq)+ V^{'}(\vq))} 
 \label{kqw1qzc0-2} \,.
\eea
This confirms that at $q_{z}c=0$, the $\omega_{+}$ mode is the even mode. Physically, this corresponds to an in-phase charge oscillation within the bilayer. Its intensity vanishes at $\vq_{\parallel}=(0,0)$ because of charge conservation. Consequently, the other charge excitation mode, the $\omega_{-}$ mode, is an out-of-phase mode in the bilayer system, a finding consistent with prior literature \cite{griffin89}. The reason why the $\omega_{-}$ mode vanishes for $q_z c=0$ lies in its out-of-phase character. That is, when infinitesimally small external electric field with $q_z c=0$ is applied to the system, it is uniform along the $c$ direction and cannot couple to the out-of-phase charge oscillation. However, it is crucial to note that this separation into in-phase and out-of-phase modes only strictly valid at $q_{z}c=0$. 

Away from this specific momentum, the spectrum changes significantly. 
The upper boundary of the particle-hole continuum acquires a finite energy due to the presence of interlayer hopping $t_{z}$. Above the continuum, two distinct modes emerge: the higher-energy $\omega_{+}$ mode and the lower-energy $\omega_{-}$ mode \cite{griffin89}, as seen in Figs.~\ref{kqw-00}(b)--(d). These modes should not be confused with the even and odd modes, because the LRC now couples them. A key observation is the gapless dispersion of the $\omega_{-}$ mode, which extends smoothly into the continuum spectrum around $\vq=(0,0)$. At $q_{z}c = 2n\pi$, as shown in \fig{kqw-00}(e), the $\omega_{-}$ mode becomes gapped. In contrast, the $\omega_{+}$ mode loses its spectral weight at $\vq_{\parallel}=(0,0)$ at this momentum, similar to the case at $q_{z}c=0$.  A comparison between Figs.~\ref{kqw-00}(a) and (e) reveals that the plasmon excitations at $q_{z}c=0$ are not generic. The behavior at $q_{z}c = 2n\pi$ for $n\ne 0$ are more representative: both $\omega_{\pm}$ modes are present, but the intensity of the $\omega_{+}$ mode vanishes at $\vq_{\parallel}=(0,0)$ while the $\omega_{-}$ mode retains a finite spectral weight there.

\begin{figure}[t]
\centering
\includegraphics[width=12cm]{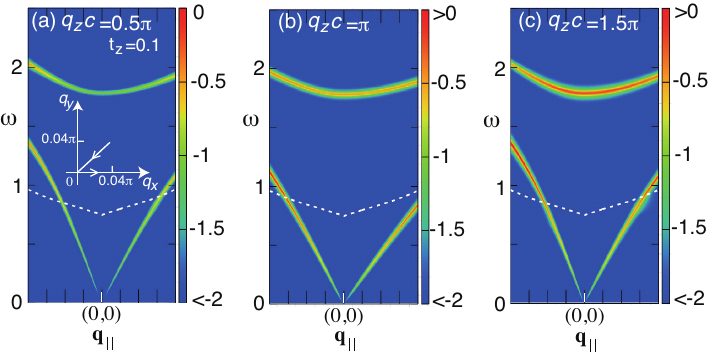}
\caption{Intensity maps of $\log_{10} | {\rm Im}\kappa(\vq, \omega)|$  in the vicinity of $\vq_{\parallel}=(0,0)$ along $(0.04\pi, 0.04\pi)$--$(0, 0)$--$(0.04\pi, 0)$ [see the inset in (a)] for three choices of $q_{z}c$: (a) $q_{z}c=0.5\pi$, (b) $q_{z}c=\pi$, and (c) $q_{z}c=1.5\pi$. The $\omega_{-}$ mode is gapless at $\vq_{\parallel}=(0,0)$ in spite of the presence of $t_{z}=0.1$. The white dotted curve denotes the upper boundary of the particle-hole continuum. To keep an appropriate contrast, the same color is used below -2 and above 0. Adapted from Ref.~\cite{yamase25} (\copyright\, 2025 American Physical Society). 
}
\label{kqw-000}
\end{figure}

The gapless nature of the $\omega_{-}$ mode is further highlighted in Figs.~\ref{kqw-000}(a)--(c), which magnify the region near $\vq_{\parallel}=(0,0)$. The intensity of the $\omega_{-}$ mode gradually weakens as the energy decreases, vanishing at $\omega=0$. A notable feature is that the $\omega_{-}$ mode crosses the continuum smoothly without any abrupt change, extending all the way down to zero energy. This behavior is due to the very small spectral weight of the continuum itself in this region, which makes the difference in spectral weight across the continuum boundary almost imperceptible. 

The right panels in \fig{kqw-00} show maps of the square of the modulus of the determinant $| \mathfrak{det} |^{2}$.  Plasmons appear as prominent peaks in 
Im$\kappa(\vq, \omega)$, which correspond to the minima of this denominator. As $q_{z}c=0$ [\fig{kqw-00}(a')], there are two minima for a given $\qp$. However, the finite spectral weight is realized only along the $\omega_{+}$ mode and zero along a possible $\omega_{-}$ mode. In contrast, at $q_{z}c=2\pi$ [\fig{kqw-00}(e')], both $\omega_{+}$ and $\omega_{-}$ modes are realized along the minimum and have the finite spectral weight except for the $\omega_{+}$ mode at $\vq_{\parallel}=(0,0)$ in \fig{kqw-00}(e). In \fig{kqw-00}(d), the $\omega_{+}$ mode has strong intensity only near $\vq_{\parallel}=(0,0)$ and the intensity is suppressed as going away from $\vq_{\parallel}=(0,0)$ although the denominator of \eq{kqw1} has a minimum there [\fig{kqw-00}(d')]. This comes from the $\vq_{\parallel}$ dependence of the numerator of \eq{kqw1}. The intensity of $\omega_{-}$ mode remains visible even inside the continuum as shown in Figs.~\ref{kqw-00}(b)--(d) and Figs.~\ref{kqw-000}(a)--(c). However, the modulus of \eq{detk} acquires additional contributions from the mixture with the particle-hole excitations [Figs.~\ref{kqw-00}(b')--(d')]. As a result, the peak position of the Im$\kappa(\vq, \omega)$ deviates slightly from the minimum of \eq{detk}.

\begin{figure}
\centering
\includegraphics[width=8cm]{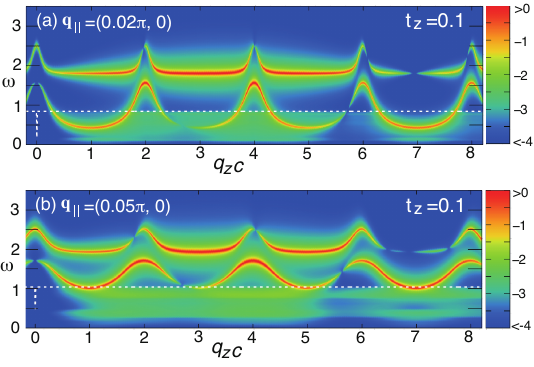}
\caption{$q_{z}c$ dependences of the intensity maps of $\log_{10} | {\rm Im}\kappa(\vq, \omega)|$ at (a) $\vq_{\parallel}=(0.02\pi, 0)$ and (b) $\vq_{\parallel}=(0.05\pi, 0)$ for $t_{z}=0.1$. To get a reasonable contrast, the same color is used above 0 and below -4. The white dotted line is the upper boundary of the continuum spectrum and exhibits a sharp drop at $q_{z}c=0$ because of the vanishing of the $\omega_{-}$ mode there. Adapted from Ref.~\cite{yamase25} (\copyright\, 2025 American Physical Society). 
}
\label{kqw-qzc}
\end{figure}
So far, we have focused on specific values of $q_{z}c$. To gain a comprehensive understanding, we now clarify the full $q_{z}c$ dependence of the $\omega_{\pm}$ modes. Figure~\ref{kqw-qzc}(a) shows results at a small in-plane momentum $\vq_{\parallel}=(0.02\pi, 0)$. The $\omega_{+}$ mode consistently exists above the continuum. A prominent peak appears at $q_{z}c = 2n\pi$. The energy of the $\omega_{+}$ mode decreases rapidly as it moves away from $q_{z}c = 2n\pi$, remaining nearly constant until the next peak at $q_{z}c = 2(n+1) \pi$. The $\omega_{-}$ mode, which is completely absent at $q_{z}c=0$, quickly gain intensity as $q_{z}c$ increases. It exhibits a dispersive behavior with an intensity peak at $q_{z}c = 2n\pi (n \ne 0)$, and it crosses the upper boundary of the continuum. 

Figure~\ref{kqw-qzc}(b) is the same plot, but for a larger in-plane momentum $\vq_{\parallel}=(0.05\pi, 0)$. The $\omega_{+}$ mode shows a weak dependence on $\vq_{\parallel}$ and is thus similar to the result in \fig{kqw-qzc}(a). On the other hand, the $\omega_{-}$ mode now exhibits a cosinelike dispersion along the $q_{z}c$ direction  and is situated almost entirely above the continuum. 

In both plots, the intensity of the $\omega_{\pm}$ modes displays a  characteristic $q_{z}c$ dependence. The $\omega_{+}$ mode loses intensity around $q_{z}c=7\pi$, and both modes show ``nodes'' or minima in their intensity at certain points. These features are not results of the denominator of Im$\kappa(\vq, \omega)$, which retains $2\pi$ periodicity along the $q_{z}c$ direction, but rather originate from the suppression of the numerator of \eq{kqw1}. Consequently, the intensity loses $2\pi$ periodicity along the $q_{z}$  direction. This is a direct consequence of the geometrical feature that our unit cell contains a basis at $\vq=(0,0,d)$. Therefore, the charge correlation function acquires a phase ${\rm e}^{i q_z d}$, which works to destroy 2$\pi$ periodicity of the plasmon intensity along the $q_z$ direction. 

\subsection{Comparison with RIXS data in Y-based cuprates}\label{RIXS-Y-base}
A comparison of our theoretical results with available experimental data for Y-based cuprates \cite{bejas24} provides a compelling test of our model. Previous analyses of these RIXS data on plasmon excitations employed a LRC derived from an electron-gas model \cite{fetter74,griffin89}. While this approach may not be strictly justified for a system near half-filling such as in cuprates (electron-liquid systems), the original aim was to investigate whether the experimental features were fundamentally tied to the $\vq^{-2}$ singularity of the LRC---a key characteristic of plasmon physics in the long-wavelength limit. 

\begin{figure}[tbh]
\centering
\includegraphics[width=8cm]{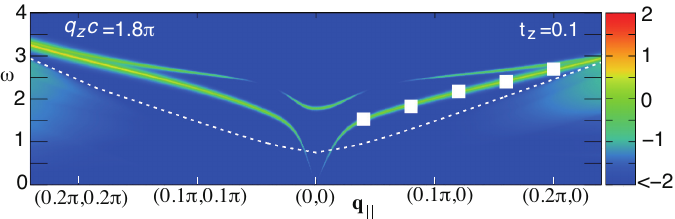}
\caption{Comparison with the plasmon energy (solid squares) reported in Y-based cuprate superconductors in Ref.~\cite{bejas24}; $t$ is assumed to be 233 meV. The experimental data are superimposed on the intensity map of $\log_{10} | {\rm Im}\kappa(\vq, \omega)|$ at $q_{z}c=1.8\pi$, and $t_{z}=0.1$. The white dotted curve denotes the upper boundary of the particle-hole continuum. To keep an appropriate contrast, the same color is used below -2. Adapted from Ref.~\cite{yamase25} (\copyright\, 2025 American Physical Society). 
}
\label{hepting-fit}
\end{figure}

The present theory extends the Fetter model \cite{fetter74,griffin89} by deriving the lattice LRC [Eqs.~(\ref{Vq}) and (\ref{Vqp})] for a bilayer structure, making it applicable across a wide range of electron densities. This provides a more robust framework for interpreting the experimental data. As shown in \fig{hepting-fit}, our model can reproduce the experimental data points with a choice of an effective hopping parameter $t=233$ meV. This value is somewhat smaller than those obtained in {\it ab initio} calculations \cite{hybertsen90,andersen95,markiewicz05}, a discrepancy that is expected when comparing an effective parameter to experimental data \cite{nag24}. While our finding shares some common ground with those of Ref.~\cite{bejas24}, we reveal important distinctions. 

Experimental data were taken at $q_{z}c=1.8\pi$ in the superconducting phase. Assuming that the superconducting gap does not essentially affect the plasmon dispersion, we apply our theory and find that the spectral weight of the $\omega_{-}$ mode is significantly higher than that of the $\omega_{+}$ mode, and the experimental data points closely follow the dispersion of the $\omega_{-}$ mode as shown in \fig{hepting-fit} \cite{yamase25}. This conclusion is independently supported by a separate study on the density response in the superconducting phase \cite{sellati25}. However,  they interpreted $\omega_{-}$ mode as (ghost) Josephson plasmons \cite{sellati25}. This difference of interpretation can be tested by performing RIXS studies above $T_{c}$, where Josephson plasmons vanish \cite{sellati25} whereas our metallic plasmons survive \cite{yamase25}. 

The assignment of the mode contrasts with the analysis in Ref.~\cite{bejas24}, which assumed $2\pi$  periodicity for $q_{z}c$ and concluded that the data corresponded to the $\omega_{+}$ mode at $q_{z}c=0.2\pi$. As our analysis in \fig{kqw-qzc} explicitly demonstrates, the plasmon intensity does not have  $2\pi$ periodicity along the $q_{z}c$ direction and exhibits a strong, non-trivial $q_{z}c$ dependence. We emphasize that $q_{z}c=0$ is a singular point where the $\omega_{-}$ mode vanishes entirely, behavior that does not hold for $q_{z}c = 2n \pi (n \ne 0)$. The complex modulation of the peak intensity is a direct result of the numerator of \eq{kqw1}, which is sensitive to the factor $q_{z}d$ and thus to the specific geometry of the bilayer lattice. 

While we have neglected interbilayer hopping $t_z^{'}$, a role of $t_z^{'}$ is to yield a gap in the gapless mode at $\qp=(0,0)$ for $q_{z}c \neq 2 n \pi$. This gap would be small, but will be detected in RIXS near future. In addition, there is a caveat about the value of $t_z$. In this review, we have presented results for $t_z=0.1t$ as a realistic value of Y-based cuprates (Fig. 28). If we would employ $t_z \gtrsim 0.16t$ \cite{yamase25}, the energy hierarchy of $\omega_{\pm}$ mode changes and $\omega_{-}$ mode has higher energy than the $\omega_{+}$ mode. Furthermore, it is $\omega_{+}$ mode instead that would exhibit a gapless mode at $\qp=(0,0)$. Since the value of $t_z=0.16t$ is rather close to $t_z=0.1t$, the assignment of the actual mode of bilayer systems should be performed carefully, calling for a more comprehensive experimental data.

These considerations suggest a practical procedure for future RIXS measurements to definitely identify the plasmon modes. By conducting measurements near $q_{z}c=2n\pi (n \ne 0)$ and small $\qp$, any detected collective charge excitations should be primarily associated with the $\omega_{-}$ mode, as the spectral weight of the $\omega_{+}$ mode is predicted to be negligible there [see Fig.~\ref{kqw-00}(e)]. Subsequently, by tracing the dispersion of the $\omega_{-}$ mode as $q_{z}c$ is varied away from $2n\pi$, it may also be possible to  observe the $\omega_{+}$ mode, which is expected to appear as a separate signal above or below the $\omega_{-}$ mode depending on the value of $t_{z}$; see Ref.~\cite{yamase25} for this $t_{z}$ dependence.  
 
A more quantitative comparison with experiments will require incorporating the effects of strong electronic correlations, which are especially significant in cuprates. The next step is to integrate the bilayer-lattice LRC derived here [Eqs.~(\ref{Vq}) and (\ref{Vqp})] into the framework of the well-established $t$-$J$-$V$ model \cite{greco16}, a task that has been just published \cite{yamase26}.

\subsection{Contrast to nearest-neighbor interaction} 
To provide a clear contrast with the LRC, we now examine a simplified case where the interaction is restricted to nearest-neighbor sites. Since our general formalism for the susceptibility, captured by Eqs.~(\ref{HI}) and (\ref{biRPA}), is valid for any functional form of the interaction, we may consider a short-range Coulomb interaction, yet beyond the on-site Hubbard interaction, defined as: 
\bea
&&V(\vq)= 2 V_{xy}(\cos q_{x} + \cos q_{y}) \,, \\
&& V^{'}(\vq) = V_{z} {\rm e}^{-i q_{z} d} \, .
\eea
Here, $V_{xy}$ represents the in-plane nearest-neighbor interaction and $V_{z}$ is the interlayer interaction. Assuming for simplicity that there is no interbilayer interaction, the resulting total susceptibility can be expressed as: 
\be
\kappa(\vq, \omega) = \cos^{2} \frac{q_{z}d}{2}  \kappa_{\rm even}(\vq,  \omega)+  \sin^{2} \frac{q_{z}d}{2} \kappa_{\rm odd}(\vq, \omega) \, ,
\label{short-eo}
\ee
where the even and odd mode susceptibilities are given by 
\bea
&& \kappa_{\rm even}(\vq, \omega) = \frac{\kappa^{0}_{\rm even}(\vq, \omega)} {1-(V(\vq) + V_{z}) \kappa^{0}_{\rm even}(\vq, \omega) }
\label{short-e} \,, \\
&&  \kappa_{\rm odd}(\vq, \omega) = \frac{\kappa^{0}_{\rm odd}(\vq, \omega)} {1-(V(\vq) - V_{z}) \kappa^{0}_{\rm odd}(\vq, \omega) }
\label{short-o} \,.
\eea
In contrast to the coupled plasmon modes that arise from the LRC [as discussed in Sec.~\ref{bilayerLRC}], the charge excitations in this short-range model are perfectly decoupled into distinct even and odd modes. This decoupling is a direct consequence of the short-range nature of the interaction, which lacks the long-wavelength singularity ($\vq^{-2}$) that is essential for coupling the modes and generating plasmons. 

A similar analytical approach can be applied to short-range magnetic interaction, such as those defined by $J(\vq)=2 J ( \cos q_{x} +  \cos q_{y})$  and $J^{'}(\vq) = J_{z} {\rm e}^{-i q_{z}d}$. This yields the dynamical magnetic susceptibility, which exhibits exactly the same  functional form as Eqs.~(\ref{short-eo}), (\ref{short-e}), and (\ref{short-o}). This mathematical similarity highlights the fundamental role of the underlying lattice structure. The resulting bilayer modulation has been observed experimentally in inelastic neutron scattering on Y-based cuprates \cite{pailhes03,pailhes06}, providing a crucial validation of these theoretical models.

\section{Superconductivity} \label{sc-section}
Having established the experimental presence of the LRC in both single-layer and bilayer cuprates, elucidated the resulting renormalization of one-particle properties, and compared  with short-range interaction, yet beyond the on-site Hubbard interaction, we now turn to the critical question of how these charge fluctuations relate to the mechanism of high-temperature superconductivity. 

While a broad consensus points to spin fluctuations as the key pairing mechanism for high-$T_{c}$ superconductivity, the role of charge fluctuations---and specifically, the Coulomb interaction that underlies them---remains a subject of debate. Instead of framing this in general as a competition between charge and spin fluctuations, we propose to investigate the importance of the screened nearest-neighbor Coulomb interaction as a competing element within a spin-fluctuation-driven framework. We shall demonstrate that this screened interaction plays a crucial role in the overall pairing mechanism, providing a new perspective on the complex interplay between charge and spin degrees of freedom in cuprates.  To arrive at this insight we first need to recognize the self-restraint effect of superconductivity in the spin-fluctuation mechanism.

\subsection{Model and formalism}
We now describe the theoretical framework used to study the superconducting instability. Our approach is based on a one-band model that incorporates both kinetic energy and a general spin-spin interaction on a square lattice.  

\subsubsection{Model Hamiltonian} 
The Hamiltonian for our system is defined by two primary terms: 
\be
H = \sum_{\vk} \xi_{\vk} c^\dagger_{\vk \sigma} c_{\vk \sigma} + 
\frac{1}{8N_{s}}\sum_{\vk \vk' \vq} \sum_{\sigma_{i}} g(\vk, \vk', \vq) \boldsymbol{\sigma}_{\sigma_{1} \sigma_{2}} \cdot 
\boldsymbol{\sigma}_{\sigma_{3} \sigma_{4}} c^\dagger_{\vk \sigma_{1}} c_{\vk+\vq \sigma_{2}} 
c^\dagger_{\vk' + \vq \sigma_{3}} c_{\vk' \sigma_{4}} \,.  
\label{model}  
\ee
The first term describes kinetic energy of electrons with a dispersion given by 
\be
\xi_{\vk} = -2 t (\cos k_{x} + \cos k_{y}) - 4 t' \cos k_{x} \cos k_{y} 
-2 t'' (\cos 2 k_{x} + \cos 2 k_{y})  -\mu \,,
\ee
where $t$, $t'$,  and $t''$ are the hopping integrals for first, second, and third nearest neighbors, respectively, and $\mu$ is the chemical potential.  $c^{\dagger}_{\vk \sigma}$ and $c_{\vk \sigma}$ are the creation and annihilation operators of electron with momentum $\vk$ and spin $\sigma$; lattice constants are set unity.   

The second term represents a general SU(2) symmetric two-particle interaction and $N_{s}$ is the total number of lattice sites on a square lattice. This term describes the effective spin-spin interaction of itinerant electrons and is well-established phenomenological description for systems close to a spin-density-wave (SDW) instability. Such an interaction was successfully employed to study superconductivity in a variety of systems \cite{nakajima73,miyake86,moriya90}. Microscopically, it is often derived as a low-energy effective magnetic interaction, for example, from the repulsive Hubbard interaction in a functional renormalization group study \cite{husemann09,eberlein14}. The interaction is proportional to the scaler product of Pauli matrices $\boldsymbol{\sigma}$, ensuring its SU(2) symmetry. Note that if we take the limit of $\vq  \rightarrow {\bf 0}$,  the interaction is reduced to the well-known SU(2)-symmetric Landau interaction function in the spin-antisymmetric channel. The  precise functional form of $g(\vk, \vk', \vq)$ depends on microscopic details of the system. In the case of a conventional SDW state described by the order parameter such as $\bra  {\bf S} (\vq) \ket = \frac{1}{2} \sum_{\vk, \alpha, \beta} \bra c^{\dagger}_{\vk \alpha} \boldsymbol{\sigma}_{\alpha \beta}  c_{\vk+\vq \beta} \ket$, we can make the practical and widely used approximation  $g(\vk, \vk', \vq) \approx g(\vq)$. For our analysis, we will specifically consider  the functional from 
\be
g(\vq) = 2 g ( \cos q_{x} + \cos q_{y})\,,
\label{gq}
\ee
which describes a dominant nearest-neighbor spin interaction in real space. For $g>0$, this form favors an SDW state at the momentum $\vq= (\pi, \pi)$ and is relevant for many interesting cases. Results for other choices of $g(\vq)$ is given in Ref.~\cite{yamase23}.

\subsubsection{Eliashberg theory and pairing gap equations} 
To investigate the superconducting instability, we employ the Eliashberg theory 
\cite{eliashberg60,schrieffer,marsiglio20}. This is a powerful, self-consistent framework that accounts for the effects of retardation and dissipation arising from the mediating fluctuations. The theory provides a set of two coupled equations for the pairing gap $\Delta (\vk, {\rm i} k_{n})$ and the quasiparticle  renormalization function $Z (\vk, {\rm i} k_{n})$ \cite{misc-Z}, where ${\rm i} k_{n}= {\rm i} (2n+1) \pi T$ is fermionic Matsubara frequency at temperature $T$. 

Numerical solutions to the Eliashberg equations can be computationally demanding, especially at low temperatures where a large number of Matsubara 
frequencies must be included even if $Z$ is set to unity. To overcome this challenge and to retain a fine resolution of momentum along the Fermi surface, we adopt a simplification: we project the momentum dependence of the functions onto the Fermi surface. This is a standard approach in both conventional electron-phonon \cite{schrieffer,marsiglio20} and spin-fluctuation \cite{radtke92,millis92} mechanisms. This allows us to perform calculations down to very low temperatures including the effect of the renormalization function  $Z$, while accurately capturing the anisotropic gap formation induced by the magnetic interactions. 

\begin{figure}[ht]
\centering
\includegraphics[width=5cm]{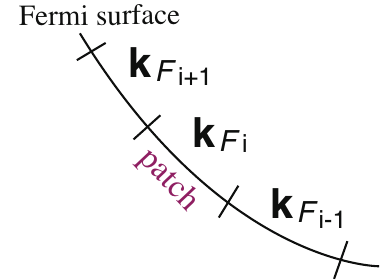}
\caption{Fermi surface patches. The Fermi surface is divided into small pieces and each patch is assigned to $\vk_{F i}$.
}
\label{FS-patch}
\end{figure}

The resulting linearized Eliashberg equations for the superconducting instability are: 
\bea
&& \Delta(\vk_{F}, {\rm i} k_{n}) Z(\vk_{F}, {\rm i} k_{n}) = - \pi T \sum_{\vk_{F}', n'} N_{\vk_{F}'} \frac{\Gamma_{\vk_{F} \vk_{F}'} ({\rm i}k_{n}, {\rm i}k_{n}')}{| k_{n}' |} \Delta(\vk_{F}', {\rm i}k_{n}')  \label{eliashberg1} \,, \\
&& Z(\vk_{F}, {\rm i} k_{n}) =1 - \pi T \sum_{\vk_{F}', n'} N_{\vk_{F}'} \frac{k_{n}'}{k_{n}} \frac{\Gamma^{Z}_{\vk_{F} \vk_{F}'} ({\rm i}k_{n}, {\rm i}k_{n}')}{| k_{n}' |} \label{eliashberg2} \,.
\eea
In contrast to the conventional treatment \cite{eliashberg60,schrieffer,marsiglio20}, we divide the Fermi surface in these equations into discrete patches, each specified by a discrete momentum $\vk_{F}$ [see  \fig{FS-patch}]. $N_{\vk_{F}}$ represents the momentum-resolved density of states on the corresponding Fermi surface patch. 

The pairing interaction $\Gamma_{\vk_{F} \vk_{F}'} ({\rm i}k_{n}, {\rm i}k_{n}')$ is derived from the magnetic interaction in \eq{model} and is computed as: 
\be
\Gamma_{\vk_{F} \vk_{F}'} ({\rm i}k_{n}, {\rm i}k_{n}') = - \frac{1}{4} \bra \hat{g}(\vk - \vk', {\rm i} k_{n} - {\rm i} k_{n}' ) 
+ 2 \hat{g}(\vk + \vk', {\rm i} k_{n} + {\rm i} k_{n}' ) \ket_{\vk_{F} \vk_{F}'} \,, 
\label{Gamma}
\ee
where the first term on the right-hand side is due to longitudinal spin fluctuations---$\sigma^{z}$ component in \eq{model}--- and the second term  from the transverse spin fluctuations. The $\bra \cdots \ket_{\vk_{F} \vk_{F}'}$ denote the average over the Fermi surface patches $\vk_{F}$ and $\vk_{F}'$. 

The spin fluctuation propagator $\hat{g}(\vq, {\rm i} q_{m})$ is calculated within the RPA as 
\be
\hat{g}(\vq, {\rm i}q_{m}) = g^{*}(\vq) - \frac{g(\vq) \chi_{0}(\vq, {\rm i} q_{m})g(\vq)}{1+g(\vq)\chi_{0}(\vq, {\rm i}q_{m})} \,,
\label{Vhat}
\ee
where $\chi_{0}(\vq, {\rm i}q_{m})$ is the Lindhard function, which describes a simple bubble diagram. The first term $g^{*}(\vq)$ is instantaneous and represents the bare interaction [$g^{*}(\vq)=g(\vq)$]. However, as we will discuss in Sec.~6.2.4 from a physical perspective, it will be regraded as a renormalized version of the bare interaction due to screening effects from the Coulomb repulsion [see \eq{V*}]. The second term, which depends on frequency, accounts for the crucial retardation effects on the pairing and is typically the focus of studies on spin-fluctuation-mediated superconductivity \cite{scalapino12}. Applicability of  the present model to cuprates will be discussed in Sec.~6.2.5.

At the same level of approximation, the term $\Gamma^{Z}_{\vk_{F} \vk_{F}'} ({\rm i}k_{n}, {\rm i}k_{n}')$ in \eq{eliashberg2} is calculate as  
\be
\Gamma^{Z}_{\vk_{F} \vk_{F}'} ({\rm i}k_{n}, {\rm i}k_{n}') = \frac{1}{4} \bra 3 \hat{g}(\vk - \vk', {\rm i} k_{n} - {\rm i} k_{n}' ) 
-  2 g(\vk - \vk') \ket_{\vk_{F} \vk_{F}'} \,. 
\ee
After evaluating the renormalization function $Z$ from \eq{eliashberg2}, we solve the eigenvalue problem defined by \eq{eliashberg1} numerically. A superconducting instability occurs when the maximum eigenvalue $\lambda$ exceeds unity. The corresponding eigenvector then describes the structure of the pairing gap.

\subsection{Results} 
We now present our numerical results for the superconducting instability, focusing on the interplay between different scattering processes. Since it is vital to distinguish between the retarded and instantaneous interactions---second and first terms on the right hand side in \eq{Vhat}, we first study the former. In Sec.~6.2.1 we find a new concept---self-restraint effect of superconductivity. Its intuitive understanding is given in Sec.~6.2.2. The typical gap and renormalization function is presented in Sec.~6.2.3. In Sec.~6.2.4, we study instantaneous interactions by including possible renormalization from the nearest-neighbor Coulomb interaction [see \eq{V*}] and reveal $T_c$ and gap function. In Sec.~6.2.5, we discuss the relevance to cuprates. In Sec.~6.2.6 we conclude a potential role of screened Coulomb interaction in cuprate about $T_c $ as well as an importance of the functional form of $g(\vq)$ in other superconductors. In Sec.~6.2.7, the interplay of superconductivity and LRC is briefly discussed. We use representative parameters for the kinetic energy: $t'/t=-0.25$, $t''/t=0$,  with an electron density fixed at 0.85. Unless otherwise noted, the interaction strength is set to $g/t=0.95$.  All energy scales are measured in units of $t$. 

\subsubsection{Self-restraint effect of superconductivity}
\begin{figure}[ht]
\centering
\includegraphics[width=7cm]{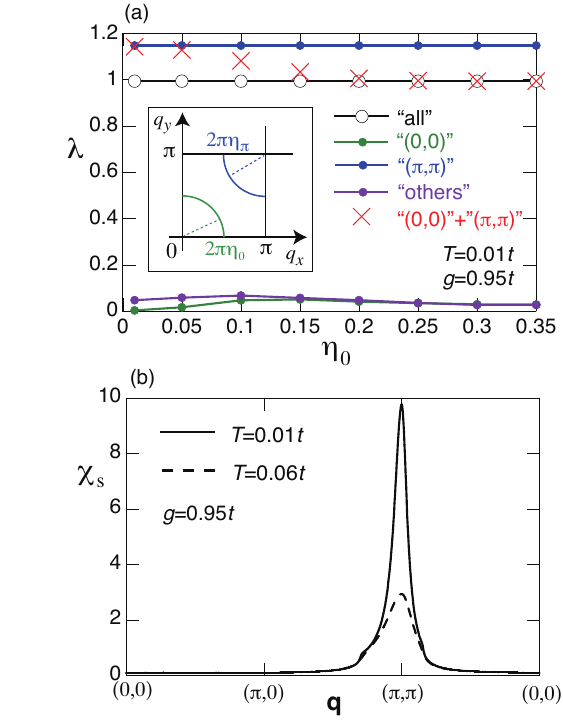}
\caption{(a) $\eta_{0}$ dependence of eigenvalue $\lambda$ for different scattering processes at $T=0.01(\approx T_{c})$ and $\eta_{\pi}=0.25$. The inset defines the scattering processes: ``(0,0)'' and ``$(\pi, \pi)$'' correspond to a region $\sqrt{q_{x}^{2}+ q_{y}^{2}} \leq 2 \pi \eta_{0}$ and $\sqrt{( |q_{x}|-\pi )^{2} + ( |q_{y}|-\pi )^{2}} \leq 2 \pi \eta_{\pi}$, respectively, and ``others'' denotes the remaining scattering processes. (b) Momentum dependence of the static spin susceptibility at two temperatures. Adapted from Ref.~\cite{yamase23}  [\copyright\, 2023 The Author(s)]. 
}
\label{process}
\end{figure}

Our analysis begins by isolating the effects of the retarded interaction, for which we set  $g^{*}(\vq)$ to zero in \eq{Vhat}. To understand how different scattering processes influence the pairing instability, we define distinct momentum regions in the first Brillouin zone as illustrated in the inset of \fig{process}(a). The regions are defined as: ``(0,0)''---processes with a momentum transfer $| \vq | \leq 2 \pi \eta_{0}$, ``$(\pi, \pi)$''---processes with a momentum transfer $\sqrt{( |q_{x}|-\pi )^{2} + ( |q_{y}|-\pi )^{2}} \leq 2 \pi \eta_{\pi}$, and ``others''---all remaining processes. We then compute the maximum eigenvalue $\lambda$ of the Eliashberg equations [Eqs.~(\ref{eliashberg1}) and (\ref{eliashberg2})] at $T=0.01$ as a function of the parameter $\eta_{0}$, with a fixed value of $\eta_{\pi}=0.25$. As a benchmark, we also calculate the eigenvalue for the ``all'' processes, which corresponds to including all scattering channels. 

Figure~\ref{process}(a) shows that the eigenvalue for the ``all'' and ``$(\pi, \pi)$'' processes remains independent of $\eta_{0}$ as expected. The largest eigenvalue is obtained from the ``$(\pi, \pi)$'' channel, confirming that the superconductivity is predominantly driven by spin fluctuations centered around the antiferromagnetic momentum $\vq=(\pi, \pi)$. This result is consistent with the significant spectral weight of the static spin susceptibility around this momentum as shown in \fig{process}(b). 

A key finding is the dramatic suppression of the eigenvalue when both ``$(\pi, \pi)$'' and ``(0,0)''  scattering channels are included as shown by the ``(0,0)''+``$(\pi,\pi)$'' curve. As $\eta_{0}$ increases, the eigenvalue is suppressed by more than 15 \%. This suppression is significant, considering that the spin susceptibility in the small momentum region around $\vq=(0,0)$ is very small [see \fig{process}(b)] and thus naturally the eigenvalue for ``(0,0)'' is very small in \fig{process}(a). We refer to this suppression as the  {\it self-restraint effect} of superconductivity. Notably, this effect is not due to competing ferromagnetic fluctuations at $\vq=(0,0)$, as the spin susceptibility has no peak in this region. Instead, it arises from the scattering processes not relevant to the dominant pairing channel at $(\pi, \pi)$. 

\begin{figure}[t]
\centering
\includegraphics[width=8cm]{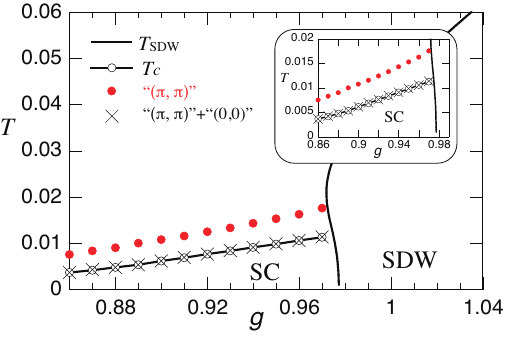}
\caption{Phase diagram of superconductivity and SDW in the plane of the interaction strength $g$ and temperature $T$. ``$(\pi, \pi)$'' and ``$(0, 0)$''+``$(\pi, \pi)$'' indicate $T_{c}$ obtained by considering only scattering processes specified by ``$(\pi, \pi)$'' and those by ``$(0, 0)$'' and ``$(\pi, \pi)$'', respectively.  The inset magnifies the superconducting phase. Adapted from Ref.~\cite{yamase23}  [\copyright\, 2023 The Author(s)]. 
}
\label{phase}
\end{figure}

To quantify the impact of this self-restraint effect, we construct the phase diagram in the plane of interaction strength $g$ and temperature $T$. Based on our finding from \fig{process}(a), we choose fixed parameters $\eta_{0}=0.25$ and $\eta_{\pi}=0.25$, which are sufficient to capture the saturated effects of the respective scattering regions. The value of $\eta_{\pi}=0.25$ roughly corresponds to the region where the spin susceptibility is large---this region shows a weak temperature dependence as seen in \fig{process}(b), although the peak height depends on temperature. 

Performing comprehensive computations by changing $T$ and the interaction strength $g$, we construct the phase diagram as shown in \fig{phase}. 
The SDW phase appears at large values of $g$, with its transition always occurring at the momentum $\vq=(\pi, \pi)$ as expected. The reentrant behavior of the critical line at low $T$ around $g=0.97$--$0.98$ is due to the sharpening of the Fermi surface. The superconducting critical temperature  $T_{c}$ increases monotonically as the system approaches the SDW phase, a direct consequence of the enhanced spin fluctuations. 

The most striking results are revealed by comparing the different curves for $T_{c}$. The critical temperature obtained by considering only the ``$(\pi, \pi)$'' scattering processes is suppressed by nearly a factor of two when the ``(0,0)'' scattering processes are included. This demonstrates that the spin-fluctuation-mediated superconductivity suffers from a substantial self-restraint effect, with $T_{c}$ being significantly reduced by what would appear to be an insignificant contribution from the small-momentum ``(0,0)''  scattering. As the curve for ``(0,0)''+``$(\pi, \pi)$'' is nearly identical to the curve ``all'', it  confirms that the dominant suppression of $T_{c}$ comes from the interplay between these two channels. 

The inset of \fig{phase} provides a magnified view of the superconducting phase, further highlighting the pronounced difference in $T_{c}$ with and without ``(0,0)'' scattering channel. This finding underscores the importance of a comprehensive treatment of all scattering processes, even though some scattering processes have small contributions to the pairing interaction.

\subsubsection{Intuitive understanding of the self-restraint effect} 
\begin{figure}
\centering
\includegraphics[width=7cm]{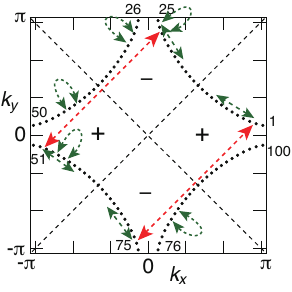}
\caption{Intuitive understanding of the self-restrained effect. The Fermi surface (black dotted curve) is divided into pieces. Red arrows depict scattering between opposite signs of the $d$-wave gap mediated by major antiferromagnetic fluctuations at $\vq \approx (\pi, \pi)$. Green arrows depict scattering between same-sign regions of the gap mediated by fluctuations at small $\vq$. These processes introduce phase frustration, which suppresses the overall pairing tendency.  
}
\label{intuitive}
\end{figure}

The physical origin of the self-restraint effect lies in a phase frustration of the pairing gap, as schematically illustrated in \fig{intuitive}. As is well known, spin fluctuations mediate a repulsive pairing interaction, meaning that the pairing kernel $\Gamma_{\vk_{F} \vk_{F}'}$ in \eq{eliashberg1} is positive [$\hat{g}$ in \eq{Gamma} is thus negative]. To satisfy the Eliashberg equation, the pairing gap $\Delta$ must change sign between two Fermi surface points connected by a dominant repulsive interaction. This is precisely why the spin-fluctuation mechanism naturally favors an anisotropic state such as $d$-wave gap, where the sign change along the Fermi surface across the lines $k_{y}=\pm k_{x}$. 

However, a key point is that spin fluctuations are not confined to the antiferromagnetic momentum $\vq \approx (\pi, \pi)$. Instead, they exist across the entire momentum space as a ``tail'' of the spectral weight. In particular, fluctuations at small momentum transfer [$\vq \approx (0,0)$] tend to connect nearby points on the Fermi surface. Like the stronger fluctuations at $(\pi,\pi)$, these small-$\vq$ fluctuations are also repulsive and thus also favor a sign change in the gap function. This creates a direct conflict, or ``frustration'', because a $d$-wave gap function has the same sign for nearby point on the Fermi surface except for the nodal regions. 

One might consider that such frustration could be negligible because the spectral weight of spin fluctuations is very small at low momentum. Our results directly challenge this assumption. The phase frustration from these seemingly weak fluctuations is, in fact, strong enough to significantly suppress the superconducting instability, leading to the substantial reduction in $T_{c}$ shown in \fig{phase}. While the small-momentum fluctuations themselves would only lead to a tiny eigenvalue (as seen for the ``(0,0)'' curve in \fig{process}), their inclusion in the full calculation has a disproportionately large impact by introducing a destructive interference.

The self-restraint effect is an intrinsic feature of repulsive pairing interactions. In contrast, for attractive pairing interaction such as those mediated by electron-phonon coupling \cite{bardeen57,schrieffer}  or orbital nematic fluctuations \cite{yamase13b}, the pairing interaction is attractive (negative), which favors a gap with a constant sign, e.g., $s$-wave. In such a cases, all scattering processes, regardless of momentum transfer, work constructively to enhance the pairing instability. There is no phase frustration, and thus no self-restraint effect. This fundamental difference highlights a critical distinction between attractive and repulsive pairing mechanisms. This explanation does not depend on details of models and approximations. 

\subsubsection{Pairing gap and renormalization function}
\begin{figure}
\centering
\includegraphics[width=7cm]{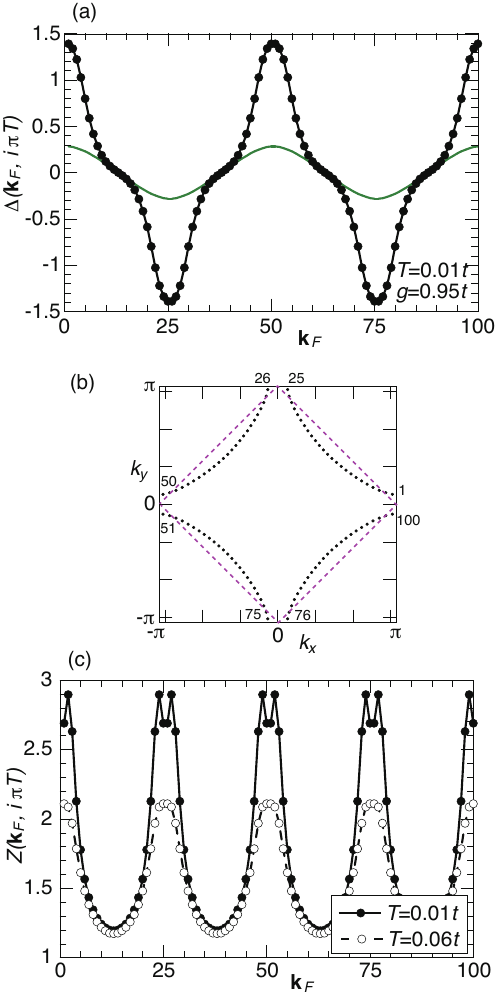}
\caption{Momentum dependence of the pairing gap and  the renormalization function at the lowest Matsubara frequency along the Fermi surface. (a) The pairing gap at $T=0.01$ $(\approx T_{c})$. The curve in green shows a fit to the simple form $\Delta(\vk_{F}) = - 0.144(\cos k_{x} - \cos k_{y})$. (b) The Fermi surface, with the Fermi momenta $\vk_{F}$ numbered from 1 to 100. The  dashed line denotes the magnetic zone boundary. The ``$(\pi,\pi)$'' scattering processes are optimized between two $\vk_{F}$ points such as 2-74, 24-52, 27-99, and 49-77. (c) The renormalization function  at $T=0.01$ and $0.06$.  Adapted from Ref.~\cite{yamase23}  [\copyright\, 2023 The Author(s)]. 
}
\label{gap}
\end{figure}

The momentum dependence of the pairing gap $\Delta(\vk_{F}, {\rm i}k_{n})$ at the lowest Matsubara frequency shown in \fig{gap}(a) provides a clear signature of the spin-fluctuation-mediated pairing mechanism. As expected from the $B_{1}$ representation of the $C_{4v}$ point group symmetry, the gap exhibits $d_{x^{2}-y^{2}}$-wave symmetry with four nodes on the Fermi surface. However, a key insight from our analysis is that the gap is not described by the simple form $\Delta(\vk_{F}) \propto \cos k_{x} - \cos k_{y}$. As illustrated by the green curve in \fig{gap}(a), which shows a fit of this simple form to the nodal regions, the actual gap has a stronger momentum dependence from higher-order harmonics with $d$-wave symmetry, resulting in a substantial enhancement of the gap magnitude in the antinodal regions around $\vk \approx (\pi, 0)$ and $(0, \pi)$. This enhanced anisotropy is a direct consequence of the antiferromagnetic spin fluctuations being maximized when two Fermi surface momenta are connected by the wavevector $\vq \approx (\pi, \pi)$, as schematically depicted in \fig{gap}(b).

We now turn our attention to the quasiparticle renormalization function $Z(\vk_{F}, {\rm i}k_{n})$ shown in \fig{gap}(c). The strong spin fluctuations also have a profound effect on the quasiparticle spectrum, leading to a large momentum dependence of $Z$. As expected, $Z$ becomes particularly large in the antinodal regions around $\vq \approx (\pi, 0)$ and $(0,\pi)$ where the scattering is strongest. Interestingly, when the system is very close to the SDW phase, the low-energy spin fluctuations acquire a very sharp peak at $\vq=(\pi, \pi)$. This sharp peak in the scattering can lead to a subtle but distinct dip structure in $Z$ at the magnetic zone boundary, where the scattering wavevector deviates slightly from the peak momentum of the fluctuation at $\vq=(\pi, \pi)$ [see also \fig{gap}(b)]. This feature is  particularly noticeable at $T=0.01t$. Upon moving away from the SDW phase, for example, by increasing the temperature, the spin fluctuation peak broadens, and the dip structure vanishes, leaving a single broad peak in $Z$ at the antinodal region, as shown by the result for $T=0.06t$ in \fig{gap}(c).

It is worth noting the influence of the self-restraint effect on the shape of the pairing gap and the renormalization function. We have confirmed  by performing similar calculations that including only the dominant ``$(\pi, \pi)$'' scattering processes does not yield a significant change in the momentum dependence of either $\Delta$ or $Z$. This is a crucial distinction from the two-band model, where the so-called $s_{\pm}$-wave pairing symmetry is realized. In that case, the self-restraint effect is known to lead to a large momentum dependence of the pairing gap \cite{yamase20}. Our results therefore highlight that while the self-restraint effect is a fundamental characteristic of repulsive pairing mechanism, its influence on the specific shape of the gap and the quasiparticle spectrum is highly dependent on the underlying band structure and the nature of the pairing channels.

\subsubsection{Role of instantaneous magnetic interaction and screened Coulomb interaction} 
Next we consider the effect of the instantaneous magnetic interaction, represented by $g^{*}(\vq)$ in \eq{Vhat}, which is independent of frequency. It is important to acknowledge a subtle aspect here: the screened Coulomb interaction is also instantaneous and generally acts to suppress  the effect of $g(\vq)$---this is what we wish to discuss as  a role of Coulomb interaction beyond the Hubbard interaction. We therefore model the effective instantaneous interaction as a  superposition of a constant term and a term proportional to the momentum-dependent interaction: 
\be
g^{*}(\vq) = g_{0} + r  g(\vq) \,.
\label{V*} 
\ee
Here, the constant $g_{0}$ represents the effect from the on-site Coulomb interaction. This term can be understood as playing the role of the pseudo-Coulomb interaction $\mu^{*}$ \cite{schrieffer} at the energy scale of the magnetic interaction. However, for the $d$-wave pairing symmetry that is the focus on the present work, the effect of $g_{0}$ is entirely cancelled upon  momentum summation in the Eliashberg equation, \eq{eliashberg1}. The new aspect we consider is the nearest-neighbor screened Coulomb  interaction, which is modeled by the positive parameter $r (<1)$. A smaller value of $r$ indicates a stronger suppression of the pairing interaction by this nearest-neighbor repulsion. 

To numerically solve the Eliashberg equations,  Eqs.~(\ref{eliashberg1}) and (\ref{eliashberg2}),  with the including of the instantaneous interaction  $g^{*}(\vq)$, it is necessary to introduce a cutoff energy for the Matsubara frequency, a standard practice in the literature \cite{morel62,schrieffer,marsiglio20}. We choose a cutoff $i\omega_{c} \approx i 20 t$, a magnitude approximately twice the band width. Since this choice of $\omega_{c}$ is not unique, the parameter  $r$ should be regarded as a phenomenological factor that describes the net  suppression of the pairing instability due to the nearest-neighbor Coulomb repulsion. This technical aspect, however, does not alter our core conclusions. 

\begin{figure}[t]
\centering
\includegraphics[width=7cm]{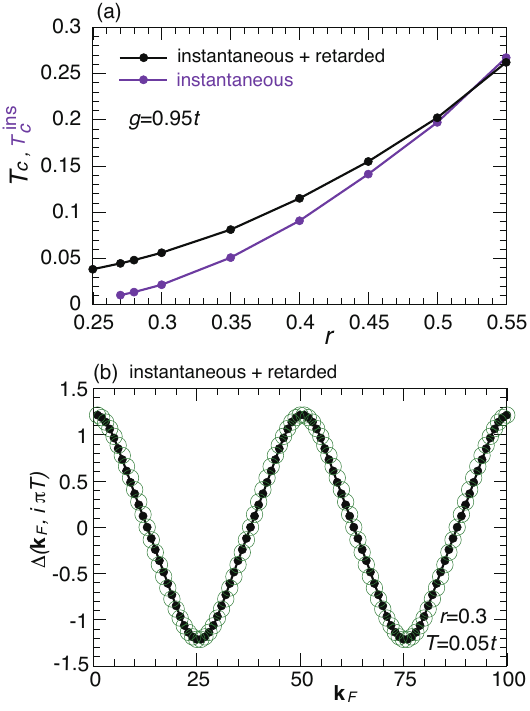}
\caption{Results in the presence of the nearest-neighbor instantaneous interaction. (a) $T_{c}$ as a function of $r$---the effect of the Coulomb repulsion [see \eq{V*}] in two different cases: only the instantaneous interaction, i.e., only $g^{*}(\vq)$ in \eq{Vhat}, and both instantaneous and retarded interactions, i.e., the full expression of \eq{Vhat}. (b) Momentum dependence of the pairing gap at the lowest Matsubara frequency at $T=0.05 (\approx T_{c})$ along the Fermi surface [see \fig{gap}(b)] when both instantaneous and retarded interactions are considered. The green circles correspond to the simple $d$-wave gap described by $\Delta(\vk) = -1.237\,(\cos k_{x} - \cos k_{y})/2$. Adapted from Ref.~\cite{yamase23}  [\copyright\, 2023 The Author(s)]. 
}
\label{phasei3}
\end{figure}

Figure~\ref{phasei3}(a) illustrates the superconducting critical temperature $T_{c}$ as a function of the parameter $r$. As expected, $T_{c}$ is substantially suppressed as $r$ decreases, which corresponds to an increase of the Coulomb screening. In the limit of $r=0$, our results reproduce the value $T_{c}=0.0098t$ at $g=0.95t$, where only the retardation effect remains (see \fig{phase}). This demonstrates a key point: the instantaneous interaction  $g^{*}(\vq)$ can significantly boost $T_{c}$, and the final value of the critical temperature is highly sensitive to the degree of Coulomb repulsion parameterized by $r$. 

The instantaneous interaction also has a dramatic effect on the pairing gap. In contrast to the results for a purely retarded interaction shown in \fig{gap}(a),  
the pairing gap is now well-characterized by the simple form of $\cos k_{x} - \cos k_{y}$, as shown in \fig{phasei3}(b) even for reasonably small instantaneous interaction at $r=0.3$. This indicates that the pairing gap is formed predominantly by a nearest-neighbor interaction in real space, a direct proof that the enhanced anisotropy of $\Delta(\vk_{F})$ observed previously was a consequence of the retarded part of the interaction.  

The quasiparticle renormalization function $Z$, on the other hand, remains  essentially the same as \fig{gap}(c) where only the retarded interaction was considered. This is because the instantaneous part of the interaction is cancelled out after the Matsubara summation in \eq{eliashberg2}, which determines $Z$. This indicates that the momentum dependence of quasiparticle renormalization is determined solely by the retarded interaction. 

While the instantaneous interaction is dominant even for a small $r$, this does not imply that the retardation effect is irrelevant. As shown In \fig{phasei3}(a), we also plot $T_{c}^{\rm ins}$ obtained by considering only the instantaneous interaction. For $r \lesssim 0.3$,  $T_{c}^{\rm ins}$  becomes negligible, yet the full calculation for $T_{c}$ retains a value a few times higher. This implies that both the retardation and instantaneous interactions work constructively to achieve a relatively high $T_{c}$, especially in the regime where the instantaneous interaction is suppressed by the screened Coulomb repulsion, i.e., for small values of $r$ in \fig{phasei3}(a). 

It should be noted that these results are consequences of the special form of $g(\vq)$ [see \eq{gq}] and not general features of the instantaneous interaction. This is because $g(\vq)$ acts as a repulsive pairing interaction for momentum transfer $\vq \approx (\pi, \pi)$ but becomes attractive for small momentum transfer $\vq \approx (0,0)$ due to its sign change. This is in contrast to the retarded interaction, which is always repulsive [see \eq{Vhat}].  Because this specific, yet realistic, instantaneous interaction has both attractive and repulsive components, it avoids the phase frustration of the pairing gap (see also \fig{intuitive}), and is therefore free from the self-restraint effect---all scattering processes work constructively, unlike the case shown in \fig{process}.

\subsubsection{Relevance to cuprates}
As we have shown in \fig{gap}(a), the pairing gap predicted from the retarded interaction deviates significantly from the simple $d$-wave form factor. This feature is not unique to the Eliashberg theory \cite{lenck94}; similar results have also been obtained from functional renormalization group studies \cite{reiss07,jwang14} and second-order renormalized perturbation theory \cite{neumayr03}. Intuitively, this can be understood by considering the spatial nature of the interaction. The simple $d$-wave pairing gap, characterized by the form $\cos k_{x} - \cos k_{y}$, is associated with a dominant pairing interaction between nearest-neighbor sites in real space.  This is consistent with our result in  \fig{phasei3}(b), where the instantaneous interaction $g(\vq) \propto \cos k_{x} + \cos k_{y}$ [see \eq{V*}]  is dominant. In contrast, the retarded spin fluctuations typically induce weaker, but non-negligible, pairing interactions beyond the nearest-neighbor sites in real space. This gives rise to higher-order harmonics in the $d$-wave symmetry in momentum space as obtained in \fig{gap}(a). 

\begin{figure}[ht]
\centering
\includegraphics[width=14cm]{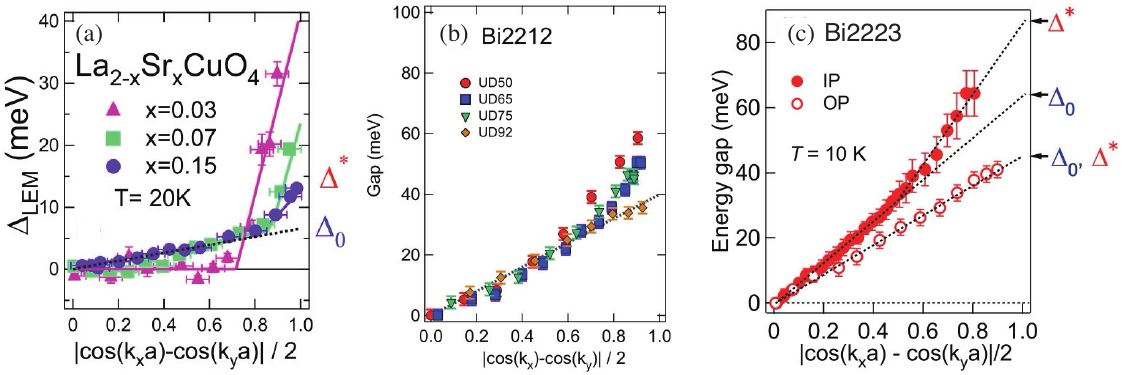}
\caption{Momentum dependences of the pairing gap in cuprate superconductors: (a) ${\rm La_{2-x}Sr_{x}CuO_{4}}$, (b) ${\rm Bi_{2}Sr_{2}CaCu_{2}O_{8+\delta}}$ (Bi2212), and (c) ${\rm Bi_{2}Sr_{2}Ca_{2}Cu_{3}O_{10+\delta}}$ (Bi2223). The simple $d$-wave pairing gap is described by the dashed line and its deviation from that is due to the pseudogap, which develops at large $| \cos k_{x} - \cos k_{y} |/2$. In (b), ``UD50'' stands for underdoped Bi2212 with $T_{c}=50$ K. The deviation from the dash line becomes pronounced at a smaller $T_{c}$, where the pseudogap develops more. In (c) ``IP'' (``OP'') stands for {\it outer} ({\it inner}) Fermi surface and ``IP'' is considered to be in an underdoped region, where the pseudogap develops. Adapted from Ref.~\cite{yoshida12} (\copyright\, 2014 The Physical Society of Japan). 
}
\label{exp-sc-gap}
\end{figure}

A crucial piece of experimental evidence that supports our findings is the measured  momentum dependence of the $d$-wave superconducting gap in cuprates. Experiments have consistently shown that the gap follows the simple form of $\cos k_{x} - \cos k_{y}$ almost perfectly \cite{ding96,sdchen22}, not only in the optimally-doped and overdoped regions but also in the underdoped region when the pseudogap effect is accounted for separately \cite{yoshida12}; see \fig{exp-sc-gap}. This implies that the instantaneous interaction $g(\vq) \propto  ( \cos q_{x} + \cos q_{y})$, which is free from the self-restraint effect, plays a crucial role in the mechanism of high $T_{c}$ superconductivity in cuprates. 

Our finding aligns with the seminal insight of Anderson \cite{anderson07}, who highlighted the vital role of the instantaneous pairing interaction rather than the retarded spin fluctuations in understanding high-$T_{c}$ cuprate superconductors. This insight holds true within our model as long as the Coulomb repulsion moderately suppresses the effect of the instantaneous pairing interaction. However, when the suppression is severe [e.g., for $r \lesssim 0.3$ in \fig{phasei3}(a)], a constructive interplay between both the retarded and instantaneous interactions is necessary to achieve a high $T_{c}$. It is important to note, however, that the present theory is based on the Eliashberg framework and thus does not properly account for strong correlation effects, except for the cancellation of the large on-site Coulomb repulsion. The application of this work to cuprates may therefore be most accurate in the overdoped region, where correlation effects are believed to be weaker.

\subsubsection{Role of screened Coulomb interaction} 
As \fig{phasei3}(a) shows, $T_{c}$ is highly sensitive to the degree of Coulomb repulsion. The effective strength of the Coulomb interaction may vary between materials, suggesting that it could be an important factor contributing to the material dependence of $T_{c}$ within the cuprate family. Since the material dependence of $T_{c}$ is frequently discussed in terms of the magnitude of $t'$ \cite{pavarini01} and the number of CuO$_{2}$ layers \cite{iyo07}, a fascinating direction for future work would be to investigate how these parameters are related to the effective Coulomb repulsion. In particular, the charge dynamics becomes very different between single-layer and multi-layer cuprates, where the effective Coulomb interaction may also be different. This difference may then  alter the degrees of screening of spin instantaneous interaction and could work positively to increase $T_{c}$ if its screening becomes weaker.

From the perspective of our theory, a distinguishing feature of cuprate superconductors is the specific form of $g(\vq)$ [see \eq{gq}], which makes it possible to achieve high $T_{c}$. The spin-fluctuation mechanism of superconductivity is believed to be a general phenomenon applicable to other systems as well  \cite{scalapino12}, but they are typically characterized by a much lower $T_{c}$ than that of cuprates. This difference may arise primarily from a different functional form of $g(\vq)$, rather than the different strength of the screening of Coulomb interaction. For example, as shown in Ref.~\cite{yamase23}, other systems may be better described by Lorentz-type or Hubbard-like interactions, which can reproduce a relatively low $T_{c}$. 

\subsubsection{A role of true LRC}
A truly LRC itself would not be effective to forming the Cooper pairs. However, it yields low-energy plasmons (plasmon band) in layered system (see \fig{plasmon2}). An important role of such plasmons to superconductivity was discussed by several authors \cite{ruvalds87,cui91,ishii93,malozovsky93,varshney95,pashitskii08,falter94,bauer09,kresin88,bill03} as mentioned in Sec.~2.1. Rather, we consider the screening of the LRC is important to the nearest-neighbor spin instantaneous interaction and thus to superconductivity.

\section{Applicability to other materials} 
While this review has focused on cuprate superconductors, the theoretical frameworks employed---specifically the large-$N$ theory and the Eliashberg formalism---possess a much broader applicability. These approaches can be extended to various other correlated materials. 

Our large-$N$ formalism for the $t$-$J$-$V$ model is particularly well-suited for studying the charge dynamics of doped Mott insulators. This approach is advantageous because it naturally separates the charge degrees of freedom, which are treated at leading order in the $1/N$ expansion, from the spin degrees of freedom, which are active at higher order. This allows for an exclusive analysis of charge dynamics and their feedback effects on electron properties. For instance, assuming that infinite-layer nickelates are  strongly correlated materials, the large-$N$ theory of the $t$-$J$-$V$ model has been successfully applied to RNiO$_{2}$ (R=La, Pr, Nd) to predict their plasmon dispersion \cite{zinni23}. The large-$N$ technique was also applied to quarter-filled layered organic molecular crystals to study the dynamical charge fluctuations as well as the one-particle spectral function \cite{merino03}. Application to irridates, manganites, and vanadates would also be an interesting direction.

The RPA, which is a fundamental technique of condensed matter theory, is a versatile first-step approach for capturing the overall feature of spin and charge dynamics and other physical properties across many material classes. 

The Eliashberg theory, on the other hand, is well-established and widely applicable framework for superconductivity itself. Its scope is not limited to cuprates but extends to a diverse range of systems where bosonic fluctuations mediate pairing. Prominent examples include the spin-fluctuation-mediated superconductivity in iron-based pnictides/chalcogenides, as well as heavy fermion materials \cite{scalapino12}. The theory is also highly relevant for understanding pairing in organic conductors, where both spin and charge fluctuations are believe to play a role \cite{mckenzie97}. The core strength of Eliashberg theory lies in its ability to self-consistently account for retardation and dissipation effects, making it a powerful tool for a variety of unconventional superconducting mechanisms beyond the simple BCS  paradigm.

\section{Perspectives}
{\it Limitations of Traditional Models and the Essential Role of the LRC}. 
Cuprate superconductors are widely understood as doped Mott insulators, a viewpoint that has led to a dominant force on the physics derived from short-range interactions, such as those described by the Hubbard and $t$-$J$ models  \cite{anderson87}. While these models have provided invaluable insights, we contend that it is premature to conclude that they capture the complete physics of the high-$T_{c}$ mechanism. The central argument of this review is that the LRC is not merely a secondary effect, but an indispensable component for accurately describing the charge dynamics of cuprate superconductors. Given that both charge and spin dynamics are fundamentally important to the high-$T_{c}$ mechanism, we hope this work inspires a crucial conceptual shift toward models that explicitly account for the LRC and nearest-neighbor Coulomb interaction.  

{\it Screened Instantaneous Spin Exchange as a Pathway to High-$T_{c}$}. 
How is the nearest-neighbor Coulomb interaction related to the high-$T_{c}$ mechanism? Our work proposes  a specific answer: it plays a vital role by mediating a screening effect on the instantaneous nearest-neighbor spin-spin interaction. In our model, this screening is phenomenologically captured by the parameter $r$, and we have shown that it can be essential for achieving a high-$T_{c}$, even in the small-$r$ regime where the repulsion from the screened Coulomb interaction is significant. This finding suggests a new perspective on the intricate balance between repulsive pairing channels, highlighting how the nearest-neighbor Coulomb interaction can constructively regulate the total interaction. 

{\it Towards Quantifying Screening in Real Materials}. 
 To quantify the screening effect in real materials, we can exploit modern experimental techniques like RIXS, which can directly probe charge dynamics. Multilayer cuprates, which generally  exhibits higher critical temperatures than their single-layer counterparts, offer a perfect test for this theory. Through an accurately  theoretical comparison of RIXS data with our large-$N$ theory of the $t$-$J$-$V$ model, we can precisely extract the dielectric constant---a direct measure of the Coulomb repulsion's screening effect. The final step would be to investigate the correlation between the material's highest $T_{c}$ and its measured dielectric constant. If a strong correlation is found, it would provide compelling evidence for our theoretical framework, a new window of the quest to understand the high-$T_{c}$ mechanism. 
 
 {\it Broader Implications for Correlated Electron Systems}. 
Beyond cuprates, the principles developed in this review have broad implications for other correlated materials. Universal concepts are the interplay between charge and spin dynamics, the crucial role of long-range interactions, and the subtle yet powerful effects of self-restraint of  superconductivity. We believe this review will be instrumental in exploring unconventional superconductivity and other exotic electronic phases in a wide range of materials where the conventional short-range paradigm may be insufficient. The future of research in this field lies in embracing these complex interactions to reveal the full picture of correlated electron systems.

\section*{Acknowledgements}
The author is deeply grateful for the invaluable contribution of numerous colleagues. The core ideas presented in this review are the result of close collaborations. Specifically, the content section~2 is based on collaboration primarily with M. Bejas and A. Greco, as well as with B. Keimer, M. Hepting, A. Nag, S. Nakata, H. Suzuki, Ke-Jin Zhou. Sections~3 and 4 are indebted to the collaboration with M. Bejas and A. Greco. The author also wishes to thank  many individuals who provided valuable insights and stimulating discussions that helped shape this work. Fruitful and intense theoretical discussions were held with M. Bejas, L. Benfatto, C. Falter, A. Greco, P. Horsch, P. Jakubczyk,  L. Manuel, W. Metzner, G. Khaliullin, M. Nieszporski, A. Ole\'s, T. Sch\"{a}fer, R. Zeyher, and L. Zinni. Moreover, the author benefited from insightful experimental discussions with N. P. Armitage,  A. V. Boris, A. Fujimori, M. Fujita, M. Hepting, M. Horio, S. Ideta, B. Keimer,  W. C. Lee, A. Nag, S. Nakata, H. Suzuki, H. Yamaguchi, and Ke-Jin Zhou.  Special thanks are extended to the Max-Planck-Institute for Solid State Research in Stuttgart for their warm hospitality. 
This work was financially supported by JSPS KAKENHI Grants No.~JP18K18744 and JP20H01856, as well as the World Premier International Research Center Initiative (WPI), MEXT, Japan.


\begin{thebibliography}{219}
\bibitem [{mis({\natexlab{a}})}]{misc-ph}%
  \BibitemOpen
  \href@noop {} {} ({\natexlab{a}}),\ \bibinfo {note} {it is possible to define
  a particle-hole transformation with $\vQ=(\pi,\pi,\pi)$. Even in this case,
  our conclusions are not modified.}\BibitemShut {Stop}%
\bibitem [{mis({\natexlab{c}})}]{misc-IP}%
  \BibitemOpen
  \href@noop {} {} ({\natexlab{c}}),\ \bibinfo {note} {there can be a different
  IP in that Im$\Sigma_{\rm ch}$ exhibits a peak at $\omega=0$ and
  Re$\Sigma_{\rm ch}$ has a positive slope there, as actually obtained in a
  theoretical study of electronic nematic fluctuations \cite{yamase12}.
  However, we do not find this kind of IP in \fig{incoherent}(a).}\BibitemShut
  {Stop}%
\end{thebibliography}


%

\appendix

\section{Modeling of the pseudogap}
Our phenomenological model for the pseudogap, as represented by \eq{selfPG}  in the main text, is based on the framework of Ref.~\cite{norman07}. It can be considered a simplified yet powerful approach that is versatile enough to capture the essential features of the pseudogap across different microscopic scenarios. The self-energy we employ to model the pseudogap's influence on the electronic spectrum is given by 
\be
\Sigma_{\rm pg}(\vk, \omega) = \frac{c_{\vk}^{2}} {\omega + \tilde\varepsilon_{\vk} + i \Gamma} \,.
\ee
This form of the self-energy can be justified from a number of distinct physical origins. 

In the case of a commensurate density-wave state such as spin- or charge-density-wave with a nesting momentum $\vQ=(\pi, \pi)$, the parameter $c_{\vk}$ corresponds to the magnitude of the density-wave gap. In this scenario, the energy dispersion $\tilde\varepsilon_{\vk}$ is given by 
\be
\tilde\varepsilon_{\vk}= - \varepsilon_{\vk+ \vQ} \,.
\ee

Alternatively, this form can also be connected to the phenomenological Yang-Rice-Zhang (YRZ) model \cite{yang06}. In this framework, the parameter  $c_{\vk}$ directly controls the magnitude of the pseudogap, and the dispersion $\tilde\varepsilon_{\vk}$ is given by the nearest-neighbor term of the tight-binding dispersion 
\be
\tilde\varepsilon_{\vk}= - 2 t (\cos k_{x} + \cos k_{y}) \,. 
\ee

A third possibility is that the pseudogap is a consequence of strong short-range $d$-wave pairing fluctuations. In this case,  $c_{\vk}$ is the usual $d$-wave pairing gap and $\tilde\varepsilon_{\vk}$ is simply given by the electron dispersion itself,  
\be
\tilde\varepsilon_{\vk}= \varepsilon_{\vk} \,.
\ee

The key physical consequence of this self-energy form is the existence of a Luttinger surface, defined by the condition $\tilde\varepsilon_{\vk}=0$. On this surface, the self-energy diverges as $\omega \rightarrow 0$ and $\Gamma \rightarrow +0$. Consequently, the spectral function is expected to be strongly suppressed at zero energy wherever the Fermi surface intersects the Luttinger surface. In hole-doped cuprates, this condition is typically met near antinodal regions, when the Fermi surface passes close to the magnetic zone boundary. This consideration naturally leads to the simplified form presented in \eq{selfPG}, particularly after allowing for a momentum dependence of the scattering rate $\Gamma$.

While our simplified model has successfully reproduced experimental data (see \fig{PG-fit}), this does not necessarily imply that the pseudogap should be explained by one of the three specific scenarios mentioned above. This is because the functional from of our self-energy $\Sigma_{\rm pg}(\vk, \omega)$  is remarkably general. Similar forms can be obtained in other microscopic scenarios. In this sense, our model serves as a universal phenomenological tool to capture the core physics of the pseudogap, independent of its specific microscopic origin. 

\end{document}